\documentclass[11pt]{article}
\usepackage[a4paper,left=2.2cm,right=2.2cm,top=2.5cm,bottom=2.5cm]{geometry}
\usepackage{indentfirst}
\usepackage[numbers]{natbib}
\usepackage{amsmath,amsthm,amssymb,gensymb}
\usepackage{tikz,tikz-feynman,subfigure}
\usepackage[export]{adjustbox}
\usepackage{hyperref}
\usepackage[normalem]{ulem}
\usepackage{xcolor}

\newcommand{\email}[1]{\footnote{{\em } \texttt{#1}}}

\newcommand{\ChPT}{\rm\chi PT}
\newcommand{\mL}{\mathcal{L}}
\newcommand{\mO}{\mathcal{O}}
\newcommand{\M}{\mathcal{M}}

\newcommand{\GeV}{\mathrm{GeV}}
\newcommand{\MeV}{\mathrm{MeV}}

\newcommand{\ct}{c_{\theta}}
\newcommand{\st}{s_{\theta}}
\newcommand{\cALP}{c_{\rm ALP}}

\newcommand{\LO}{\mathrm{LO}}
\newcommand{\tree}{\mathrm{tree}}
\newcommand{\uni}{\mathrm{uni}}

\newcommand{\asc}{a_\mathrm{sc}}

\allowdisplaybreaks

\begin{document}

\title{\Large \bf Axion-like particle production from kaon decays within U(3) chiral theory }
\author{\small Jin-Bao Wang$^a$\email{wangjinbao@hebtu.edu.cn},\, Zhi-Hui Guo$^{a}$\email{zhguo@hebtu.edu.cn},\, Hai-Qing Zhou$^b$\email{zhouhq@seu.edu.cn} \\[0.5em]
{ \small\it ${}^a$ Department of Physics and Hebei Key Laboratory of Photophysics Research and Application, } \\ 
{\small\it Hebei Normal University,  Shijiazhuang 050024, China} \\ [0.3em]
{ \small\it ${}^b$ School of Physics, Southeast University, Nanjing 211189, China }
}
\date{}

\maketitle

\begin{abstract}
Kaon decays provide important constraints on the axion-like particle (ALP), particularly through the $K\to\pi a$ processes. In this work, we compute the $K\to\pi a$ decay amplitudes, as well as the $K\to\pi\pi$ decay amplitudes, at leading order within U(3) chiral perturbation theory. The $\pi\pi$ final-state interaction effects in both processes are implemented by employing the chiral unitarization approach. The relevant weak low-energy constants are determined by fitting to the experimental data on $K\to\pi\pi$ decays. Our numerical analysis shows that the explicit $\eta_0$ contribution to the $K\to\pi a$ amplitudes is small, whereas the new weak operator unique to the U(3) theory can give a sizable contribution. The $\pi\pi$ final-state interactions are found to be insignificant in the $K^\pm\to\pi^\pm a$ decays, but give pronounced effects in the $K^0/\bar{K}^0\to\pi^0a$ channels. Constraints on the ALP couplings are derived from the most recent NA62 upper limit on the $K^+\to\pi^+ +\,invisible$ decay. For completeness, we also include constraints from the KOTO search for $K_L\to\pi^0+\, invisible$, the NA62 measurement of $K^+\to\pi^+\gamma\gamma$, and the NA48 measurement of $K_S\to\pi^0\gamma\gamma$. 
\end{abstract}

\section{Introduction}

Axion provides an elegant solution to the strong CP problem via Peccei-Quinn (PQ) mechanism~\cite{Peccei:1977hh,Peccei:1977ur} and is also a compelling candidate for the dark matter. 
In the PQ mechanism, axion originates from the spontaneous breaking of a global $\rm U(1)_{PQ}$ symmetry at a high scale $\Lambda\equiv 4\pi f_a$, with $f_a$ the axion decay constant. In the conventional QCD axion scenario, the mass of axion is fixed by $f_a$ through $m_{a,{\rm QCD}}=m_\pi F_\pi/(2f_a)$. Additionally, the axion-like particle (ALP) represents a general pseudoscalar state that arises in well-motivated extension of the QCD axion case, with the ALP mass disentangled from its decay constant $f_a$. For simplicity, we will collectively designate both ALP and axion as ALP throughout this work. 

Below the high scale $\Lambda$, ALP generally couples to the quarks, leptons, and gauge bosons of SM through dimension-5 or higher operators suppressed by $1/\Lambda^n$, with $n\geq 1$. Due to their pseudo-Nambu-Goldstone boson (pNGB) nature, the ALP mass $m_a$ can be much smaller than the symmetry-breaking scale, making them typical weakly interacting light particles. 
The search for ALPs spans astrophysics~\cite{Raffelt:1990yz,Payez:2014xsa,Jaeckel:2017tud}, cosmology~\cite{Chang:1993gm,DEramo:2022nvb,Jung:2025dyo}, and particle physics~\cite{Mimasu:2014nea,Brivio:2017ije,Li:2025ski}. 
In the absence of clear signals of new physics at the multi-TeV scale at the LHC, increasing attention has been directed toward the intensity frontier, such as NA62~\cite{NA62:2025upx}, KOTO~\cite{KOTO:2024zbl}, BESIII~\cite{BESIII:2022mxl}, Belle~II~\cite{Belle-II:2018jsg}, Continuous Electron Beam Accelerator Facility (CEBAF)~\cite{Arrington:2021alx}, Super Tau-charm Facility~\cite{Petrov:2026nxj}, etc, where ALPs may be abundantly produced. 

In particular, flavor-changing rare meson decays can provide important constraints on the ALP parameter space from MeV scale up to a few GeV. Especially, for $m_a\lesssim300$~MeV, the decay $K^{\pm}\to\pi^{\pm} a$ offers the most stringent constraints from particle physics studies ~\cite{Bauer:2021wjo,Bauer:2021mvw}, benefiting from the high precision of experiments~\cite{NA62:2025upx} and the theoretically clean SM prediction~\cite{Cirigliano:2011ny} for the $K\to\pi\nu\bar{\nu}$ decay. In the pioneering work~\cite{Georgi:1986df}, the branching ratio of $K^{\pm}\to\pi^{\pm}a$ decay was calculated. However, the ALP terms in the weak effective Lagrangian were not fully incorporated in the former work, leading to a residual dependence of the amplitude on the unphysical parameters of the chiral rotation used to eliminate the ALP-gluon coupling. Many works ~\cite{Bjorkeroth:2018dzu,Ertas:2020xcc,Gori:2020xvq} followed these kinds of $K\to\pi a$ amplitudes to carry out various phenomenological analyses. This issue has recently been resolved in Ref.~\cite{Bauer:2021wjo}, where the correct representation of the flavor-changing quark currents is implemented in SU(3) chiral perturbation theory ($\chi$PT)~\cite{Gasser:1984gg}, yielding the proper decay amplitudes that are independent of the chiral rotation scheme. Importantly, it is pointed out in Ref.~\cite{Bauer:2021wjo} that the correct chiral amplitude after including the previously omitted contributions enhances the $K^{\pm}\to\pi^{\pm}a$ branching ratio by nearly a factor of 40,  compared to earlier estimates. Given the importance of $K^{\pm}\to\pi^{\pm}a$ decay, this process is then further computed in Ref.~\cite{Cornella:2023kjq} by incorporating the complete next-to-leading order (NLO) corrections in SU(3) $\ChPT$, including the novel operators that arise in the presence of the ALP. The resulting corrections could be sizable, but they are obscured by large uncertainties, which are primarily driven by our incomplete knowledge of the relevant low-energy constants (LECs).

So far, the study of the $K\to \pi a$ decay has been confined to the SU(3) chiral framework, which does not explicitly include the singlet $\eta_0$ (main component of the $\eta'$) but instead integrates it out, encoding its effects into the LECs. Recently, it has been demonstrated in Refs.~\cite{Gao:2022xqz,Gao:2024vkw} that the axion mixing with $\eta_8$ and $\eta_0$ can be even more important than the $a$-$\pi^0$ mixing in the U(3) $\ChPT$~\cite{Herrera-Siklody:1996tqr,Kaiser:2000gs}, which explicitly incorporates $\eta_0$ as the ninth pNGB through the large-$N_C$ expansion. Therefore, it is quite interesting to investigate to what extent the $a$-$\eta_0$ mixing can affect the $K\to\pi a$ amplitude, apart from the $a$-$\eta_8$ and $a$-$\pi^0$ ones that appear in the SU(3) case. Moreover, a consistent matching to QCD in the large-$N_C$ limit, as often employed in the meson weak decays, requires the explicit inclusion of the $\eta_0$~\cite{Peris:1994dh,Herrera-Siklody:1996tqr,Kaiser:2000gs}. These considerations can be systematically addressed within the framework of U(3) $\ChPT$. In addition, new weak effective operators arise in U(3) $\ChPT$~\cite{Gerard:2005yk,Gerard:2012ud}, and whether such new operators could give rise to sizable contributions to the $K\to\pi a$ decay remains unexplored. In this work, we will calculate the $K\to\pi a$ decay at leading order (LO) in U(3) $\ChPT$ and quantify the contributions from both the $\eta_0$ and the new U(3) operators. With the new U(3) $K\to\pi a$ amplitudes, we further determine the ALP couplings by taking the experimental information from the $K^+\to\pi^++\,{ invisible}$ decay recently measured by NA62~\cite{NA62:2025upx}, as well as the $K_L\to\pi^0+\,{ invisible}$ measurement from KOTO~\cite{KOTO:2024zbl}, the measurement of $K^+\to\pi^+\gamma\gamma$ and $K_S\to\pi^0\gamma\gamma$ by NA62~\cite{NA62:2023olg} and NA48~\cite{NA48:2003ydp}, respectively.

The rest of the paper is organized as follows. In Sec.~\ref{sec.lag}, the relevant effective Lagrangians, including the QCD and weak U(3) chiral Lagrangians, and the $a$-$\pi^0$-$\eta$-$\eta'$ mixing, are discussed in detail. We proceed to calculate the $K\to\pi a$ and $K\to\pi\pi$ amplitudes in Sec.~\ref{sec.amp}, where the unitarization is implemented to account for the $\pi\pi$ final-state interactions. Phenomenological studies are given in Sec.~\ref{sec.pheno}, which include the determination of the weak chiral couplings from the $K\to\pi\pi$ decays, the comparison of the SU(3) and U(3) results and the derivation of the ALP bounds from the $K\to\pi a$ decays. We give a short summary and conclusions in Sec.~\ref{sec.sum}. The discussions about the $N_C$ orders of the $K\to\pi a$ amplitudes and the $a\to\gamma\gamma$ decays are relegated to the Appendices.

\section{Effective Lagrangians}\label{sec.lag}

\subsection{The effective Lagrangians at quark level}

The dynamics of the $K\to \pi a$ decay is governed by the QCD, electroweak (EW) and ALP interactions simultaneously. Consequently, at the quark level (QL) the relevant effective Lagrangians to the $K\to \pi a$ decay contain three parts
\begin{equation}
\begin{aligned}
\mL_{\rm QL}^{K\to\pi a}= \mL_{\rm QCD}+\mL_{\rm EW\text{-}eff}^{|\Delta S| = 1} +\mL_{\rm ALP}  \,,\label{eq_L_tot}
\end{aligned}
\end{equation}
where $\mL_{\rm QCD}$ is the standard QCD Lagrangian, $\mL_{\rm EW\text{-}eff}^{|\Delta S| = 1}$ denotes the effective EW four-fermion interactions with strangeness change $|\Delta S|=1$ and $\mL_{\rm ALP}$ stands for the ALP effective Lagrangian describing the ALP-SM particles interactions. 
Below the chiral symmetry breaking scale of QCD ($\Lambda_\chi$), the general ALP effective Lagrangian, including the free ALP part and ALP-SM couplings up to dimension-5, is given by~\cite{Georgi:1986df,Bauer:2020jbp,Bauer:2021wjo,Bauer:2021mvw} 
\begin{equation}
\begin{aligned}
\mL_{\rm ALP}^{D\leq 5}=&\frac{1}{2}\partial_{\mu}a\partial^{\mu}a - \frac{1}{2}m_{a,0}^2a^2 
+\frac{\partial_{\mu}a}{f}\left(\bar{q}_Lk_Q\gamma^{\mu}q_L + \bar{q}_R k_q\gamma^{\mu}q_R\right)
-c_{GG}\frac{\alpha_s}{4\pi}\frac{a}{f}G_{\mu\nu}^a\widetilde{G}^{\mu\nu,a}-c_{\gamma\gamma}\frac{\alpha}{4\pi}\frac{a}{f}F_{\mu\nu}\widetilde{F}^{\mu\nu}\,,\label{eq_L_ALP}
\end{aligned}
\end{equation}
where $\widetilde{G}^{\mu\nu,a}\equiv\frac{1}{2}\varepsilon^{\mu\nu\rho\sigma}G_{\rho\sigma}^a$ and $\widetilde{F}^{\mu\nu}\equiv\frac{1}{2}\varepsilon^{\mu\nu\rho\sigma}F_{\rho\sigma}$ $(\varepsilon_{0123}=+1)$ denote the dual field-strength tensors of gluons and photons, respectively. For the kaon decays, the active quark flavors are $u$, $d$, and $s$, collected in the column vector $q = (u,d,s)^T$. The left- and right-handed components of the quark fields are defined as $q_{R/L} = P_{R/L}q$ with $P_{R/L} = \frac{1\pm \gamma_5}{2}$ and $ \gamma_5 \equiv i\gamma^0\gamma^1\gamma^2\gamma^3$. 
The $3\times3$ Hermitian matrices $k_Q$ and $k_q$ in Eq.~\eqref{eq_L_ALP} can be written as  
\begin{equation}
k_Q = \begin{pmatrix}[k_U]_{11}&&\\&[k_D]_{11}&[k_D]_{12}\\&[k_D]_{21}&[k_D]_{22}\end{pmatrix}\,,
\quad
k_q = \begin{pmatrix}[k_u]_{11}&&\\&[k_d]_{11}&[k_d]_{12}\\&[k_d]_{21}&[k_d]_{22}\end{pmatrix}\,,\label{eq_kQ_and_kq}
\end{equation}
which correspond to the left- and right-handed quark flavor spaces, respectively, and encode the flavor structures of the ALP-quark couplings. For the diagonal ALP-quark couplings, it is convenient to reshuffle them in the vector and axial-vector bases as  
\begin{equation}
c_{uu}^{v/a} = [k_u]_{11} \pm [k_U]_{11}\,,
\quad
c_{dd}^{v/a} = [k_d]_{11} \pm [k_D]_{11}\,,
\quad
c_{ss}^{v/a} = [k_d]_{22} \pm [k_D]_{22}\,.
\end{equation}
We do not consider the ALP-lepton interactions in the $K\to\pi a$ decay, and therefore they are not shown explicitly in Eq.~(\ref{eq_L_ALP}). The ALP field approximately respects a shift symmetry $a\to a+ {\rm const.}$, which is softly broken by the ALP-gluon coupling and the bare mass term $m_{a,0}$. The latter provides another contribution to the ALP mass $m_a$ in addition to the QCD anomaly, thereby decoupling $m_a$ from the scale $f$. It is noted that the relation between the parameter $f$ in Eq.~\eqref{eq_L_ALP} and the commonly used ALP decay constant $f_a$~\cite{DiLuzio:2020wdo} is $f=-2c_{GG}f_a$. 

The ALP-gauge boson couplings are scale invariant up to two-loop level, whereas the ALP-fermion couplings run under the SM renormalization group (RG) evolution~\cite{Bauer:2020jbp,Bauer:2021mvw}. 
This implies that even if the ALP couplings are flavor conserving at the UV scale when $\Lambda>\mu_w$, with $\mu_w$ the electroweak scale, nonzero $[k_D]_{12}=[k_D]_{21}^*$, corresponding to $s\leftrightarrow d$ flavor transitions, can be generated radiatively below $\mu_w$~\cite{Bauer:2020jbp,Bauer:2021mvw}.
In Eq.~(\ref{eq_L_ALP}), the matrices $k_Q$ and $k_q$ are defined at the QCD chiral symmetry breaking scale, which encode the dynamics both at the UV model and the RG evolution. The off-diagonal matrix elements of $k_Q$ and $k_q$ directly give the contact interacting vertices for the $K\to\pi a$ decays. While, the diagonal matrix elements of $k_Q$ and $k_q$ can only enter the $K\to\pi a$ processes after taking into account the $s\leftrightarrow d$ transition caused by the SM weak interaction. 

At the scale of $\Lambda_\chi$, the $|\Delta S| = 1$ nonleptonic SM weak interaction is described by a set of four-fermion operators, with the effective Lagrangian given by~\cite{Pich:1990mw}
\begin{equation}
\mL_{\rm EW-eff}^{|\Delta S| = 1}=-\frac{G_F}{\sqrt{2}}V_{ud}^*V_{us}\sum_{i=1}^6c_i(\mu_0)\hat{Q}_i+\mathrm{h.c.}\,,
\label{eq_L_weak_quark}
\end{equation}
where $G_F$ is the Fermi constant, $V_{ij}$ are the Cabibbo-Kobayashi-Maskawa (CKM) matrix elements and the explicit four-fermion operators read 
\begin{equation}
\begin{array}{lll}
\hat{Q}_1 = 4\left(\bar{d}_L\gamma^{\mu}s_L\right)\left(\bar{u}_L\gamma_{\mu}u_L\right)\,,
&&
\hat{Q}_2 = 4\left(\bar{d}_L\gamma^{\mu}u_L\right)\left(\bar{u}_L\gamma_{\mu}s_L\right)\,,
\\
\hat{Q}_3 = 4\left(\bar{d}_L\gamma^{\mu}s_L\right)\left(\bar{q}_L\gamma_{\mu}q_L\right)\,,
&&
\hat{Q}_4 = 4\left(\bar{d}_L\gamma^{\mu}q_L\right)\left(\bar{q}_L\gamma_{\mu}s_L\right)\,,
\\
\hat{Q}_5 = 4\left(\bar{d}_L\gamma^{\mu}s_L\right)\left(\bar{q}_R\gamma_{\mu}q_R\right)\,,
&&
\hat{Q}_6 = -8\left(\bar{d}_Lq_R\right)\left(\bar{q}_Rs_L\right)\,.\label{eq_Q_1-6}
\end{array}
\end{equation}
Additional operators induced by electroweak penguin effects are neglected in this work.
In the strict large-$N_C$ limit, the Wilson coefficients are found to be $c_2 = 1$ and $c_{1,3\text{--}6}=0$~\cite{Pich:1990mw}.

Due to the nonperturbative nature of QCD in the low-energy regime, the effective Lagrangians in Eq.~\eqref{eq_L_tot} can not be directly used for the computation of the $K\to\pi a$ decay. Instead, they will be matched onto the U(3) $\ChPT$, which enables us to perform the calculation at the hadron level.

\subsection{QCD U(3) chiral Lagrangian}

According to the large-$N_C$ reasoning, the U$(1)_A$ anomaly of QCD is $1/N_C$ suppressed when $N_C\rightarrow \infty$. Since the QCD U$(1)_A$ anomaly is responsible for the large mass of the singlet $\eta_0$ at $N_C=3$, this immediately implies that the massive $\eta_0$ will become Goldstone boson in the chiral and large-$N_C$ limits. 
As a result, the singlet $\eta_0$, together with the octet $\pi, K$ and $\eta_8$, can be included in the U(3) $\ChPT$ as a nonet through 
\begin{equation}
U(x)=\exp\left(\frac{i\phi(x)}{F}\right)\,,
\end{equation}
with  
\begin{equation}
\phi(x)=\sum_{i=0}^8\lambda_i\phi_i(x)=\begin{pmatrix}
\pi^3+\frac{1}{\sqrt{3}}\eta_8+\sqrt{\frac{2}{3}}\eta_0&\sqrt{2}\pi^+&\sqrt{2}K^+\\
\sqrt{2}\pi^-&-\pi^3+\frac{1}{\sqrt{3}}\eta_8+\sqrt{\frac{2}{3}}\eta_0&\sqrt{2}K^0\\
\sqrt{2}K^-&\sqrt{2}\bar{K}^0&-\frac{2}{\sqrt{3}}\eta_8+\sqrt{\frac{2}{3}}\eta_0
\end{pmatrix}\,.
\end{equation}
The $\delta$ counting scheme that employs the simultaneous expansions of momentum, light-quark masses and $1/N_C$ is typically introduced in U(3) $\ChPT$, in order to systematically calculate the amplitudes involving $\pi, K, \eta_8$ and $\eta_0$ order by order~\cite{Herrera-Siklody:1996tqr,Kaiser:2000gs}. The quantity $F$ corresponds to the meson decay constant at LO in the $\delta$ counting. 

The explicit symmetry breaking effects via the quark masses and $\mathrm{U}(1)_A$ anomaly, together with EW currents, can be conveniently incorporated into the chiral effective Lagrangians through the external source method~\cite{Gasser:1984gg}, organized order by order in the $\delta$ counting scheme. 
In the U(3) framework, the QCD Lagrangian with external sources takes the form~\cite{Kaiser:2000gs}
\begin{equation} 
\begin{aligned}
\mL_{\rm QCD}^{\rm ext}=&
\mL_{\rm QCD}^0
+\bar{q}\gamma^{\mu}(v_{\mu}+a_{\mu}\gamma_5)-\bar{q}(s-i\gamma_5p)q-\theta\omega
\\=&
\mL_{\rm QCD}^0
+\bar{q}_R\gamma^{\mu}r_{\mu}q_R+\bar{q}_L\gamma^{\mu}\ell_{\mu}q_L-\bar{q}_R(s+ip)q_L-\bar{q}_L(s-ip)q_R-\theta\omega\,,\label{eq_L_QCD_ext}
\end{aligned}
\end{equation}
where $\mL_{\rm QCD}^0$ denotes the QCD Lagrangian in the chiral limit, and $\omega(x)=\frac{\alpha_s}{8\pi}G_{\mu\nu}^a\widetilde{G}^{\mu\nu,a}$ is the gluon topological charge density, coupled to the source $\theta(x)$. The external source fields $v_{\mu}(x)$, $a_{\mu}(x)$, $s(x)$, and $p(x)$ are $3\times 3$ Hermitian matrices, with $r_\mu=v_\mu+a_\mu$ and $\ell_\mu=v_\mu-a_\mu$. 
The U(3) chiral Lagrangian is constructed from the meson fields $\pi$, $K$, $\eta_8$, $\eta_0$ and the external fields introduced above, constrained by the local $G_{\chi}=\mathrm{U(3)}_L\otimes\mathrm{U(3)}_R$ symmetry together with the discrete P and CP symmetries. 
Explicit chiral symmetry breaking due to the light quark masses $m_q$ is introduced by taking the scalar source $s(x)=\mathrm{diag}(m_u,m_d,m_s)$, while the U$(1)_A$ anomaly is incorporated by a nonvanishing topological susceptibility, viz. the two-point correlation of $\omega$. These effects generate nonzero masses for the $\pi$, $K$, $\eta_8$ and $\eta_0$, rendering them pNGBs.

U(3) $\ChPT$ provides a suitable framework for the cases with nonzero $\langle a_{\mu} \rangle$ and $\theta$, where $\langle\cdots\rangle=\mathrm{Tr}(\cdots)$ denotes the trace in the flavor space. Therefore, it is not necessary to eliminate the ALP-gluon coupling by means of a chiral rotation of the quark fields, as done in the SU(2)/SU(3) cases~\cite{Georgi:1986df,DiLuzio:2020wdo,Bauer:2021wjo}. Instead, the ALP-gluon coupling can be introduced directly through effective operators involving $\omega$. Nevertheless, following Ref.~\cite{Bauer:2021wjo}, we still perform a field redefinition of the quark fields
\begin{equation}
q(x)\to \exp\left(-i\kappa_q\gamma_5c_{GG}\frac{a(x)}{f}\right)q(x)\,,
\label{eq_redef_quark}
\end{equation}
where $\kappa_q$ is chosen, for simplicity, to be a real diagonal matrix, $\kappa_q=\mathrm{diag}\left(\kappa_u,\kappa_d,\kappa_s\right)$. Applying this redefinition to Eq.~(\ref{eq_L_tot}), one obtains
\begin{equation}
\begin{aligned}
\mL_{\rm QCD} + \mL_{\rm ALP}=&\mL_{\rm QCD}^0+\frac{1}{2}\partial_{\mu}a\partial^{\mu}a-\frac{1}{2}m_{a,0}^2a^2+\frac{\partial_{\mu}a}{f}\bar{q}_R\gamma^{\mu}\hat{k}_q(a) q_R + \frac{\partial_{\mu}a}{f}\bar{q}_L\gamma^{\mu}\hat{k}_Q(a)q_L
\\
&-\bar{q}_R \hat{M}(a) q_L - \bar{q}_L\hat{M}^{\dagger}(a)q_R-2\hat{c}_{GG}\frac{a}{f}\omega-\hat{c}_{\gamma\gamma}\frac{\alpha}{4\pi}\frac{a}{f}F_{\mu\nu}\widetilde{F}^{\mu\nu}\,,\label{eq_L_QCD_ALP_kappaq}
\end{aligned}
\end{equation}
where
\begin{subequations}
	\begin{align}
	\hat{k}_q(a)&=e^{i\kappa_q c_{GG}\frac{a}{f}}k_qe^{-i\kappa_qc_{GG}\frac{a}{f}}+c_{GG}\kappa_q\,,
	\\
	\hat{k}_Q(a)&=e^{-i\kappa_qc_{GG}\frac{a}{f}}k_Qe^{i\kappa_q c_{GG}\frac{a}{f}}-c_{GG}\kappa_q\,,
	\\
	\hat{M}(a)&=Me^{2i\kappa_qc_{GG}\frac{a}{f}}\,,
	\\
	\hat{c}_{GG}&=\left[1-\mathrm{Tr}(\kappa_q)\right]c_{GG}\,,
	\\
	\hat{c}_{\gamma\gamma}&=c_{\gamma\gamma}-2N_C\mathrm{Tr}\left(\kappa_qQ_q^2\right)c_{GG}\,.
	\end{align}
\end{subequations}
Here, $M=\mathrm{diag}\left(m_u,m_d,m_s\right)$ and $Q_q=\mathrm{diag}\left(\frac{2}{3},-\frac{1}{3},-\frac{1}{3}\right)$ are the diagonal quark mass matrix and the quark electric charge matrix, respectively. It is pointed out that the quark redefinition in Eq.~\eqref{eq_redef_quark} is not intended to eliminate the ALP-gluon coupling here, therefore $\langle \kappa_q \rangle$ is not required to be unity. Furthermore, owing to the arbitrariness of the parameters in $\kappa_q$, Eq.~(\ref{eq_L_QCD_ALP_kappaq}) represents an infinite set of physically equivalent forms. Upon matching Eq.~(\ref{eq_L_QCD_ALP_kappaq}) onto U(3) $\ChPT$ and computing physical amplitudes, the final results must be independent of the parameters in $\kappa_q$, as required by the reparameterization invariance of physical observables. This requirement provides a nontrivial check of our calculation.

Comparing Eq.~(\ref{eq_L_QCD_ALP_kappaq}) with Eq.~(\ref{eq_L_QCD_ext}), the ALP dressed external sources are identified as
\begin{equation}\label{eq.alpsource}
r_{\mu}=\frac{\partial_{\mu}a}{f}\hat{k}_q(a)\,,
\quad
\ell_{\mu}=\frac{\partial_{\mu}a}{f}\hat{k}_Q(a)\,,
\quad
s+ip=\hat{M}(a)\,,
\quad
\theta=2\hat{c}_{GG}\frac{a}{f}\,.
\end{equation}
The LO QCD chiral Lagrangian in U(3) $\ChPT$ is given by
\begin{equation}
\mL_{\rm eff}^{(0)}(\kappa_q)=\frac{F^2}{4}\left\langle D_{\mu}UD^{\mu}U^{\dagger} + \chi U^{\dagger}+U\chi^{\dagger} \right\rangle-\frac{F^2M_0^2}{12}X^2\,,\label{eq_L_eff_0}
\end{equation}
where
\begin{subequations}
	\begin{align}
	&D_{\mu}U=\partial_{\mu}U-ir_{\mu}U+iU\ell_{\mu}\,,\quad
	D_{\mu}U^{\dagger}=\partial_{\mu}U^{\dagger}-i\ell_{\mu}U^{\dagger}+iU^{\dagger}r_{\mu}\,,
	\\
	&\chi=2B_0(s+ip)\,,
	\\
	&X=\psi+\theta\,,\quad
	\psi\equiv-i\mathrm{Tr}\log U=\frac{\sqrt{6}\eta_0}{F}\,.
	\end{align}
\end{subequations}
Compared with the SU(3) case, a new LEC $M_0$ appears at LO, which represents the $\mathrm{U}(1)_A$ anomaly contribution to the $\eta_0$ mass. The singlet field $\psi$ is affected only by the $\mathrm{U(1)}_A$ component of $G_{\chi}$, and $X$ is invariant under the $G_\chi$ chiral transformation provided that $\theta$ transforms as $\theta \to \theta + i \mathrm{Tr}\log V_R V_L^{\dagger}$~\cite{Kaiser:2000gs}, with $(V_{L},V_R)\in G_{\chi}$. However, after promoting $\theta$ as the ALP field $a$, $X$ will introduce explicit symmetry breaking associated with the $\mathrm{U}(1)_A$ anomaly, since the ALP field does not transform under chiral rotations. This procedure is similar as the situation when taking the scalar external source $s(x)={\rm diag}(m_u,m_d,m_s)$ that explicitly breaks the chiral symmetry. The $N_C$ scaling of the LO LECs is given by $F\sim\mO(N_C^{1/2})$, $B_0\sim\mO(N_C^0)$, and $M_0^2\sim\mO(N_C^{-1})$~\cite{Kaiser:2000gs}. 

In the U(3) $\delta$ counting scheme, the expansions of momentum, light-quark masses, and $N_C^{-1}$ are uniformly denoted by a single expansion parameter as $p^2\sim\mO(\delta)\,, m_{u,d,s}\sim\mO(\delta)\,, N_C^{-1}\sim\mO(\delta)$. 
Accordingly, the LO QCD U(3) chiral Lagrangian $\mL_{\rm eff}^{(0)}$ is of $\mO(\delta^0)$, as indicated by the superscript ``(0)''. The ALP mass $m_a$ also enters the amplitudes, and it is reasonable to assign the same $\delta$ counting for $m_a^2\sim \mO(\delta)$ as the pNGB mass squared, since the $K\to \pi a$ is focused in our study and the value of $m_a$ then lies in the range between 0 and $m_K-m_\pi$.

\subsection{Weak U(3) chiral Lagrangian}

The effective $\ChPT$ realization of the four-fermion operators $\hat{Q}_i$ in Eq.~(\ref{eq_Q_1-6}) must reproduce the same flavor structure as the original operators, which can be decomposed into $(8_L,1_R)$ and $(27_L,1_R)$ representations of the chiral group $G_{\chi}$~\cite{Pich:1990mw}. Moreover, the effective operators are required to satisfy the so-called CPS symmetry~\cite{Kambor:1989tz}: CP-even (odd) operators are even (odd) under the  $s\leftrightarrow d$ permutation. The effective chiral Lagrangian describing the $|\Delta S|=1$ nonleptonic weak interactions has been constructed up to $\mO(p^4)$ in SU(3) $\ChPT$~\cite{Kambor:1989tz,Ecker:1992de}, while in the U(3) case, to our knowledge, only the LO studies exist~\cite{Gerard:2005yk,Gerard:2012ud}. For completeness, we briefly review below the LO U(3) chiral effective Lagrangian for $|\Delta S|=1$ nonleptonic weak interactions. 

Up to $\mO(p^2)$, all the Hermitian and Lorentz-invariant chiral structures transforming as $O_L\to V_L O_L V_L^{\dagger}$ under $G_{\chi}$ are given by
\begin{equation}
X^nL_{\mu}L^{\mu}\,,
\quad
X^nD_{\mu}\psi L^{\mu}\,,
\quad
X^nD_{\mu}\theta L^{\mu}\,,
\quad
X^n\left(\chi^{\dagger}U+U^{\dagger}\chi\right)\,,
\quad
X^ni\left(\chi^{\dagger}U-U^{\dagger}\chi\right)\,,\label{eq_O_L}
\end{equation}
where
\begin{equation}
L_{\mu}\equiv iU^{\dagger}D_{\mu}U\,,
\quad
D_{\mu}\psi \equiv -\langle L_{\mu} \rangle=\partial_{\mu}\psi - \langle r_{\mu} - \ell_{\mu} \rangle\,,
\quad
D_{\mu}\theta \equiv \partial_{\mu}\theta + \langle r_{\mu} - \ell_{\mu} \rangle\,.
\end{equation}
Projecting these structures onto $\lambda_{6}$ and $\lambda_7$ (the standard Gell-Mann matrices), one obtains the operators describing the $|\Delta S|=1$ transitions. To ensure the CPS symmetry, the structures in Eq.~(\ref{eq_O_L}) need to be CP-even~\cite{Kambor:1989tz}, which requires the power $n$ of $X$ in the last term of Eq.~(\ref{eq_O_L}) to be odd, while $n$ in the other terms must be even. Because the operators with one additional power of $X$ are suppressed by the factor of $N_C^{-1}$, we retain only the lowest power of $X$ and the $(8_L,1_R)$ part of the weak chiral effective Lagrangian takes the form 
\begin{subequations}
\begin{align}
&\mL_{(8_L,1_R)}=\mL_{(8_L,1_R)_{+}}+\mL_{(8_L,1_R)_{-}}\,,
\\
&\begin{aligned}
\mL_{(8_L,1_R)_{\pm}}=&c_1^{\pm}\left\langle\lambda_{6/7}L_{\mu}L^{\mu}\right\rangle
+c_2^{\pm}\left\langle\lambda_{6/7}\left(\chi^{\dagger}U+U^{\dagger}\chi\right)\right\rangle
+c_3^{\pm}Xi\left\langle\lambda_{6/7}\left(\chi^{\dagger}U-U^{\dagger}\chi\right)\right\rangle
\\
&+c_4^{\pm}D_{\mu}\psi \left\langle\lambda_{6/7}L^{\mu}\right\rangle
+c_5^{\pm}D_{\mu}\theta \left\langle\lambda_{6/7}L^{\mu}\right\rangle\,,
\end{aligned}
\end{align}
\end{subequations}
where $c_i^{\pm}$ stand for the real weak chiral LECs. The term $\mL_{(8_L,1_R)_{+}}$ is CP-even and dominant, while $\mL_{(8_L,1_R)_{-}}$ induces small CP-odd phases. The Lagrangian $\mL_{(8_L,1_R)}$ can be explicitly recast in terms of the operators mediating the $s\leftrightarrow d$ transitions 
\begin{equation}
\begin{aligned}
\mathcal{L}_{(8_L,1_R)}=&G_8\left\langle Q^2_3L_{\mu}L^{\mu}\right\rangle
+G_8'\left\langle Q^2_3\left(\chi^{\dagger}U+U^{\dagger}\chi\right) \right\rangle
+G_8^P X i \left\langle Q^2_3 \left(\chi^{\dagger}U-U^{\dagger}\chi\right) \right\rangle
\\
&+G_8^{\psi}D_{\mu}\psi\left\langle Q^2_3 L^{\mu} \right\rangle 
+ G_8^{\theta} D_{\mu}\theta \left\langle Q^2_3 L^{\mu} \right\rangle
+\mathrm{h.c.}\,,\label{eq_L_8L1R}
\end{aligned}
\end{equation}
where $\left(Q^a_b\right)_{ij}=\delta_{ai}\delta_{bj}$ and they satisfy ${Q^a_b}^\dagger={Q^a_b}^T=Q^b_a$. For an arbitrary $3\times 3$ matrix $O$, one has $\langle Q^a_b O \rangle = O_{ba}$. The relations between the weak chiral LECs $G_i$ and $c_i^{\pm}$ are
\begin{equation}
G_8=c_1^+-ic_1^-\,,\quad G_8'=c_2^+-ic_2^-\,,\quad G_8^P=c_3^+-ic_3^-\,,\quad G_{8}^{\psi}=c_4^+-ic_4^-\,,\quad
G_8^{\theta}=c_5^+-ic_5^-\,.
\end{equation}
The $N_C$ orders of the weak chiral LECs are $\{G_8,G_8^{\psi}\}\sim \mO(N_C^2)$ and $\{G_8',G_8^P,G_8^\theta\}\sim \mO(N_C)$, and the discussion related to the assignment of the $N_C$ order is given at the end of this section. The Lagrangian in Eq.~(\ref{eq_L_8L1R}) retains terms up to $\mO(p^2)$ and keeps the leading contribution in the $1/N_C$ expansion for each octet structure. In the $\delta$ expansion, $\mathcal{L}_{(8_L,1_R)}$ is counted as $\mO(\delta^{-1})$. Nevertheless, this will not cause the proliferation of weak chiral interacting vertices for a given amplitude, since each weak chiral interacting vertex contains a suppression factor of $G_F$. In this work, we only consider the $K\to \pi a$ amplitude at the level of $\mO(G_F)$. In Eq.~(\ref{eq_L_8L1R}), the operators associated with $G_8$ and $G_8^{\psi}$ constitute the full LO part of $\mathcal{L}_{(8_L,1_R)}$, whereas the $G_8'$, $G_8^P$ and $G_8^{\theta}$ operators are, in principle, of NLO in the $\delta$ expansion. However, the NLO part in Eq.~(\ref{eq_L_8L1R}) is not complete, as the operators of $\mO(p^4N_C^2)$ also belong to the same order but are not included. The $G_8'$, $G_8^P$ and $G_8^{\theta}$ operators have the same momentum order as the $G_8$ and $G_8^{\psi}$ ones, and it is the $1/N_C$ expansion that pushes them to NLO in the $\delta$ counting. For comparison with the SU(3) results, where only the momentum expansion is considered, we tentatively include the partial NLO contributions associated with $G_8'$, $G_8^P$ and $G_8^{\theta}$ in this calculation. 

The operator associated with $G_8$ is the most well-known one, first introduced by Cronin~\cite{Cronin:1967jq} in the context of nonleptonic kaon decays. The operators in $\mL_{(8_L,1_R)}$ induce $|\Delta I|=1/2$ transitions, in which the large value of $G_8$ accounts for the so-called ``$|\Delta I|=1/2$'' rule in $|\Delta S| = 1$ nonleptonic weak decays.  The operator associated with $G_8'$ in Eq.~(\ref{eq_L_8L1R}) is known as the ``weak mass term''. It has no observable effect in pure meson processes at LO~\cite{Bernard:1985wf,Crewther:1985zt,Leurer:1987ih}, as it can be absorbed into the quark mass term in $\mL_{\rm eff}^{(0)}$ through a redefinition of $U$. However, as emphasized in Ref.~\cite{Cornella:2023kjq}, this is no longer the case in the presence of ALPs. The $G_8'$ operator provides a nonvanishing contribution to the $K\to\pi a$ decay already at LO. We will discuss this point in detail in the next section. It has been pointed out in Ref.~\cite{Gerard:2005yk} that the $G_8^P$ operator can be absorbed into the $G_8^{\psi}$ term using the equation of motion (EOM). However, this statement holds only when $\left[\ell_{\mu},L^{\mu}\right]=0$. In the presence of general external fields, the LO EOM satisfied by the pNGB nonet reads
\begin{equation}
\quad U^{\dagger}D_{\mu}D^{\mu}U-D_{\mu}D^{\mu}U^{\dagger}U-U^{\dagger}\chi+\chi^{\dagger}U+i\frac{2}{3}M_0^2X=0\,,
\end{equation}
from which one obtains the operator relation 
\begin{equation}
\begin{aligned}
 X i \left\langle Q^2_3\left(\chi^{\dagger}U-U^{\dagger}\chi\right) \right\rangle=&
-2\partial_{\mu}\left(X\left\langle Q^2_3L^{\mu} \right\rangle\right)
+2 X i \left\langle Q^2_3\left[\ell_{\mu}, L^{\mu}\right] \right\rangle
\\
&+2 D_{\mu}\psi \left\langle Q^2_3L^{\mu} \right\rangle
+2 D_{\mu}\theta \left\langle Q^2_3L^{\mu} \right\rangle\,.\label{eq.g8prel}
\end{aligned}
\end{equation}
When the ALPs are not involved and only the pure meson sector is focused, i.e., with vanishing $\left[\ell_{\mu},L^{\mu}\right]$, it is clear that the left-hand side of Eq.~\eqref{eq.g8prel} can be written in terms of the operators accompanying $G_8^{\psi}$ and $G_8^{\theta}$ in Eq.~\eqref{eq_L_8L1R}. 
Nevertheless, the ALP operators in Eq.~\eqref{eq.alpsource} lead to $\left[\ell_{\mu}, L^{\mu}\right]\neq 0$, and therefore $G_8^P$ cannot be completely absorbed into $G_8^{\psi}$ and $G_8^{\theta}$. For the pure meson processes, $G_8^{\psi}$ and $G_8^P$ enter the physical amplitudes only through the fixed combination of $G_8^{\psi}+2G_8^P$. The $G_8^{\theta}$ operator has already been introduced in SU(3) analysis of the $K^{\pm}\to\pi^{\pm}a$ decay~\cite{Cornella:2023kjq}. However, this operator does not contribute to pure meson processes, and therefore the corresponding LEC $G_8^\theta$ cannot be determined from experimental data. In Ref.~\cite{Cornella:2023kjq}, some tentative values of $G_8^\theta$ are assigned in the phenomenological discussions. 
In the U(3) framework, the inclusion of $\eta_0$ renders the singlet sector dynamical, leading to an additional singlet operator associated with $G_8^{\psi}$, whose value can be fixed from weak meson decays~\cite{Gerard:2005yk}. It is noted that even in the limit where the $\eta_0$ decouples, the $G_8^{\psi}$ operator continues to contribute to the $K\to\pi a$ decay, since the ALP is also a singlet. 

The LO weak U(3) $\ChPT$ Lagrangian in the $(27_L,1_R)$ sector is given by
\begin{equation}
\mL_{(27_L,1_R)} = G_{27}^{1/2} O_{27}^{1/2}
+G_{27}^{3/2} O_{27}^{3/2} + \mathrm{h.c.}\,,\label{eq_L_27L1R}
\end{equation}
where $O_{27}^{1/2}$ and $O_{27}^{3/2}$ correspond to the straightforward U(3) extensions of the corresponding SU(3) operators~\cite{Bernard:1985wf} 
\begin{subequations}\label{eq_O27}
\begin{align}
&O_{27}^{1/2}=\left\langle Q^2_3L_{\mu} \right\rangle \left\langle Q^1_1L^{\mu} \right\rangle + \left\langle Q^1_3L_{\mu} \right\rangle \left\langle Q^2_1L^{\mu} \right\rangle + 2\left\langle Q^2_3L_{\mu} \right\rangle \left\langle Q^2_2L^{\mu} \right\rangle - 3 \left\langle Q^2_3L_{\mu} \right\rangle \left\langle Q^3_3L^{\mu} \right\rangle\,,
\\
&O_{27}^{3/2}=\left\langle Q^2_3L_{\mu} \right\rangle \left\langle Q^1_1L^{\mu} \right\rangle + \left\langle Q^1_3L_{\mu} \right\rangle \left\langle Q^2_1L^{\mu} \right\rangle - \left\langle Q^2_3L_{\mu} \right\rangle \left\langle Q^2_2L^{\mu} \right\rangle\,,
\end{align}
\end{subequations}
which induce $|\Delta I|= 1/2$ and $|\Delta I|=3/2$ transitions, respectively. The weak chiral LECs $G_{27}^{1/2}$ and $G_{27}^{3/2}$ scale as $\mO(N_C^2)$, therefore $\mL_{(27_L,1_R)}$ is of order $\mO(\delta^{-1})$. When considering exact SU(3) symmetry, $G_{27}^{1/2}$ and $G_{27}^{3/2}$ satisfy the relation $G_{27}^{3/2}=5G_{27}^{1/2}$. It is worth noting that, although $\langle L_{\mu} \rangle \neq 0$ in the U(3) framework, the singlet component of $L_{\mu}$ exactly cancels in the expressions of Eq.~(\ref{eq_O27}).

Combining Eqs.~(\ref{eq_L_8L1R}) and (\ref{eq_L_27L1R}), one obtains the U(3) weak chiral effective Lagrangian describing the $|\Delta S|=1$ nonleptonic transition
\begin{equation}
\mL_{\rm weak}^{\mathrm{U(3)}\,\ChPT} = \mL_{(8_L,1_R)} + \mL_{(27_L,1_R)}\,,\label{eq_L_weak}
\end{equation}
which provides the mesonic realization of $\mL_{\rm EW-eff}^{|\Delta S| = 1}$ in Eq.~(\ref{eq_L_weak_quark}). In the strict large-$N_C$ limit, $\mL_{\rm EW-eff}^{|\Delta S| = 1}$ reduces to
\begin{equation}
\left.\mL_{\rm EW-eff}^{|\Delta S| = 1}\right|_{N_C\to\infty}
=-\frac{4G_F}{\sqrt{2}}V_{ud}^*V_{us}\big(\bar{d}_L\gamma^{\mu}u_L\big)\big(\bar{u}_L\gamma_{\mu}s_L\big)+\mathrm{h.c.}\,.\label{eq_L_weak_quark_LN}
\end{equation}
The non-factorizable gluon exchanges are suppressed by $1/N_C$. As a result, the two color-singlet quark currents factorize and can be matched independently onto the hadronic currents. At LO, the left-handed quark current is represented in terms of meson fields as
\begin{equation}
\bar{q}_{L,i}\gamma^{\mu}q_{L,j}\to -i\frac{F^2}{2}\left(U^{\dagger}D^{\mu}U\right)_{ji}=-\frac{F^2}{2}(L^{\mu})_{ji}\,,
\end{equation}
with $i,j$ denoting flavor indices. Consequently, in the large-$N_C$ limit, the weak U(3) chiral Lagrangian in Eq.~\eqref{eq_L_weak} similarly reduces to 
\begin{equation}
\left.\mL_{\rm weak}^{\mathrm{U(3)}\,\ChPT}\right|_{N_C\to\infty}=-\frac{F^4G_F}{\sqrt{2}}V_{ud}^*V_{us}(L^{\mu})_{31}(L_{\mu})_{12}+\mathrm{h.c.}\,,\label{eq_L_weak_LN}
\end{equation}
which implies that the leading weak chiral LECs scale as $\mO(N_C^2)$. The operators that cannot be written as products of two left-handed currents will not contribute to the leading order of the $1/N_C$ expansion. From this simple analysis, we acquire the $N_C$ scaling of different weak chiral LECs 
\begin{equation}
\{G_8,\,G_8^{\psi},\,G_{27}^{1/2},\,G_{27}^{3/2}\}\sim \mathcal{O}(N_C^2)\,,
\quad
\{G_8',\,G_8^P,\,G_8^{\theta}\}\sim \mathcal{O}(N_C)\,.
\end{equation}

It is typical to define the dimensionless weak couplings $g_i$ via~\cite{Cirigliano:2011ny}
\begin{equation}\label{eq.defgw}
G_i=G_W g_i = G_W\left(\bar{g}_i-\tau\widetilde{g}_i\right)\,,\quad G_W\equiv -\frac{F^4G_F}{\sqrt{2}}V_{ud}^*V_{us}\,,
\end{equation}
where $\tau = V_{td}^*V_{ts}/V_{ud}^*V_{us}$ is a ratio of CKM matrix elements. In the standard parameterization of the CKM matrix, $V_{ud}$ and $V_{us}$ are real, and $\tau$ carries a nontrivial CP-odd phase. Since $G_W$ behaves as $ \mO(N_C^2)$, the large-$N_C$ scaling of $g_i$ is given by
\begin{equation}
\{g_8,\,g_8^{\psi},\,g_{27}^{1/2},\,g_{27}^{3/2}\}\sim\mathcal{O}(N_C^0)\,,
\quad
\{g_8',\,g_8^P,\,g_8^{\theta}\}\sim\mathcal{O}(N_C^{-1})\,.
\end{equation}
Keeping only the leading-$N_C$ contribution in Eq.~(\ref{eq_L_weak}), the weak chiral Lagrangian becomes
\begin{equation}
\begin{aligned}
\left.\mL_{\rm weak}^{\mathrm{U(3)}\,\ChPT}\right|_{N_C\to\infty}=&-\frac{F^4G_F}{\sqrt{2}}V_{ud}^*V_{us}\Big\{
\left(g_8+g_{27}^{1/2}+g_{27}^{3/2}\right)(L_{\mu})_{31}(L^{\mu})_{12}
\\
&+\left(g_{27}^{1/2}+g_{27}^{3/2}-g_8^{\psi}\right)(L_{\mu})_{32}(L^{\mu})_{11}
\\
&+\left(g_8+2g_{27}^{1/2}-g_{27}^{3/2}-g_8^{\psi}\right)(L_{\mu})_{32}(L^{\mu})_{22}
\\
&+\left(g_8-3g_{27}^{1/2}-g_8^{\psi}\right)(L_{\mu})_{32}(L^{\mu})_{33}
\Big\}+\mathrm{h.c.}\,.\label{eq_L_weak_LN_h}
\end{aligned}
\end{equation}
Comparing Eqs.~(\ref{eq_L_weak_LN}) and (\ref{eq_L_weak_LN_h}), we obtain
\begin{equation}
g_8\big|_{N_C\to\infty}=\frac{3}{5}\,,\quad
g_8^{\psi}\big|_{N_C\to\infty}=\frac{2}{5}\,,\quad
g_{27}^{1/2}\big|_{N_C\to\infty}=\frac{1}{15}\,,\quad
g_{27}^{3/2}\big|_{N_C\to\infty}=\frac{1}{3}\,.\label{eq_gi_LN}
\end{equation}
The leading-$N_C$ values of $g_8$, $g_{27}^{1/2}$ and $g_{27}^{3/2}$, which are also present in the SU(3) case, agree with those obtained in Ref.~\cite{Pich:1990mw}. Substituting these values into Eq.~(\ref{eq_L_weak_LN_h}) yields
\begin{equation}
\left.\mathcal{L}_{\mathrm{weak}}^{\mathrm{U(3)}\,\ChPT}\right|_{N_C\to\infty}=-\frac{F^4G_F}{\sqrt{2}}V_{ud}^*V_{us}
\left[(L_{\mu})_{31}(L^{\mu})_{12}+\left(\frac{2}{5}-g_8^{\psi}\right)(L_{\mu})_{32}\langle L^{\mu}\rangle\right]+\mathrm{h.c.}\,.\notag
\end{equation}
In the SU(3) case with $\langle L_{\mu} \rangle = 0$, the correct large-$N_C$ behavior is automatically reproduced regardless of the presence of $g_8^{\psi}$. In contrast, in the U(3) case with $\langle L_{\mu} \rangle \neq 0$, the $g_8^{\psi}$ term is essential to ensure the correct large-$N_C$ behavior.

\subsection{Elimination of tadpole vertices in the weak mass term}

The Lagrangian $\mL_{\rm weak}^{\mathrm{U(3)}\,\ChPT}$ in Eq.~(\ref{eq_L_weak}) gives the effective realization of $\mL_{\rm EW-eff}^{|\Delta S| = 1}$ in Eq.~(\ref{eq_L_weak_quark}). However, recall that we are working in the basis after the quark-field redefinition in Eq.~(\ref{eq_redef_quark}). This redefinition introduces an additional ALP-dependent phase to the weak chiral effective Lagrangian 
\begin{equation}
\begin{aligned}
\mL_{\rm weak}^{\mathrm{U(3)}\,\ChPT}(\kappa_q)=&e^{i(\kappa_s-\kappa_d)c_{GG}\frac{a}{f}}\Big\{
G_8\left\langle Q^2_3L_{\mu}L^{\mu}\right\rangle
+G_8'\left\langle Q^2_3\left(\chi^{\dagger}U+U^{\dagger}\chi\right) \right\rangle
\\
&+G_8^PXi\left\langle Q^2_3\left(\chi^{\dagger}U-U^{\dagger}\chi\right) \right\rangle
+G_8^{\psi}D_{\mu}\psi\left\langle Q^2_3L^{\mu} \right\rangle
\\
&+G_8^{\theta}D_{\mu}\theta \left\langle Q^2_3L^{\mu}\right\rangle
+G_{27}^{1/2}O_{27}^{1/2}
+G_{27}^{3/2}O_{27}^{3/2}
\Big\}+\mathrm{h.c.}\,.\label{eq_L_weak_kappaq}
\end{aligned}
\end{equation}
Besides the overall phase factor, the ALP field also appears in the external fields $r_{\mu}$, $\ell_{\mu}$, $\chi$, and $\theta$.

The weak mass term accompanied by $G_8'$ generates the so-called ``tadpole'' vertices 
\begin{equation*}
\mL_{\rm weak}^{\mathrm{U(3)}\,\ChPT}\supset2 \sqrt{2}iB_0(m_s-m_d)\left(\frac{G_8'}{F}\bar{K}^0-\frac{G_8'^{*}}{F}K^0\right)\,,
\end{equation*}
which are linear in $K^0$ and $\bar{K}^0$ fields. In the absence of external fields, the entire weak mass term can be absorbed into the quark mass term in $\mL_{\rm eff}^{(0)}$. However, this is no longer the case in the presence of the ALP. Expanding the weak mass term to zeroth order in the ALP field, we obtain
\begin{equation}
\begin{aligned}
\mL_{\rm weak}^{\mathrm{U(3)}\,\ChPT}(\kappa_q)\supset&
G_8'e^{i(\kappa_s-\kappa_d)c_{GG}\frac{a}{f}}\left\langle Q^2_3\left(\chi^{\dagger}U+U^{\dagger}\chi\right) \right\rangle
+G_8'^*e^{i(\kappa_d-\kappa_s)c_{GG}\frac{a}{f}}\left\langle Q^3_2\left(\chi^{\dagger}U+U^{\dagger}\chi\right) \right\rangle
\\
=& \dashuline{\left\langle \left[(c_2^+\lambda_6+c_2^-\lambda_7)\chi(0)\right]U+U^{\dagger}\left[\chi(0)(c_2^+\lambda_6+c_2^-\lambda_7)\right] \right\rangle}+\mO\left(\frac{a}{f}\right)\,,\label{eq_weak_mass}
\end{aligned}
\end{equation}
where $\chi(0)=2B_0 M$ and the dashed-underlined operators can generate tadpole vertices. They can be eliminated by a redefinition of $U$, which takes the form
\begin{equation}
U\to V_RUV_L^{\dagger}\,,\quad
V_{R,L}\simeq 1-i\theta_{R,L} \in \mathrm{U}(3)_{R,L}\,,\label{eq_redef_U}
\end{equation}
with $\theta_{R,L}\sim \mathcal{O}(G_F)$. 
Keeping terms up to $\mO(G_F)$, it suffices to consider the modification of $\mL_{\rm eff}^{(0)}$ under the redefinition 
\begin{equation}
\mL_{\rm eff}^{(0)} \to \mL_{\rm eff}^{(0)} + \delta\mL_{\rm eff}^{(0)}\,,
\end{equation}
with 
\begin{equation}
\begin{aligned}
\delta\mathcal{L}_{\mathrm{eff}}^{(0)}=&
\frac{F^2}{2}\left\langle \left[\theta_R,\,r_{\mu}\right]UD^{\mu}U^{\dagger}-\left[\theta_L,\,\ell_{\mu}\right]D^{\mu}U^{\dagger}U \right\rangle
+\frac{F^2M_0^2}{6}\langle\theta_R-\theta_L\rangle X
\\
&-\frac{F^2}{2}c_{GG}\frac{a}{f}\left\langle \left[\theta_R\chi(0)\kappa_q-\chi(0)\kappa_q\theta_L\right]U^{\dagger}-\left[\theta_L\chi(0)\kappa_q-\chi(0)\kappa_q\theta_R\right]U \right\rangle 
\\
&\dashuline{+i\frac{F^2}{4}\left\langle \left[\theta_R\chi(0)-\chi(0)\theta_L\right]U^{\dagger}+\left[\theta_L\chi(0)-\chi(0)\theta_R\right]U\right\rangle} \,.\label{eq_d_L_eff_0}
\end{aligned}
\end{equation}
Requiring the dashed-underlined term in Eq.~(\ref{eq_d_L_eff_0}) to cancel the corresponding one in Eq.~(\ref{eq_weak_mass}) determines $\theta_{R,L}$, but does not uniquely fix their values. A convenient choice can be taken as
\begin{subequations}\label{eq_theta_R,L}
	\begin{align}
	&\theta_R=\frac{2}{F^2}\left(\frac{1}{y_{sd}}-y_{sd}\right)\left(c_2^-\lambda_6-c_2^+\lambda_7\right)\,,
	\\
	&\theta_L=\frac{2}{F^2}\left(\frac{1}{y_{sd}}+y_{sd}\right)\left(c_2^-\lambda_6-c_2^+\lambda_7\right)\,,
	\end{align}
\end{subequations}
with $y_{sd}=\frac{m_s-m_d}{m_s+m_d}$. It can be straightforwardly verified that this choice satisfies the relation 
\begin{equation*}
i\left[\theta_R\chi(0)-\chi(0)\theta_L\right]+\chi(0)\frac{4}{F^2}\left(c_2^+\lambda_6+c_2^-\lambda_7\right)=0\,.
\end{equation*}

As a result, after performing the redefinition of $U$ as specified in Eqs.~(\ref{eq_redef_U}) and (\ref{eq_theta_R,L}), the dashed-underlined terms in Eqs.~(\ref{eq_d_L_eff_0}) and (\ref{eq_weak_mass}) will cancel each other. We then reshuffle the remaining terms in $\delta\mL_{\rm eff}^{(0)}$ into $\mL_{\rm weak}^{\mathrm{U(3)}\,\ChPT}$ and obtain the final effective chiral Lagrangians for calculating the $K\to\pi a$ decay amplitude. The LO QCD chiral Lagrangian with ALP takes the form 
	\begin{equation}
	\mL_{\rm eff}^{\chi}(\kappa_q)=\frac{1}{2}\partial_{\mu}a\partial^{\mu}a-\frac{1}{2}m_{a,0}^2a^2+\mL_{\rm eff}^{(0)}(\kappa_q)\,,\label{eq_L_eff_chi}
	\end{equation}
	where $\mL_{\rm eff}^{(0)}(\kappa_q)$ is given in Eq.~(\ref{eq_L_eff_0}).
The relevant weak chiral Lagrangian is given by 
	\begin{equation}\label{eq.lagwxpt}
	{\mL_{\rm weak}^{\mathrm{U(3)}\,\ChPT}}'(\kappa_q)=\mL_{(8_L,1_R)}'(\kappa_q)+\mL_{(27_L,1_R)}(\kappa_q)+\mL_{G_8'}(\kappa_q)\,,
	\end{equation}
with 
	\begin{subequations}
	\begin{align}
	&\begin{aligned}
	\mL_{(8_L,1_R)}'(\kappa_q)=&e^{i(\kappa_s-\kappa_d)c_{GG}\frac{a}{f}}\big[
	G_8\left\langle Q^2_3L_{\mu}L^{\mu} \right\rangle
	+G_8^PXi\left\langle Q^2_3\left(\chi^{\dagger}U-U^{\dagger}\chi\right) \right\rangle
	\\
	&+G_8^{\psi}D_{\mu}\psi\left\langle Q^2_3L_{\mu}\right\rangle
	+G_8^{\theta}D_{\mu}\theta\left\langle Q^2_3L_{\mu}\right\rangle
	\big]+\mathrm{h.c.}\,,
	\end{aligned}
	\\
	&\mL_{(27_L,1_R)}(\kappa_q)=e^{i(\kappa_s-\kappa_d)c_{GG}\frac{a}{f}}\left(
	G_{27}^{1/2}O_{27}^{1/2}+G_{27}^{3/2}O_{27}^{3/2}\right)+\mathrm{h.c.}\,,
	\\
	&\begin{aligned}
	\mL_{G_8'}(\kappa_q)=&-2iG_8'c_{GG}\frac{a}{f}\left\langle Q^2_3\left[\chi(0)\kappa_qU-U^{\dagger}\chi(0)\kappa_q\right] \right\rangle
	\\
	&+G_8'i(\kappa_s-\kappa_d)c_{GG}\frac{a}{f}\left\langle Q^2_3\left[\chi(0)U+U^{\dagger}\chi(0)\right] \right\rangle
	\\
	&+iG_8'\Big\langle Q^2_3\Big\{\left(\frac{1}{y_{sd}}-y_{sd}\right)\left[r_{\mu},\,UD^{\mu}U^{\dagger}\right]
	\\
	&-\left(\frac{1}{y_{sd}}+y_{sd}\right)\left[\ell_{\mu},\,D^{\mu}U^{\dagger}U\right]\Big\} \Big\rangle
	\\
	&-iG_8'c_{GG}\frac{a}{f}\Big\langle Q^2_3\Big[\left(\frac{1}{y_{sd}}-y_{sd}\right)\left(\chi(0)\kappa_qU^{\dagger}+U\chi(0)\kappa_q\right)
	\\
	&-\left(\frac{1}{y_{sd}}+y_{sd}\right)\left(\chi(0)\kappa_qU+U^{\dagger}\chi(0)\kappa_q\right)
	\Big] \Big\rangle+\mathrm{h.c.}\,,
	\end{aligned}
	\end{align}
	\end{subequations}
where $\mL_{G_8'}$ only contains the operators involving ALP couplings.

\subsection{The $a$-$\pi^3$-$\eta$-$\eta'$ mixing}

Apart from the direct $K\pi a$ couplings, another kind of important contribution to the $K\to\pi a$ decay is given by the interacting vertices of the $K\pi\pi^3$, $K\pi\eta_{8}$ and $K\pi\eta_0$ types that are driven by the SM weak transitions. The latter vertices should be further combined with the $a$-$\pi^3$-$\eta$-$\eta'$ mixing, in order for them to contribute to the $K\to\pi a$ decays. In this part, we compute the mixing of these fields. 

The calculation of the $a$-$\pi^3$-$\eta$-$\eta'$ mixing has been pursued up to NLO in the $\delta$ counting within the U(3) $\ChPT$ in Refs.~\cite{Gao:2022xqz,Gao:2024vkw}. Nevertheless, only the model-independent axion operator $aG\tilde{G}$ is included in the former two references. In this work, we further take into account the model dependent ALP operators, such as those involving $k_Q$ and $k_q$ in Eq.~\eqref{eq_L_ALP}, in addition to the model independent one. 
The quadratic terms of the QCD chiral Lagrangian $\mL_{\rm eff}^{\chi}$ in Eq.~\eqref{eq_L_eff_chi} up to $\mO(a/f)$ read
\begin{align}
\left.\mL_{\rm eff}^{\chi}\right|_{\rm quadratic}=&
\frac{1}{2}\partial_{\mu}a\partial^{\mu}a-\frac{1}{2}m_{a,0}^2a^2
+\partial_{\mu}\pi^+\partial^{\mu}\pi^--m_{0\pi}^2\pi^+\pi^- +\frac{1}{2}\partial_{\mu}\pi^3\partial^{\mu}\pi^3-\frac{1}{2}m_{0\pi}^2\pi^3\pi^3
\notag\\
&+\partial_{\mu}K^+\partial^{\mu}K^--m_{0K^{\pm}}^2K^+K^-
+\partial_{\mu}K^0\partial^{\mu}\bar{K}^0-m_{0K^{0}}^2K^0\bar{K}^0
\notag\\
&+\frac{1}{2}\partial_{\mu}\eta_8\partial^{\mu}\eta_8+\frac{1}{2}\partial_{\mu}\eta_0\partial^{\mu}\eta_0
-\frac{1}{2}m_{0\eta_8}^2\eta_8^2-\frac{1}{2}m_{0\eta_0}^2\eta_0^2+\frac{2\sqrt{2}}{3}\Delta_8\eta_8\eta_0
\notag\\
&+\frac{1}{\sqrt{3}}\delta_I\pi^3\eta_8+\sqrt{\frac{2}{3}}\delta_I\pi^3\eta_0
-\epsilon_fk_{aK^0}\partial_{\mu}a\partial^{\mu}K^0-\epsilon_fk_{a\bar{K}^0}\partial_{\mu}a\partial^{\mu}\bar{K}^0
\notag\\
&-\epsilon_fk_{a\pi^3}\partial_{\mu}a\partial^{\mu}\pi^3
-\epsilon_fk_{a\eta_8}\partial_{\mu}a\partial^{\mu}\eta_8
-\epsilon_fk_{a\eta_0}\partial_{\mu}a\partial^{\mu}\eta_0
\notag\\
&+\epsilon_fm_{a\pi^3}^2a\pi^3
+\epsilon_fm_{a\eta_8}^2a\eta_8
+\epsilon_fm_{a\eta_0}^2a\eta_0\,.\label{eq_L_quadratic}
\end{align}
The masses and the mixing of the pure pNGB sector are generated by the symmetry breakings from the light-quark masses and the $\mathrm{U}(1)_A$ anomaly, which read 
\begin{equation}
\begin{aligned}
&m_{0\pi}^2=2B_0\hat{m}\,,\quad m_{0K^{\pm}}^2=B_0(m_u+m_s)\,,\quad m_{0K^{0}}^2=B_0(m_d+m_s)\,,
\\
&m_{0K}^2=B_0(\hat{m}+m_s)=\frac{1}{2}\left(m_{0K^{\pm}}^2+m_{0K^{0}}^2\right)\,,
\\
&m_{0\eta_8}^2=\frac{2}{3}B_0(\hat{m}+2m_s)=\frac{1}{3}\left(4m_{0K}^2-m_{0\pi}^2\right)\,,
\\
&m_{0\eta_0}^2=M_0^2+\frac{2}{3}B_0(2\hat{m}+m_s)=M_0^2+\frac{1}{3}\left(2m_{0K}^2+m_{0\pi}^2\right)\,,
\\
&\Delta_8=B_0(m_s-\hat{m})=m_{0K}^2-m_{0\pi}^2\,,\quad \delta_I=B_0\epsilon_I\,, 
\end{aligned}
\end{equation}
with $\hat{m}=\frac{m_u+m_d}{2}$ and $\epsilon_I=  m_d-m_u$. 
The mixing terms involving the ALP are given by
\begin{equation}
\begin{aligned}
&\epsilon_f\equiv\frac{F}{f}\,,
\\
&k_{a\pi^3}=\frac{1}{2}\left[\left(c_{uu}^a-c_{dd}^a\right)+2c_{GG}\left(\kappa_u-\kappa_d\right)\right]\,,
\\
&k_{a\eta_8}=\frac{1}{2\sqrt{3}}\left[\left(c^a_{uu}+c^a_{dd}-2c^a_{ss}\right)+2c_{GG}\left(\kappa_u+\kappa_d-2\kappa_s\right)\right]\,,
\\
&k_{a\eta_0}=\frac{1}{\sqrt{6}}\left[\left(c^a_{uu}+c^a_{dd}+c^a_{ss}\right)+2c_{GG}\left(\kappa_u+\kappa_d+\kappa_s\right)\right]\,,
\\
&k_{aK^0}=\frac{1}{\sqrt{2}}(k_d-k_D)_{21}\,,\quad k_{a\bar{K}^0}=\frac{1}{\sqrt{2}}(k_d-k_D)_{12}\,,
\\
&m_{a\pi^3}^2=2c_{GG}B_0(\kappa_um_u-\kappa_dm_d)\,,
\\
&m_{a\eta_8}^2=\frac{2}{\sqrt{3}}c_{GG}B_0(\kappa_um_u+\kappa_dm_d-2\kappa_sm_s)\,,
\\
&m_{a\eta_0}^2=2\sqrt{\frac{2}{3}}c_{GG}B_0(\kappa_um_u+\kappa_dm_d+\kappa_sm_s)-\frac{\sqrt{6}}{3}\hat{c}_{GG}M_0^2\,,
\end{aligned}
\end{equation}
which depend on $\kappa_q$ and are therefore not physical observables. The $\kappa_q$ dependence of various contributions must cancel out in the final physical amplitude. 

The QCD chiral Lagrangian generates mixing among the flavor neutral fields up to $\mO(\epsilon_f)$. Among these mixings, the mixing between $\eta_8$ and $\eta_0$ fields, namely, $\eta$-$\eta'$ mixing, can be particularly sensitive to the higher-order corrections, because both the U$(1)_A$ anomaly and the strange quark mass enter this mixing and they are numerically much larger than the up- and down-quark masses. 
The LO $\eta_8$-$\eta_0$ mass matrix can be exactly diagonalized by the following orthogonal transformation~\cite{Guo:2015xva} 
\begin{equation}
\begin{aligned}
&\eta_8=\ct\eta+\st\eta'\,,\quad \eta_0=\ct\eta'-\st\eta\,,
\\
&\st=\sin\theta=-\left\{ \left[\frac{3\left(m_{0\eta_0}^2-m_{0\eta_8}^2\right)+\sqrt{9\left(m_{0\eta_0}^2-m_{0\eta_8}^2\right)^2+32\Delta_8^2}}{4\sqrt{2}\Delta_8}\right]^2+1 \right\}^{-\frac{1}{2}}\,,
\\
&\ct=\cos\theta=\mathrm{sign}(\Delta_8)\sqrt{1-\st^2}\,,\label{eq_eta_8-0_rotation}
\end{aligned}
\end{equation}
where $\eta$ and $\eta'$ are the fields that have diagonal quadratic terms up to $\mO(\delta_I^0)$ and $\mO(\epsilon_f^0)$ at LO in the $\delta$ counting. The masses of the diagonalized $\eta$ and $\eta'$ are 
\begin{subequations}\label{eq_m0_eta_etap}
	\begin{align}
	m_{0\eta}^2&=\frac{1}{6}\left[3\left(m_{0\eta_0}^2+m_{0\eta_8}^2\right)-\sqrt{9\left(m_{0\eta_0}^2-m_{0\eta_8}^2\right)^2+32\Delta_8^2}\right]\,,\label{eq_m0eta}
	\\
	m_{0\eta'}^2&=\frac{1}{6}\left[3\left(m_{0\eta_0}^2+m_{0\eta_8}^2\right)+\sqrt{9\left(m_{0\eta_0}^2-m_{0\eta_8}^2\right)^2+32\Delta_8^2}\right]\,.\label{eq_m0etap}
	\end{align}
\end{subequations}
At LO in $\delta$ expansion, the physical masses of the nonet pseudoscalar mesons, ignoring $\mO(\delta_I^2)$ and electroweak corrections, are given by
\begin{equation}
m_{\pi^{\pm}}^2=m_{\pi^0}^2=m_{0\pi}^2\,,\quad
m_{K^{\pm}}^2=m_{0K^{\pm}}^2\,,\quad m_{K^{0}}^2=m_{0K^{0}}^2\,,\quad
m_{\eta}^2=m_{0\eta}^2\,,\quad m_{\eta'}^2=m_{0\eta'}^2\,,
\end{equation}
where $m_{0\pi}$ and $m_{0K}$ are fixed by the physical pion and kaon masses. 

It is demonstrated in many works~\cite{Leutwyler:1997yr,Feldmann:1999uf,Guo:2015xva} that the LO calculation by only including a single parameter $M_0$ can not well describe the masses of $\eta$ and $\eta'$. To pursue more precise descriptions, one has to go beyond LO in the $\delta$ expansion. Higher-order correction can noticeably modify the $\eta$-$\eta'$ mixing, and it leads to a more general mixing pattern, namely the two-mixing-angle scheme, which can be written as 
\begin{equation}
\begin{pmatrix}
\eta\\\eta'
\end{pmatrix}
=
\frac{1}{F}\begin{pmatrix}
F_8\cos\theta_8&-F_0\sin\theta_0\\F_8\sin\theta_8&F_0\cos\theta_0
\end{pmatrix}
\begin{pmatrix}
\eta_8\\ \eta_0
\end{pmatrix}\,.
\end{equation}
This can be equivalently expressed as
\begin{equation}
\begin{aligned}
&\eta_8=\beta_{11}\eta+\beta_{12}\eta'\,,\quad \eta_0=\beta_{22}\eta'+\beta_{21}\eta\,,
\\
&\beta_{11}=\frac{F\cos\theta_0}{F_8\cos(\theta_0-\theta_8)}\,,\quad
\beta_{12}=\frac{F\sin\theta_0}{F_8\cos(\theta_0-\theta_8)}\,,
\\
&\beta_{22}=\frac{F\cos\theta_8}{F_0\cos(\theta_0-\theta_8)}\,,\quad
\beta_{21}=-\frac{F\sin\theta_8}{F_0\cos(\theta_0-\theta_8)}\,.\label{eq_eta_8-0_twoangle}
\end{aligned}
\end{equation}
The $\eta_0$ contribution to the $K\to\pi a$ decay is expected to be sensitive to the $\eta$-$\eta'$ mixing. In this work, to incorporate such important NLO correction, we adopt the two-mixing-angle scheme of Eq.~(\ref{eq_eta_8-0_twoangle}) in the interaction vertices and use the phenomenological values of mixing-angle parameters determined in Ref.~\cite{Gu:2018swy}. The LO expression of the one-mixing angle formula can be recovered with the replacement of $\beta_{11}\to\ct$, $\beta_{22}\to\ct$, $\beta_{12}\to\st$, and $\beta_{21}\to-\st$, i.e., by taking $F_0=F_8=F$ and $\theta_0=\theta_8=\theta$ in Eq.~\eqref{eq_eta_8-0_twoangle}. 

Apart from the $\eta$-$\eta'$ mixing discussed above, the remaining mixing terms are treated as interaction vertices. These include the ALP-pNGBs and the $\pi^3$-$\eta$, $\pi^3$-$\eta'$ mixing in Eq.~(\ref{eq_L_quadratic}), as well as those contained in $\mL_{\rm weak}^{\mathrm{U(3)}\,\ChPT}$. It is noted that the weak chiral Lagrangian also generates flavor-changing mixings among different particles, such as $K^-\pi^+$, $K^0\pi^0$, etc, which are accompanied by the $G_F$ factor. All of these pertinent mixing effects will be taken into account in the calculation of the Green's functions relevant to the decay.

\section{$K\to \pi a$ decay amplitudes in U(3) $\ChPT$}\label{sec.amp}

\subsection{Calculation of the LO $K \to \pi a$ amplitudes}
\label{sec_tree-level-Ktopia-amps}

In this part, we compute the $K\to\pi a$ decay amplitude up to $\mO(\epsilon_f)$ and $\mO(G_F)$, and include isospin breaking corrections from $m_u\neq m_d$ up to $\mO(\delta_I)$. At this order, the $s\leftrightarrow d$ flavor transition in $K\to\pi a$ decay can be generated either by the direct ALP flavor-changing  coupling $\left(k_d+k_D\right)_{12,21}$ or by the SM weak interaction. Accordingly, the contributions to the decay amplitude can be separated into class-1 and class-2, in which the $s\leftrightarrow d$ transition is induced by $\left(k_d+k_D\right)_{12,21}$ and by the SM weak interaction, respectively. In the following, we calculate each contribution at LO in the U(3) $\ChPT$.

In the $\delta$ expansion, a given amputated Green function scales as $\mO(\delta^D)$, where the power $D$ is given by $D=\frac{1}{2}N+2N_L+\sum_{n}n N_n$. Here $N$ is the number of external meson lines, $N_L$ is the number of independent loops, and $N_n$ is the number of $\mO(\delta^n)$ vertices. For class-1, the SM weak interaction does not enter, and the LO contribution consists of tree diagrams built from the vertices of $\mL_{\rm eff}^{(0)}$ in Eq.~\eqref{eq_L_eff_0}. For class-2, the LO contribution is given by the tree diagrams containing one vertex of ${\mL_{\rm weak}^{\mathrm{U(3)}\,\ChPT}}'$ in Eq.~\eqref{eq.lagwxpt}  together with any number of flavor-conserving vertices in $\mL_{\rm eff}^{(0)}$. The contributions from $G_8'$, $G_8^P$, and $G_8^{\theta}$ are also included in our class-2 calculation, which are suppressed by $1/N_C$ relative to the LO contribution. 

\begin{figure}[t]
	\centering
	\subfigure[class-1\label{fig_Km_to_pim_a_c1}]
	{\includegraphics[width=0.3\linewidth]{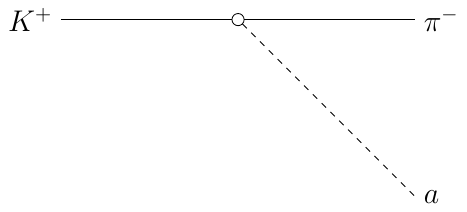}}
	\hfill
	\subfigure[class-2: (1)\label{fig_Km_to_pim_a_c2-1}]
	{\includegraphics[width=0.3\linewidth]{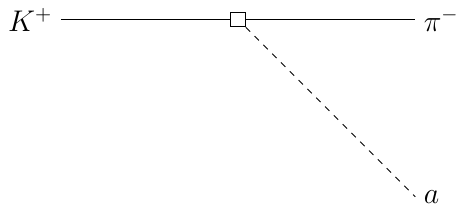}}
	\hfill
	\subfigure[class-2: (2)\label{fig_Km_to_pim_a_c2-2}]
	{\includegraphics[width=0.3\linewidth]{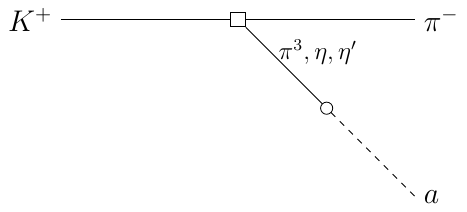}}
	\\
	\subfigure[class-2: (3)\label{fig_Km_to_pim_a_c2-3}]
	{\includegraphics[width=0.3\linewidth]{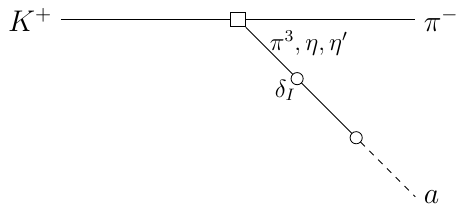}}
	\hfill
	\subfigure[class-2: (4)\label{fig_Km_to_pim_a_c2-4}]
	{\includegraphics[width=0.3\linewidth]{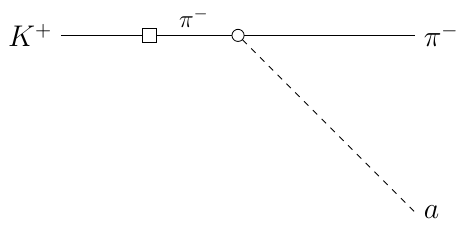}}
	\hfill
	\subfigure[class-2: (5)\label{fig_Km_to_pim_a_c2-5}]
	{\includegraphics[width=0.3\linewidth]{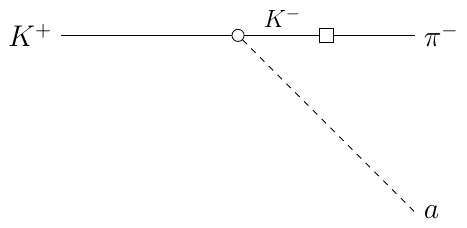}}
	\caption{Feynman diagrams for the three-point Green function $\langle 0 |\pi^- a K^+|0\rangle$ associated with the $K^-\to\pi^-a$ decay. Empty circles denote vertices from $\mathcal{L}_{\mathrm{eff}}^{(0)}$~\eqref{eq_L_eff_0}, while empty rectangles denote vertices from ${\mL_{\rm weak}^{\mathrm{U(3)}\,\ChPT}}'$~\eqref{eq.lagwxpt}. The labels $K^+$, $\pi^-$, and $a$ at the ends of the external lines denote the corresponding interpolating fields, where $K^+$ interpolates the incoming $K^-$ state. For the meaning of class-1 and class-2, see the text for details.}\label{fig_Km_to_pim_a}
\end{figure}

To obtain the $K^-\to\pi^-a$ decay amplitude, we calculate the three-point Green function involving the $K^+$, $\pi^-$ and $a$ fields. The relevant Feynman diagrams are shown in Fig.~\ref{fig_Km_to_pim_a}. From this Green function, the decay amplitude is extracted via the LSZ reduction formula, and the final result separates into two parts
\begin{equation}
A_{K^-\to\pi^-a}=A_{K^-\to\pi^-a}^{\rm AFC}+A_{K^-\to\pi^-a}^{\rm WFC}\,,\label{eq_A_Km_to_pim_a}
\end{equation}
where the ALP flavor-changing (AFC) and the SM weak flavor-changing (WFC) amplitudes correspond to the class-1 and class-2 diagrams in Fig.~\ref{fig_Km_to_pim_a}, respectively.  
The explicit form of the AFC amplitude reads
\begin{equation}
A_{K^-\to\pi^-a}^{\rm AFC}=-\frac{(k_d+k_D)_{12}}{2f}\left(m_{K^{\pm}}^2-m_{\pi}^2\right)\,,\label{eq_A_Km_to_pim_a_AFC}
\end{equation}
where the $s\to d$ transition originates from the flavor-changing ALP coupling $(k_d+k_D)_{12}$. While for the WFC amplitude, the $s\to d$ transition is induced by the SM nonleptonic weak interaction, corresponding to the class-2 contributions in Fig.~\ref{fig_Km_to_pim_a_c2-1}--\ref{fig_Km_to_pim_a_c2-5}. The amplitude $A_{K^-\to\pi^-a}^{\rm WFC}$ can be expressed in the form
\begin{equation} 
A_{K^-\to\pi^-a}^{\rm WFC}=G_W\sum_{\cALP}\frac{\cALP}{f}F_{K^-\to\pi^-a}^{\cALP}\,,\label{eq_A_Km_to_pim_a_WFC}
\end{equation}
where the flavor-changing coupling $G_W$ is defined in Eq.~\eqref{eq.defgw} and the sum runs over the flavor-conserving ALP couplings $\cALP\in\{c_{GG},c^a_{uu},c^a_{dd},c^a_{ss},c^v_{dd}-c^v_{ss}\}$. For the diagonal vector ALP-quark couplings, what enters the amplitude is the combination of $c^v_{dd}-c^v_{ss}$, as a consequence of vector-current conservation~\cite{Bauer:2021wjo}.
The dimensionless factor $F_{K^-\to\pi^-a}^{\cALP}$ corresponding to each flavor-conserving ALP coupling is given by
\begin{subequations}\small
	\begin{align}
	F^{c_{GG}}_{K^-\to\pi^-a}=&g_8 \Bigg[\frac{4 \left(m_{\pi}^2-m_{K^{\pm}}^2\right)+2\delta_I}{3F^2}x_2(M_0^2)
	+\frac{\sqrt{2}\left(3 m_a^2-2 m_{K^{\pm}}^2-m_{\pi}^2+\delta_I\right)}{3F^2}x_3(M_0^2)
	\Bigg]
	\notag\\
	&+\frac{2\left(m_{K^{\pm}}^2-m_{\pi}^2\right)}{F^2}\left[(g_8^{\psi}+2g_8^P)x_2(M_0^2)+g_8^{\theta}+2g_8^P\right]
	\\
	&+g_{27}^{1/2}\Bigg\{
	\frac{\sqrt{2}\left(m_a^2-4 m_{K^{\pm}}^2+3 m_{\pi}^2\right)}{F^2}x_3(M_0^2)
	+\frac{\sqrt{2}\delta_I}{3F^2}\left[x_3(M_0^2)+\sqrt{2}x_2(M_0^2)\right]
	\Bigg\}
	\notag\\
	&+g_{27}^{3/2}\Bigg\{\frac{\sqrt{2}\left(m_a^2-m_{K^{\pm}}^2\right) }{F^2}x_3(M_0^2)
	+\frac{\sqrt{2}\delta_I\left(m_a^2+3 m_{K^{\pm}}^2-4 m_{\pi}^2\right)}{3 F^2\left(m_a^2-m_{\pi}^2\right)}
	\left[x_3(M_0^2)+\sqrt{2}x_2(M_0^2)\right]
	\Bigg\}\,,\notag
	\end{align}
	\begin{align}
	F^{c^a_{uu}}_{K^-\to\pi^-a}=&
	g_8 \Bigg\{\frac{2 m_{K^{\pm}}^2-2 m_{\pi}^2-m_a^2}{2 F^2}
	+\frac{3 m_a^2-2 m_{K^{\pm}}^2-m_{\pi}^2}{6F^2}x_1(m_a^2)
	+\frac{2 \left(m_{\pi}^2-m_{K^{\pm}}^2\right)}{3F^2}x_2(m_a^2)
	\notag\\
	&+\frac{3 m_a^2-4 m_{K^{\pm}}^2+m_{\pi}^2}{3 \sqrt{2} F^2}x_3(m_a^2)
	+\frac{\delta_I}{m_a^2-m_{\pi}^2}\Bigg[
	\frac{m_{K^{\pm}}^2-m_a^2}{3 F^2 }x_1(m_a^2)
	\notag\\
	&+\frac{m_a^2+2 m_{K^{\pm}}^2-3 m_{\pi}^2}{3 F^2 }x_2(m_a^2)
	-\frac{m_a^2-4 m_{K^{\pm}}^2+3 m_{\pi}^2}{3\sqrt{2}F^2}x_3(m_a^2)
	\Bigg] 
	\Bigg\}
	\notag\\
	&+\frac{m_{K^{\pm}}^2-m_{\pi}^2}{F^2}\Bigg\{
	g_8^{\theta}-g_8^{\psi}
	+\frac{g_8^{\psi}+2g_8^P}{\sqrt{2}}\left(1-\frac{\delta_I}{m_a^2-m_{\pi}^2}\right)
	\left[x_3(m_a^2)+\sqrt{2}x_2(m_a^2)\right]
	\Bigg\}
	\notag\\
	&+g_{27}^{1/2}\Bigg\{\frac{4 m_{K^{\pm}}^2-m_a^2-4 m_{\pi}^2}{2 F^2}
	+\frac{m_a^2-4 m_{K^{\pm}}^2+3 m_{\pi}^2}{2 F^2 }\left[x_1(m_a^2)+\sqrt{2}x_3(m_a^2)\right]
	+\frac{\delta_I}{3 F^2}x_2(m_a^2)
	\notag\\
	&-\frac{\delta_I}{m_a^2-m_{\pi}^2}\Bigg[
	\frac{m_a^2-6 m_{K^{\pm}}^2+5 m_{\pi}^2}{3 F^2 }x_1(m_a^2)
	+\frac{m_a^2-12 m_{K^{\pm}}^2+11 m_{\pi}^2}{3 \sqrt{2} F^2}x_3(m_a^2)
	\Bigg] 
	\Bigg\}
	\notag\\
	&+g_{27}^{3/2}\Bigg\{\frac{4 m_{\pi}^4-4 m_{K^{\pm}}^2 m_{\pi}^2-m_a^4+m_{K^{\pm}}^2 m_a^2}{2 F^2 \left(m_a^2-m_{\pi}^2\right)}
	+\frac{m_a^2-m_{K^{\pm}}^2}{2F^2}\left[x_1(m_a^2)+\sqrt{2}x_3(m_a^2)\right]
	\notag\\
	&+\frac{\delta_I}{m_a^2-m_{\pi}^2}
	\Bigg[
	\frac{m_a^2+3 m_{K^{\pm}}^2-4 m_{\pi}^2}{3 F^2}x_2(m_a^2)
	-\frac{m_a^2-3 m_{K^{\pm}}^2+2 m_{\pi}^2}{3 F^2 }x_1(m_a^2)
	\notag\\
	&-\frac{m_a^2-9 m_{K^{\pm}}^2+8 m_{\pi}^2}{3 \sqrt{2} F^2}x_3(m_a^2)\Bigg] 
	\Bigg\}\,,
	\end{align}
	\begin{align}
	F^{c^a_{dd}}_{K^-\to\pi^-a}=&
	g_8 \Bigg\{\frac{m_{K^{\pm}}^2-m_{\pi}^2}{2 F^2}
	+\frac{3 m_a^2-2 m_{K^{\pm}}^2-m_{\pi}^2}{6 F^2}x_1(m_a^2)
	+\frac{2 \left(m_{\pi}^2-m_{K^{\pm}}^2\right) }{3 F^2 }x_2(m_a^2)
	\notag\\
	&+\frac{3 m_a^2-4 m_{K^{\pm}}^2+m_{\pi}^2}{3\sqrt{2}F^2}x_3(m_a^2)
	+\frac{\delta_I}{m_a^2-m_{\pi}^2}\Bigg[
	\frac{2 m_a^2-m_{K^{\pm}}^2-m_{\pi}^2}{3F^2}x_1(m_a^2)
	\notag\\
	&+\frac{m_a^2-2 m_{K^{\pm}}^2+m_{\pi}^2}{3F^2}x_2(m_a^2)
	+\frac{5 m_a^2-4 m_{K^{\pm}}^2-m_{\pi}^2}{3\sqrt{2}F^2}x_3(m_a^2)
	\Bigg] 
	\Bigg\}
	\notag\\
	&+\frac{m_{K^{\pm}}^2-m_{\pi}^2}{F^2}\Bigg\{
	g_8^{\theta}-g_8^{\psi}+y_{sd}g_8'
	+\frac{g_8^{\psi}+2g_8^P}{\sqrt{2}}\left(1+\frac{\delta_I}{m_a^2-m_{\pi}^2}\right)
	\left[x_3(m_a^2)+\sqrt{2}x_2(m_a^2)\right]
	\Bigg\}
	\notag\\
	&+g_{27}^{1/2}\Bigg\{\frac{3 \left(m_{K^{\pm}}^2-m_{\pi}^2\right)}{2 F^2}
	+\frac{m_a^2-4 m_{K^{\pm}}^2+3 m_{\pi}^2}{2 F^2 }\left[x_1(m_a^2)+\sqrt{2}x_3(m_a^2)\right]
	+\frac{\delta_I}{3F^2}x_2(m_a^2)
	\notag\\
	&+\frac{\delta_I}{m_a^2-m_{\pi}^2}\Bigg[
	\frac{2 \left(m_a^2-3 m_{K^{\pm}}^2+2 m_{\pi}^2\right)}{3 F^2}x_1(m_a^2)
	+\frac{5 m_a^2-12 m_{K^{\pm}}^2+7 m_{\pi}^2}{3 \sqrt{2} F^2}x_3(m_a^2)
	\Bigg] 
	\Bigg\}
	\notag\\
	&+g_{27}^{3/2}\Bigg\{\frac{3 m_{\pi}^2 \left(m_{K^{\pm}}^2-m_{\pi}^2\right)}{2 F^2 \left(m_a^2-m_{\pi}^2\right)}
	+\frac{m_a^2-m_{K^{\pm}}^2}{2F^2}\left[x_1(m_a^2)+\sqrt{2}x_3(m_a^2)\right]
	+\frac{2 \delta_I}{3 F^2 }x_1(m_a^2)
	\notag\\
	&+\frac{\delta_I}{m_a^2-m_{\pi}^2}\Bigg[
	\frac{m_a^2+3 m_{K^{\pm}}^2-4 m_{\pi}^2}{3 F^2}x_2(m_a^2)
	+\frac{5 m_a^2+3 m_{K^{\pm}}^2-8 m_{\pi}^2}{3 \sqrt{2} F^2}x_3(m_a^2)\Bigg] 
	\Bigg\}\,,
	\end{align}
	\begin{align}
	F^{c^a_{ss}}_{K^-\to\pi^-a}=&
	g_8 \Bigg\{\frac{m_a^2+m_{K^{\pm}}^2-m_{\pi}^2}{2 F^2}
	+\frac{2 m_{K^{\pm}}^2-3 m_a^2+m_{\pi}^2}{3 F^2}x_1(m_a^2)
	+\frac{2\left(m_{\pi}^2-m_{K^{\pm}}^2\right)}{3 F^2}x_2(m_a^2)
	\notag\\
	&+\frac{3 m_a^2+2 m_{K^{\pm}}^2-5 m_{\pi}^2}{3 \sqrt{2} F^2}x_3(m_a^2)
	+\frac{\delta_I}{3\sqrt{2} F^2 }\left[\sqrt{2}x_2(m_a^2)-\sqrt{2}x_1(m_a^2)-x_3(m_a^2)\right]
	\Bigg\}
	\notag\\
	&+\frac{m_{K^{\pm}}^2-m_{\pi}^2}{F^2 }\Bigg\{
	g_8^{\theta}-g_8^{\psi}-y_{sd}g_8'
	+(g_8^{\psi}+2g_8^P)
	\left[x_2(m_a^2)-\sqrt{2}x_3(m_a^2)\right]
	\Bigg\}
	\notag\\
	&+g_{27}^{1/2}\Bigg\{\frac{m_a^2-7 m_{K^{\pm}}^2+7 m_{\pi}^2}{2 F^2}
	+\frac{m_a^2-4 m_{K^{\pm}}^2+3 m_{\pi}^2}{\sqrt{2} F^2}\left[x_3(m_a^2)-\sqrt{2}x_1(m_a^2)\right]
	\notag\\
	&+\frac{\delta_I}{3\sqrt{2}F^2}\left[\sqrt{2}x_2(m_a^2)-\sqrt{2}x_1(m_a^2)-x_3(m_a^2)\right]
	\Bigg\}
	\notag\\
	&+g_{27}^{3/2}\Bigg\{\frac{m_a^2-m_{K^{\pm}}^2+m_{\pi}^2}{2 F^2}
	+\frac{m_a^2-m_{K^{\pm}}^2}{\sqrt{2}F^2}\left[x_3(m_a^2)-\sqrt{2}x_1(m_a^2)\right]
	\notag\\
	&+\frac{\delta_I}{m_a^2-m_{\pi}^2}\frac{m_a^2+3 m_{K^{\pm}}^2-4 m_{\pi}^2}{3\sqrt{2} F^2}\left[\sqrt{2}x_2(m_a^2)-\sqrt{2}x_1(m_a^2)-x_3(m_a^2)\right]
	\Bigg\}\,,
	\end{align}
	\begin{equation}
	F^{c^v_{dd}-c^v_{ss}}_{K^-\to\pi^-a}=
	\frac{m_{K^{\pm}}^2-m_a^2+m_{\pi}^2}{2 F^2}\left(g_8+g_{27}^{1/2}+g_{27}^{3/2} \right)
	+\frac{g_8' \left(m_{\pi}^2-m_{K^{\pm}}^2\right)}{F^2 y_{sd}}\,,\label{eq_F_Km_to_pim_a_cv}
	\end{equation}
\end{subequations}
where the functions $x_{1,2,3}$ are defined as
\begin{subequations}\label{eq_x}
	\begin{align}
	&x_1(m^2)=\frac{\beta_{11}^2m^2}{m_a^2-m_{\eta}^2}+\frac{\beta_{12}^2m^2}{m_a^2-m_{\eta'}^2}\,,
	\\
	&x_2(m^2)=\frac{\beta_{21}^2m^2}{m_a^2-m_{\eta}^2}+\frac{\beta_{22}^2m^2}{m_a^2-m_{\eta'}^2}\,,
	\\
	&x_3(m^2)=\frac{\beta_{11}\beta_{21}m^2}{m_a^2-m_{\eta}^2}+\frac{\beta_{12}\beta_{22}m^2}{m_a^2-m_{\eta'}^2}\,.
	\end{align}
\end{subequations}

It is noted that each diagram in Figs.~\ref{fig_Km_to_pim_a_c2-1}--\ref{fig_Km_to_pim_a_c2-3} depends on $\kappa_q$. Since the physical amplitude should be independent of $\kappa_q$, this requires that  the $\kappa_q$ dependence cancels out in the final result. We explicitly verify that the sum of different amplitudes in Eq.~\eqref{eq_A_Km_to_pim_a_WFC} indeed fulfills this requirement, confirming the correctness of our calculation. In this procedure, we need to take the one-mixing angle formula for the $\eta$-$\eta'$ system. It is also worth mentioning that, to fully account for the relations between the LO meson masses, one must explicitly substitute Eqs.~(\ref{eq_eta_8-0_rotation}) and (\ref{eq_m0_eta_etap}) into the amplitude to guarantee the exact cancellation of the $\kappa_q$.  
Furthermore, the $\eta_0$ degree of freedom can be integrated out by replacing $\beta_{ij}$ with $\st$, $\ct$ and taking the limit $M_0\to\infty$. In this limit, our amplitude reproduces the SU(3) results of Ref.~\cite{Bauer:2021mvw} (the LECs in Ref.~\cite{Bauer:2021mvw} are related to ours by $(f_{\pi})_{\rm Bauer,\,et.al}=\sqrt{2}F$, $(N_8)_{\rm Bauer,\,et.al}=2G_Wg_8/F^2$, $(N_{27}^i)_{\rm Bauer,\,et.al}=2G_Wg_{27}^i/F^2$, with $i=1/2$, $3/2$; the contributions from $g_8'$, $g_8^{P}$, $g_8^{\psi}$ and $g_8^{\theta}$ are not included in Ref.~\cite{Bauer:2021mvw}) and of \cite{Cornella:2023kjq} (the LECs in Ref.~\cite{Cornella:2023kjq} are related to ours by $(F)_{\rm Cornella,\,et.al}=\sqrt{2}F$, $(G_8)_{\rm Cornella,\,et.al}=G_8/F^4$, $(G_8')_{\rm Cornella,\,et.al}=G_8'/F^4$, $(G_{27}^i)_{\rm Cornella,\,et.al}=G_{27}^i/F^4$, and $(G_8^{\theta})_{\rm Cornella,\,et.al}=-G_8^{\theta}/F^4$, with $i=1/2$, $3/2$; the terms involving $G_8^P$ and $G_8^{\psi}$ are not included in the SU(3) result of Ref.~\cite{Cornella:2023kjq}).

\begin{figure}[t]
	\centering
	\subfigure[class-1: (1)\label{fig_Kbar0_to_pi0_a-c1-1}]
	{\includegraphics[width=0.3\linewidth]{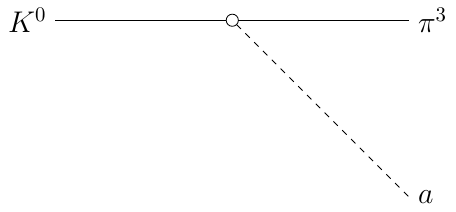}}
	\hfill
	\subfigure[class-1: (2)\label{fig_Kbar0_to_pi0_a-c1-2}]
	{\includegraphics[width=0.3\linewidth]{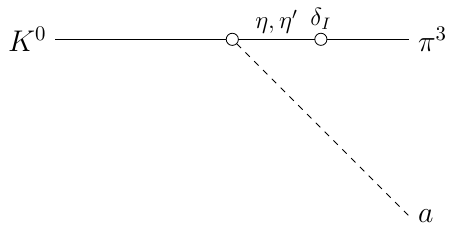}}
	\hfill
	\subfigure[class-2: (1)\label{fig_Kbar0_to_pi0_a-c2-1}]
	{\includegraphics[width=0.3\linewidth]{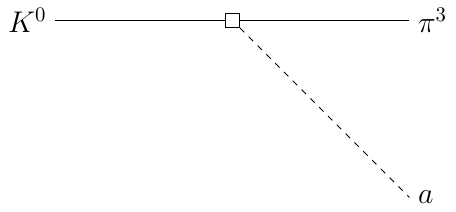}}
	\\
	\subfigure[class-2: (2)\label{fig_Kbar0_to_pi0_a-c2-2}]
	{\includegraphics[width=0.3\linewidth]{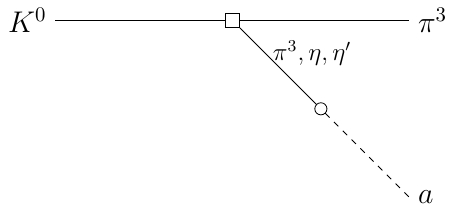}}
	\hfill
	\subfigure[class-2: (3)\label{fig_Kbar0_to_pi0_a-c2-3}]
	{\includegraphics[width=0.3\linewidth]{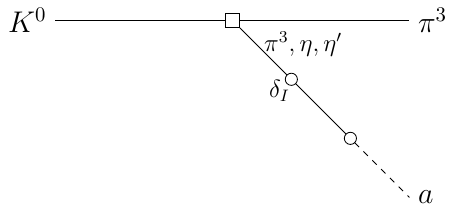}}
	\hfill
	\subfigure[class-2: (4)\label{fig_Kbar0_to_pi0_a-c2-4}]
	{\includegraphics[width=0.3\linewidth]{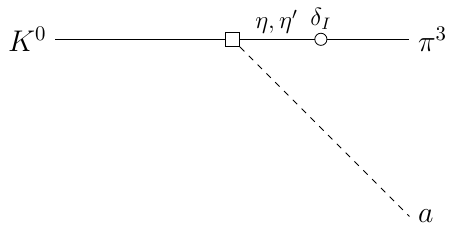}}
	\\
	\subfigure[class-2: (5)\label{fig_Kbar0_to_pi0_a-c2-5}]
	{\includegraphics[width=0.3\linewidth]{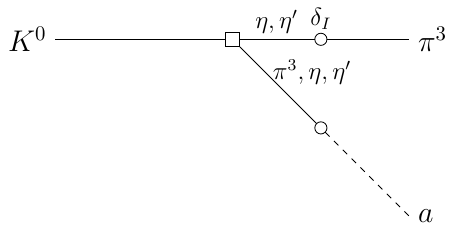}}
	\hfill
	\subfigure[class-2: (6)\label{fig_Kbar0_to_pi0_a-c2-6}]
	{\includegraphics[width=0.3\linewidth]{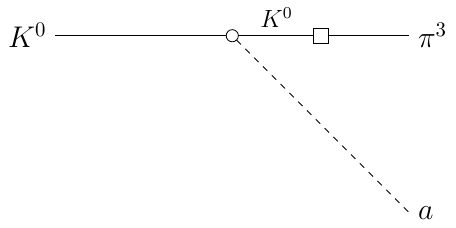}}
	\hfill
	\subfigure[class-2: (7)\label{fig_Kbar0_to_pi0_a-c2-7}]
	{\includegraphics[width=0.3\linewidth]{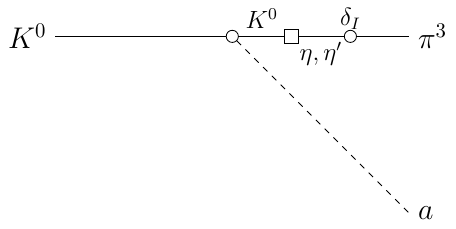}}
	\caption{Feynman diagrams for the three-point Green function $\langle 0 | \pi^3 a K^0|0\rangle$ associated with the $\bar{K}^0\to\pi^0a$ decay. Empty circles denote vertices from $\mathcal{L}_{\mathrm{eff}}^{(0)}$, while empty rectangles denote vertices from ${\mL_{\rm weak}^{\mathrm{U(3)}\,\ChPT}}'$. The labels $K^0$, $\pi^3$, and $a$ at the ends of the external lines denote the corresponding interpolating fields, where  $\pi^3$ interpolates the final $\pi^0$ state.}\label{fig_Kbar0_to_pi0_a}
\end{figure}

Similarly, the $\bar{K}^0\to\pi^0 a$ decay amplitude can be extracted from the three-point Green function involving the $K^0$, $\pi^3$, and $a$ fields. For the class-1 and class-2 contributions at LO, the relevant Feynman diagrams are shown in Fig.~\ref{fig_Kbar0_to_pi0_a}. The decay amplitude is separated into two parts as 
\begin{equation}
A_{\bar{K}^0\to\pi^0 a}=A^{\rm AFC}_{\bar{K}^0\to\pi^0 a}+A^{\rm WFC}_{\bar{K}^0\to\pi^0 a}\,,
\end{equation}
where $A^{\rm AFC}_{\bar{K}^0\to\pi^0 a}$ receives contributions from the class-1 diagrams in Figs.~\ref{fig_Kbar0_to_pi0_a-c1-1} and \ref{fig_Kbar0_to_pi0_a-c1-2}  
\begin{equation}
\begin{aligned}
A^{\rm AFC}_{\bar{K}^0\to\pi^0 a}=&\frac{\left(k_d+k_D\right)_{12}}{2\sqrt{2}f}\left(m_{K^0}^2-m_{\pi}^2\right)
\\
&\times\left\{
1+\delta_I\left[\frac{\beta_{11}(\beta_{11}+\sqrt{2}\beta_{21})}{m_{\pi}^2-m_{\eta}^2}+\frac{\beta_{12}(\beta_{12}+\sqrt{2}\beta_{22})}{m_{\pi}^2-m_{\eta'}^2}\right]
\right\}\,,\label{eq_A_Kbar0-to-pi0-a_AFC}
\end{aligned}
\end{equation}
and $A^{\rm WFC}_{\bar{K}^0\to\pi^0 a}$ receives contributions from the class-2 diagrams in Figs.~\ref{fig_Kbar0_to_pi0_a-c2-1}--\ref{fig_Kbar0_to_pi0_a-c2-7}, and can be decomposed according to flavor-conserving ALP couplings as
\begin{equation}
A^{\rm WFC}_{\bar{K}^0\to\pi^0 a}=G_W\sum_{\cALP}\frac{\cALP}{f}F_{\bar{K}^0\to\pi^0 a}^{\cALP}\,.
\label{eq_A_Kbar0-to-pi0-a_WFC}
\end{equation}
Each diagram of Figs.~\ref{fig_Kbar0_to_pi0_a-c2-1}--\ref{fig_Kbar0_to_pi0_a-c2-5} depends on $\kappa_q$ individually, and it is explicitly verified that the $\kappa_q$ dependence cancels out in the final amplitude. The full expressions for $F_{\bar{K}^0\to\pi^0 a}^{\cALP}$ are rather lengthy and are therefore not presented here. Instead, they are provided in a companion Mathematica notebook, which is publicly available on Zenodo~\cite{wang_2026_21824113}. In the limits $M_0\to\infty$ and $\delta_I\to 0$, the expressions of $F_{\bar{K}^0\to\pi^0 a}^{\cALP}$ reduces to 
\begin{subequations}
	\begin{align}
	F_{\bar{K}^0\to\pi^0a}^{c_{GG}}\big|_{M_0\to\infty,\delta_I\to 0}=&
	\frac{\sqrt{2} \left(m_{K^0}^2-m_{\pi}^2\right)}{F^2}
	\Bigg[g_8^{\psi}-g_8^{\theta}
	-\frac{4\left(m_a^2-m_{K^0}^2\right)}{3 m_a^2-4 m_{K^0}^2+m_{\pi}^2}\left(g_8-g_{27}^{3/2}\right)
	\notag\\
	&-\frac{2 g_{27}^{1/2} \left(m_a^2-4 m_{K^0}^2+3 m_{\pi}^2\right)}{3 m_a^2-4 m_{K^0}^2+m_{\pi}^2}
	\Bigg]\,,
	\end{align}
	\begin{align}
	F_{\bar{K}^0\to\pi^0a}^{c^a_{uu}}\big|_{M_0\to\infty,\delta_I\to 0}=&
	\frac{m_{K^0}^2-m_{\pi}^2}{\sqrt{2} F^2}
	\Bigg[g_8^{\psi}-g_8^{\theta}
	-\frac{4 g_8 m_a^2 \left(m_a^2-m_{K^0}^2\right)}{\left(m_a^2-m_{\pi}^2\right) \left(3 m_a^2-4 m_{K^0}^2+m_{\pi}^2\right)}
	\notag\\
	&+\frac{g_{27}^{3/2} \left[4 m_a^4+m_a^2 \left(3 m_{\pi}^2-4 m_{K^0}^2\right)-4 m_{K^0}^2 m_{\pi}^2+m_{\pi}^4\right]}{\left(m_a^2-m_{\pi}^2\right) \left(3 m_a^2-4 m_{K^0}^2+m_{\pi}^2\right)}
	\notag\\
	&-\frac{g_{27}^{1/2} \left[2 m_a^4+m_a^2 \left(3 m_{\pi}^2-8 m_{K^0}^2\right)+4 m_{K^0}^2 m_{\pi}^2-m_{\pi}^4\right]}{\left(m_a^2-m_{\pi}^2\right) \left(3 m_a^2-4 m_{K^0}^2+m_{\pi}^2\right)}
	\Bigg]\,,
	\end{align}
	\begin{align}
	F_{\bar{K}^0\to\pi^0a}^{c^a_{dd}}\big|_{M_0\to\infty,\delta_I\to 0}=&
	\frac{\left(m_{K^0}^2-m_{\pi}^2\right) (g_8^{\psi}-g_8^{\theta}-g_8' y_{sd})}{\sqrt{2} F^2}
	\notag\\
	&-\frac{1}{2 \sqrt{2} F^2 \left(m_a^2-m_{\pi}^2\right) }\Bigg\{
	\frac{3 m_a^6 \left(g_8+g_{27}^{1/2}-2 g_{27}^{3/2}\right)}{3 m_a^2-4 m_{K^0}^2+m_{\pi}^2}
	\notag\\
	&+\frac{m_a^4 \left[g_8 \left(m_{K^0}^2-7 m_{\pi}^2\right)-3 \left(g_{27}^{1/2}-2 g_{27}^{3/2}\right) \left(m_{K^0}^2+m_{\pi}^2\right)\right]}{3 m_a^2-4 m_{K^0}^2+m_{\pi}^2}
	\notag\\
	&-\frac{m_a^2}{3 m_a^2-4 m_{K^0}^2+m_{\pi}^2} \Big[g_8 \left(4 m_{K^0}^4+2 m_{K^0}^2 m_{\pi}^2-9 m_{\pi}^4\right)
	\notag\\
	&+3 g_{27}^{1/2} \left(4 m_{K^0}^4-4 m_{K^0}^2 m_{\pi}^2-m_{\pi}^4\right)
	-6 g_{27}^{3/2} m_{\pi}^2 \left(m_{K^0}^2-2 m_{\pi}^2\right)
	\Big]
	\notag\\
	&+\left(3 g_8+5 g_{27}^{1/2}-4 g_{27}^{3/2}\right) \frac{\left(4 m_{K^0}^4 m_{\pi}^2-5 m_{K^0}^2 m_{\pi}^4+m_{\pi}^6\right)}{3 m_a^2-4 m_{K^0}^2+m_{\pi}^2}
	\Bigg\}\,,
	\end{align}
	\begin{align}
	F_{\bar{K}^0\to\pi^0a}^{c^a_{ss}}\big|_{M_0\to\infty,\delta_I\to 0}=&
	\frac{\left(m_{K^0}^2-m_{\pi}^2\right) (g_8^{\psi}-g_8^{\theta}+g_8' y_{sd})}{\sqrt{2} F^2}
	\notag\\
	&+\frac{1}{2 \sqrt{2} F^2 \left(3 m_a^2-4 m_{K^0}^2+m_{\pi}^2\right)}
	\Bigg\{
	3 m_a^4 \left(g_8+g_{27}^{1/2}-2 g_{27}^{3/2}\right)
	\notag\\
	&+m_a^2 \left[\left(g_{27}^{1/2}-2 g_{27}^{3/2}\right) \left(m_{K^0}^2-4 m_{\pi}^2\right)-3 g_8 m_{K^0}^2\right]
	\notag\\
	&+\left(g_8-7 g_{27}^{1/2}+2 g_{27}^{3/2}\right) \left(4 m_{K^0}^4-5 m_{K^0}^2 m_{\pi}^2+m_{\pi}^4\right)
	\Bigg\}\,,
	\end{align}
	\begin{equation}
	F_{\bar{K}^0\to\pi^0a}^{c^v_{dd}-c^v_{ss}}\big|_{M_0\to\infty,\delta_I\to 0}=
	\frac{\left(g_8+g_{27}^{1/2}-2 g_{27}^{3/2}\right) \left(m_a^2-m_{K^0}^2-m_{\pi}^2\right)}{2 \sqrt{2} F^2}+\frac{g_8' \left(m_{K^0}^2-m_{\pi}^2\right)}{\sqrt{2} F^2 y_{sd}}\,.\label{eq_F_Kbar0-to-pi0-a-cv}
	\end{equation}
\end{subequations}
In the limit $M_0\to\infty$, our amplitude reproduces the SU(3) result of Ref.~\cite{Bauer:2021mvw}, except for a difference in the mass factor accompanying $g_{27}^{3/2}$ in $F_{\bar{K}^0\to\pi^0a}^{c^v_{dd}-c^v_{ss}}$. 
In the Appendix~\ref{sec.nckpia}, we elaborate that the $K\to\pi a$ amplitudes derived in U(3) $\ChPT$ fulfill the large-$N_C$ behaviors as expected in QCD. Finally, the decay amplitudes for the CP-conjugate processes $K^+\to\pi^+ a$ and $K^0\to\pi^0 a$ can be obtained from those for $K^-\to\pi^-a$ and $\bar{K}^0\to\pi^0a$, respectively, by reversing the overall sign, together with the replacements $(k_d+k_D)_{12}\to (k_d+k_D)_{21}$ and $G_i\to G_i^*$, with $G_i$ the weak chiral LECs. 

\subsection{Unitarized $K \to \pi a$ and $K\to \pi\pi$ amplitudes with $\pi\pi$ final-state interactions}

It is well known that, in the $K\to\pi\pi$ decay, $\pi\pi$ final-state interactions (FSIs) generate sizable strong phases and play an essential role in the long-distance enhancement associated with the $|\Delta I|=1/2$ rule. It is therefore natural to ask whether they also have a significant impact on the $K\to\pi a$ decay. To address this question, we construct the unitarized $K\to\pi a$ amplitude by incorporating $\pi\pi$ FSIs based on the LO amplitudes.

\subsubsection{Unitarized $K\to\pi\pi$ decay amplitude}

We begin by revisiting the well-studied $K\to\pi\pi$ decays, which have been extensively investigated in a series of works~\cite{Kambor:1991ah,Cirigliano:2003gt,Cirigliano:2009rr} and comprehensively reviewed in Ref.~\cite{Cirigliano:2011ny}. Within U(3) $\ChPT$, the LO ${K}\to\pi\pi$ decay amplitudes can be calculated from the weak chiral Lagrangian in Eq.~\eqref{eq.lagwxpt}, yielding
\begin{subequations}\small\label{eq_A_Ktopipi}
	\begin{align}
	&\begin{aligned}
	A_{K^-\to\pi^-\pi^0}=&
	\frac{m_{K^{\pm}}^2-m_{\pi}^2}{F^3}
	\Bigg\{-3 G_{27}^{3/2}
	\\
	&+\frac{1}{3} \left[2 G_8+3 \left(4 G_{27}^{1/2}+G_{27}^{3/2}\right)\right] \left(\frac{\beta_{11}^2 \delta_I}{m_{\pi}^2-m_{\eta}^2}+\frac{\beta_{12}^2 \delta_I}{m_{\pi}^2-m_{\eta'}^2}\right)
	\\
	&+\frac{2}{3} \left[2 G_8-3\left(G_8^{\psi}+2 G_8^P\right)\right] \left(\frac{\beta_{21}^2 \delta_I}{m_{\pi}^2-m_{\eta}^2}+\frac{\beta_{22}^2 \delta_I}{m_{\pi}^2-m_{\eta'}^2}\right)
	\\
	&+\frac{\sqrt{2}}{3}  \left[4 G_8+3 \left(4 G_{27}^{1/2}+G_{27}^{3/2}-G_8^{\psi}-2 G_8^P\right)\right] \left(\frac{\beta_{11} \beta_{21} \delta_I}{m_{\pi}^2-m_{\eta}^2}+\frac{\beta_{12} \beta_{22} \delta_I}{m_{\pi}^2-m_{\eta'}^2}\right)
	\Bigg\}\,,
	\end{aligned}
	\\
	&\begin{aligned}
	A_{\bar{K}^0\to\pi^+\pi^-}=-\frac{\sqrt{2}\left(m_{K^0}^2-m_{\pi}^2\right)}{F^3}\left(G_8+G_{27}^{1/2}+G_{27}^{3/2}\right)\,,
	\end{aligned}
	\\
	&\begin{aligned}
	A_{\bar{K}^0\to\pi^0\pi^0}=&
	-\frac{\sqrt{2} \left(m_{K^0}^2-m_{\pi}^2\right)}{F^3}
	\Bigg\{G_8+G_{27}^{1/2}-2 G_{27}^{3/2}
	\\
	&+\frac{2}{3} \left(G_8+6 G_{27}^{1/2}-3 G_{27}^{3/2}\right) \left(\frac{\beta_{11}^2 \delta_I}{m_{\pi}^2-m_{\eta}^2}+\frac{\beta_{12}^2 \delta_I}{m_{\pi}^2-m_{\eta'}^2}\right)
	\\
	&+\frac{2}{3} \left[2 G_8-3 \left(G_8^{\psi}+2 G_8^P\right)\right] \left(\frac{\beta_{21}^2 \delta_I}{m_{\pi}^2-m_{\eta}^2}+\frac{\beta_{22}^2 \delta_I}{m_{\pi}^2-m_{\eta'}^2}\right)
	\\
	&+\frac{\sqrt{2}}{3}  \left[4 G_8-3 \left(G_8^{\psi}+2 G_8^P-4 G_{27}^{1/2}+2 G_{27}^{3/2}\right)\right] \left(\frac{\beta_{11} \beta_{21} \delta_I}{m_{\pi}^2-m_{\eta}^2}+\frac{\beta_{12} \beta_{22} \delta_I}{m_{\pi}^2-m_{\eta'}^2}\right)
	\Bigg\}\,.
	\end{aligned}
	\end{align}
\end{subequations}

Throughout this work, we adopt the following phase convention for the isospin states
\begin{equation}
|\pi^+\rangle=-|I=1,I_z+1\rangle\,,
\quad
|\pi^0\rangle =|I=1,I_z=0\rangle\,,
\quad
|\pi^-\rangle=|I=1,I_z=-1\rangle\,.
\end{equation}
Accordingly, the ${K}\to\pi\pi$ decay amplitudes in such bases read
\begin{subequations}
	\begin{align}
	&A_{K^-\to(\pi\pi,I=2)}=\sqrt{2}A_{K^-\to\pi^-\pi^0}\,,
	\\
	&A_{\bar{K}^0\to(\pi\pi,I=0)}=-\frac{2}{\sqrt{3}}A_{\bar{K}^0\to\pi^+\pi^-}-\frac{1}{\sqrt{3}}A_{\bar{K}^0\to\pi^0\pi^0}\,,
	\\
	&A_{\bar{K}^0\to(\pi\pi,I=2)}=-\sqrt{\frac{2}{3}}A_{\bar{K}^0\to\pi^+\pi^-}+\sqrt{\frac{2}{3}}A_{\bar{K}^0\to\pi^0\pi^0}\,.
	\end{align}
\end{subequations}

As emphasized above, it is crucial to properly incorporate the strong $\pi\pi$ FSIs in the $K\to\pi\pi$ decays, in which the final $\pi\pi$ states can only be in the $S$ wave. 
The partial-wave expansion for $2\to 2$ scattering of spinless particles in the center-of-mass frame can be expressed as
\begin{equation}\label{eq_pwd}
T_{i\to f}(s,\cos\theta)=\left(\sqrt{2}\right)^{N_i+N_f}\sum_{J}(2J+1)P_J(\cos\theta)T_{i\to f}^{J}(s)\,,
\end{equation}
where $s$ is the square of the center-of-mass energy, $\theta$ is the scattering angle, $P_J(\cos\theta)$ is the Legendre polynomial, and $T_{i\to f}^J$ is the partial-wave amplitude with angular momentum $J$. The normalization of the partial-wave amplitude depends on whether the initial and final states are identical particles. If the initial state $i$ (or final state $f$) consists of a pair of identical particles, then $N_{i(f)}=1$; otherwise $N_{i(f)}=0$. 

To be consistent with the normalization in Eq.~(\ref{eq_pwd}), we define the isospin amplitudes $a_{\bar{K}\to(\pi\pi,I)}$ through
\begin{equation}\label{eq.ampkpipiiso}
A_{\bar{K}\to(\pi\pi,I)}=\sqrt{2}a_{\bar{K}\to(\pi\pi,I)}\,.
\end{equation}
Restricting to two-body unitarity and neglecting isospin breaking as well as CP violation in the FSIs, $a_{\bar{K}\to(\pi\pi,I)}$ satisfies the unitarity relation
\begin{equation}
\mathrm{Im}a_{\bar{K}\to(\pi\pi,I)}=\rho_{\pi\pi}(m_{K}^2)\left[T_{\pi\pi\to\pi\pi}^{I0}(m_K^2)\right]^*a_{\bar{K}\to(\pi\pi,I)}\,,\label{eq_ur_Ktopipi}
\end{equation}
where $\rho_{\pi\pi}(s)= \sqrt{\lambda(s,m_{\pi}^2,m_{\pi}^2)}/(16\pi s)$ with 
$\lambda(x,y,z)=x^2+y^2+z^2-2xy-2xz-2yz$ the K\"all\'en function and $T_{\pi\pi\to\pi\pi}^{I0}(s)$ is the $S$-wave $\pi\pi$ scattering amplitude with definite isospin $I$. In the presence of CP violation, since the CP-odd phase originates solely from the CKM ratio $\tau$, one can always write the amplitude as $A = \bar{A}-\tau \widetilde{A}$ and separately study the partial-wave unitarity for the CP-even amplitudes $\bar{A}$ and $\widetilde{A}$~\cite{Pallante:2000hk}.

\begin{figure}
	\centering
	\includegraphics[scale=0.5]{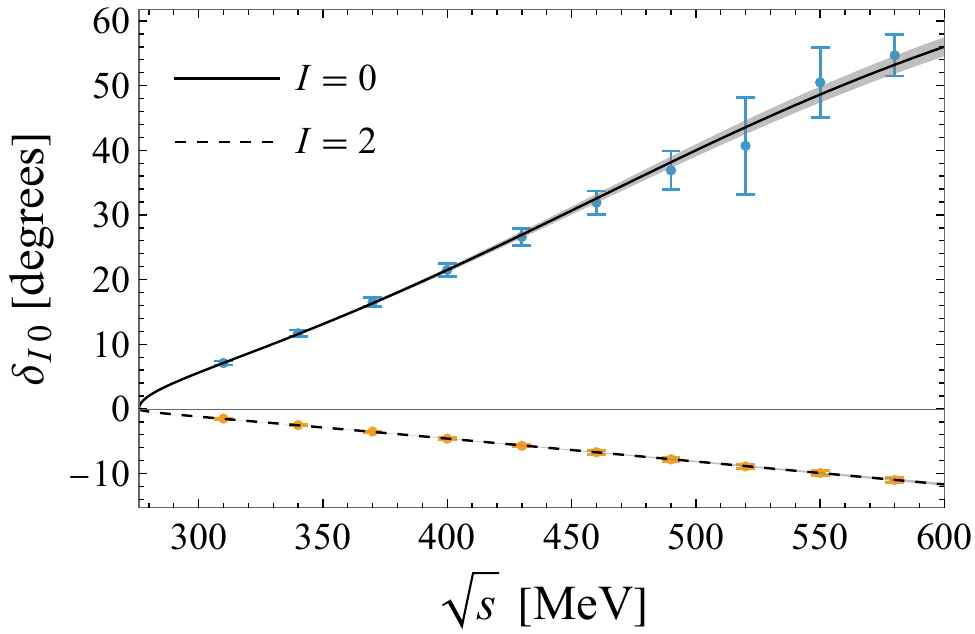}
	\caption{Fitting results for the $S$-wave $\pi\pi$ phase shifts in the energy region $\sqrt{s}<600$~MeV. The phase-shift data are taken from the Roy-equation analysis~\cite{Garcia-Martin:2011iqs}.}\label{fig_ps}
\end{figure}

The $K\to\pi\pi$ amplitudes in Eq.~\eqref{eq_A_Ktopipi} from the perturbative $\ChPT$ clearly do not fulfill the unitarity relations of Eq.~\eqref{eq_ur_Ktopipi}. Following our previous study of the $\eta\to \pi\pi a$ decay in Ref.~\cite{Wang:2024tre}, we use the chiral unitarization approach to improve the description of the $K\to\pi\pi$ amplitudes. In this approach, the unitarized $S$-wave $\pi\pi$ scattering amplitude, that satisfies elastic unitarity, can be constructed as~\cite{Oller:1998zr}
\begin{equation}
T_{\pi\pi\to\pi\pi}^{I0,\uni}(s)=\frac{T_{\pi\pi\to\pi\pi}^{I0,\LO}(s)}{1-G_{\pi\pi}(s)T_{\pi\pi\to\pi\pi}^{I0,\LO}(s)}\,,\label{eq_T_pipi_uni}
\end{equation}
where $T_{\pi\pi\to\pi\pi}^{I0,\LO}$ is the LO $\pi\pi$ $S$-wave amplitude and is given by
\begin{equation}
T_{\pi\pi\to\pi\pi}^{00,\LO}=\frac{2s-m_{\pi}^2}{2F^2}\,,
\quad
T_{\pi\pi\to\pi\pi}^{20,\LO}=\frac{2m_{\pi}^2-s}{2F^2}\,.
\end{equation}
The quantity $G_{\pi\pi}(s)$ is the two-particle one-loop function evaluated using dimensional regularization, which reads
\begin{equation}
G_{\pi\pi}(s)=\frac{-1}{(4\pi)^2}\left[\asc(\mu)+\log\frac{m_{\pi}^2}{\mu^2}-\frac{\sqrt{\lambda(s,m_{\pi}^2,m_{\pi}^2)}}{s}\log\frac{2m_{\pi}^2-s+\sqrt{\lambda(s,m_{\pi}^2,m_{\pi}^2)}}{2m_{\pi}^2}\right]\,,
\end{equation}
where $\asc$ is the corresponding subtraction constant and $\mu=770$~MeV is taken in our study. The imaginary part of $G_{\pi\pi}(s)$ is given by ${\rm Im}G_{\pi\pi}(s)=\theta(s-4m_\pi^2)\rho_{\pi\pi}(s)$. The values of the subtraction constants can be determined by fitting the $S$-wave $\pi\pi$ phase shifts. Below the inelastic threshold, the $\pi\pi$ phase shift $\delta_{I0}$ can be obtained through $e^{2i\delta_{I0}}=1+2i\rho_{\pi\pi}T_{\pi\pi\to\pi\pi}^{I0,\uni}$. By assigning different subtraction constants $\asc^{00}$ and $\asc^{20}$ for the $IJ=00$ and $IJ=20$ channels, respectively, and fitting to the phase-shift data in the energy range $\sqrt{s}<600$~MeV from the Roy-equation analysis~\cite{Garcia-Martin:2011iqs}, we obtain
\begin{equation}
\asc^{00}=-1.01^{+0.11}_{-0.10}\,,
\quad
\asc^{20}=-4.25^{+0.25}_{-0.26}\,.\label{eq_asc}
\end{equation}
The fitted results are shown in Fig.~\ref{fig_ps}, together with the Roy-equation inputs. The pole position of the $f_0(500)$ resonance on the second Riemann sheet is found to be $\sqrt{s}_{\rm pole} = (458 - 250 i)$~MeV, which is consistent with the PDG average value~\cite{ParticleDataGroup:2024cfk}.

Accordingly, the unitarized ${K}\to\pi\pi$ decay amplitude that satisfies the unitarity relation can be expressed as
\begin{equation}
a_{\bar{K}\to(\pi\pi,I)}^{\uni}=\frac{a_{\bar{K}\to(\pi\pi,I)}^{\tree}}{1-G_{\pi\pi}(m_K^2)T_{\pi\pi\to\pi\pi}^{I0,\LO}(m_K^2)}\,,\label{eq_a_Ktopipi_uni}
\end{equation}
where $a_{\bar{K}\to(\pi\pi,I)}^{\tree}$ refers to the tree-level amplitudes given in Eq.~(\ref{eq_A_Ktopipi}) after taking into account the identical-particle factor in Eq.~\eqref{eq.ampkpipiiso}. It is then straightforward to verify that the unitarized amplitudes of $a_{\bar{K}\to(\pi\pi,I)}^{\uni}$ and $
T_{\pi\pi\to\pi\pi}^{I0,\uni}(s)$ now fulfill the unitarity condition shown in Eq.~\eqref{eq_ur_Ktopipi}. 

\subsubsection{Unitarized $K\to\pi a$ decay amplitude}

\begin{figure}[t]
	\centering
	\includegraphics[width=0.3\linewidth]{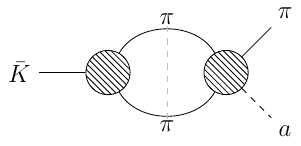}
	\caption{Schematic diagram for the absorptive part of $\bar{K}\to\pi a$ decay.}\label{fig_absorbed}
\end{figure}

Following the similar unitarized formalism as employed in the $K\to\pi\pi$ process previously, we construct the unitarized $K\to \pi a$ amplitude. 
As shown in Fig.~\ref{fig_absorbed}, the SM weak interaction vertex necessarily participates in the two-body cut of the $\bar{K}\to\pi a$ decay. Therefore, only the WFC amplitude contains the absorptive part and the AFC amplitude does not contain it. At $s=m_K^2$, only $\pi\pi$ intermediate states can be on-shell, which are expected to be the most relevant ones in the $\bar{K}\to\pi a$ decays. The absorptive part of the $\bar{K}\to\pi a$ decay amplitude is given by
\begin{equation}
\mathrm{Im}A_{\bar{K}\to\pi a}^{\rm WFC} = \sum_{I}\rho_{\pi\pi}(m_K^2)\left[T_{(\pi\pi,I)\to\pi a}^{J=0}(m_K^2)\right]^*a_{\bar{K}\to(\pi\pi,I)}\,,\label{eq_Ktopia-absorbed}
\end{equation}
where $I=2$ is allowed in the $K^-$ decay, both $I=0$ and 2 are allowed in the $\bar{K}^0$ decay, and $T_{(\pi\pi,I)\to\pi a}^{J=0}(s)$ is the $S$-wave $\pi\pi\to\pi a$ scattering amplitude.
The unitarized $\bar{K}\to\pi a$ decay amplitude and the unitarized $\pi\pi\to\pi a$ scattering amplitudes can be constructed as follows
\begin{align}
&A_{\bar{K}\to\pi a}^{\rm WFC,uni} = A_{\bar{K}\to\pi a}^{\rm WFC,tree} 
+\sum_I\frac{a_{\bar{K}\to(\pi\pi,I)}^{\tree}G_{\pi\pi}(m_K^2)T_{(\pi\pi,I)\to\pi a}^{J=0,\LO}(m_K^2)}{1-T_{\pi\pi\to\pi\pi}^{I0,\LO}(m_K^2)G_{\pi\pi}(m_K^2)}\,,\label{eq_Ktopia-WFC-uni}
\\
&T_{(\pi\pi,I)\to\pi a}^{J=0,\uni}(s)=\frac{T_{(\pi\pi,I)\to\pi a}^{J=0,\LO}(s)}{1-T_{\pi\pi\to\pi\pi}^{I0,\LO}(s)G_{\pi\pi}(s)}\,.
\end{align}
These amplitudes, together with $a_{\bar{K}\to(\pi\pi,I)}^{\uni}$, satisfy the unitarity relation in Eq.~(\ref{eq_Ktopia-absorbed}). Here, $A_{\bar{K}\to\pi a}^{\rm WFC,tree}$ is the tree-level decay amplitude computed in Sec.~\ref{sec_tree-level-Ktopia-amps}, and $T_{(\pi\pi,I)\to\pi a}^{J=0,\LO}$ is the LO $\pi\pi\to\pi a$  $S$-wave amplitude, whose full expressions are provided in the attached Mathematica notebook~\cite{wang_2026_21824113}. Following the notation in Eq.~\eqref{eq_A_Km_to_pim_a_WFC}, one can similarly define the unitarized $F_{K^-\to\pi^-a}^{\cALP}$ as well. 

In Ref.~\cite{Cornella:2023kjq}, the NLO $K^-\to\pi^-a$ decay amplitude is computed within SU(3) $\ChPT$ by neglecting the 27-plet contributions and the $\mO(\delta_I)$ terms in the $\mO(p^4)$ corrections. The absorptive part of the amplitude is found to be absent in this approximation. The $\pi\pi$ FSIs are not expected to have a significant effect in the $K^-\to\pi^-a$ decay, since this decay is expected to be dominated by the $|\Delta I|=1/2$ contribution, while only the $I=2$ component of the $\pi\pi$ intermediate states, corresponding to $|\Delta I|=3/2$, is allowed in $K^-\to\pi^-a$. In contrast, the situation is different for the $\bar{K}^0\to\pi^0 a$ decay, which involves both $\pi\pi$ intermediate states with $I=0$ and $I=2$.
We will investigate the phenomenological consequences of the $\pi\pi$ loops in $\bar{K}\to\pi a$ decay via Eq.~(\ref{eq_Ktopia-WFC-uni}) in the following section.  

\section{Phenomenological discussions}\label{sec.pheno}

\subsection{Determination of the weak chiral couplings from the $K\to \pi\pi$ decay}

In the experimental analyses, the ${K}\to\pi\pi$ decay amplitudes are usually parameterized as
\begin{subequations}\label{eq_Ktopipi-Apara}
	\begin{align}
	&A_{K^-\to\pi^-\pi^0}=\frac{3}{2}A_2^-e^{i\chi_2^-}\,,
	\\
	&A_{\bar{K}^0\to\pi^+\pi^-}=\bar{A}_0e^{i\bar{\chi}_0}+\frac{1}{\sqrt{2}}\bar{A}_2e^{i\bar{\chi}_2}\,,
	\\
	&A_{\bar{K}^0\to\pi^0\pi^0}=\bar{A}_0e^{i\bar{\chi}_0}-\sqrt{2}\bar{A}_2e^{i\bar{\chi}_2}\,,
	\end{align}
\end{subequations}
where the subscripts $0$ and $2$ label the isospin of the final $\pi\pi$ states, and $\chi_2^-$ and $\bar{\chi}_{0,2}$ are the corresponding strong phases. The small CP-odd phases are contained in the amplitudes $A_2^-$ and $\bar{A}_{0,2}$. In the CP-conserving case, these amplitudes are real and positive by definition, and in the isospin limit one has $A_2^-=\bar{A}_2$. By taking the isospin limit and neglecting CP violation, the amplitudes $\bar{A}_{0,2}$ and the phase difference $\bar{\chi}_0-\bar{\chi}_2$ can be extracted from the three ${K}\to\pi\pi$ branching ratios~\cite{Cirigliano:2011ny}:
\begin{subequations}\label{eq_Ktopipi-exp}
	\begin{align}
	&\bar{A}_0^{\rm Exp}=\left(2.704\pm 0.001\right)\times 10^{-7}~\GeV\,,\label{eq_Ktopipi-exp-a}
	\\
	&\bar{A}_2^{\rm Exp}=\left(1.210\pm 0.002\right)\times 10^{-8}~\GeV\,,\label{eq_Ktopipi-exp-b}
	\\
	&(\bar{\chi}_0-\bar{\chi}_2)^{\rm Exp}=\left(47.5\pm 0.9\right)\degree\,.
	\end{align}
\end{subequations}
The $|\Delta I|=1/2$ rule in ${K}\to\pi\pi$ decay is reflected by the large ratio of $\bar{A}_0/\bar{A}_2\simeq 22$.

The values of the weak chiral couplings $g_i$~\eqref{eq.defgw} can be determined from the measurements of ${K}\to\pi\pi$ decays. A fit to the experimental branching ratios using the NLO SU(3) ${K}\to\pi\pi$ decay amplitudes including isospin breaking effects gives~\cite{Cirigliano:2011ny}
\begin{equation}
g_8= 3.61\pm 0.28\,,\qquad g_{27}^{1/2}= 0.033\pm0.003\,.\label{eq_g_NLO-fitted}
\end{equation}
with $g_{27}^{3/2}= 5g_{27}^{1/2}$. Note that the $g_{27}$ defined in Ref.~\cite{Cirigliano:2011ny} is related to $g_{27}^{1/2}$ used here by $g_{27} = 9 g_{27}^{1/2}$.

In this work, the chiral unitarization approach in the U(3) case is employed to describe the decays of $K\to\pi\pi$ and $K\to\pi a$. In order to be self consistent in the phenomenological discussions, we revise the determination of the weak chiral LECs by taking the unitarized $K\to\pi\pi$ amplitude in Eq.~\eqref{eq_a_Ktopipi_uni}, where the $\pi\pi$ FSIs have been properly taken into account through the fit to the $\pi\pi$ phase shifts. A consistent treatment of the isospin-breaking effects in the $K\to\pi\pi$ processes would require the inclusion of electromagnetic corrections~\cite{Cirigliano:2003gt}, which are beyond the scope of the present work. We therefore determine the weak chiral LECs in the isospin limit without attempting to include the isospin-breaking effects in the $K\to\pi\pi$ decays. Through the fit to the experimental values of $\bar{A}_{0,2}^{\rm Exp}$ by adjusting $g_8$ and $g_{27}^{1/2}$, we obtain 
\begin{equation}
g_8 = 3.628\,,\qquad g_{27}^{1/2}= 0.047\,, \qquad g_{27}^{3/2} = 5g_{27}^{1/2} = 0.235\,.\label{eq_g_values}
\end{equation}
Our determination of $g_8$ is in excellent agreement with those in Ref.~\cite{Cirigliano:2011ny}. The values of $g_{27}^{1/2}$ from our study and those from the previous reference show some minor tensions. 
Additionally, the contributions of pseudoscalar poles in $K_L\to\gamma\gamma$ decay were investigated in Ref.~\cite{Gerard:2005yk}, leading to the estimate
\begin{equation}
g_8^{\psi} + 2g_8^P \simeq \frac{1}{3}g_8\,.\label{eq_g8psi_value}
\end{equation}
The weak chiral couplings $g_8'$, $g_8^P$ and $g_8^{\theta}$, which are $\mO(N_C^{-1})$ suppressed, are numerically unknown and will be simply set to zero in the phenomenological analysis. We will use Eqs.~(\ref{eq_g_values}) and (\ref{eq_g8psi_value}) as the weak-coupling inputs in the following discussion of the $K\to\pi a$ decay. It should be noted that the main difference between the values in Eqs.~(\ref{eq_g_NLO-fitted}) and (\ref{eq_g_values}) lies in $g_{27}^{1/2}$ and $g_{27}^{3/2}$. Since the $K\to\pi a$ decays are dominated by the octet contributions, using either parameter set given in Eqs.~(\ref{eq_g_NLO-fitted}) or (\ref{eq_g_values}) does not significantly affect the $K\to\pi a$ result.
For other values of all phenomenological parameters used in this work, they are summarized in Appendix~\ref{app_para-values}.

\begin{figure}[t]
\centering
\subfigure[]
{\includegraphics[width=0.3\linewidth]{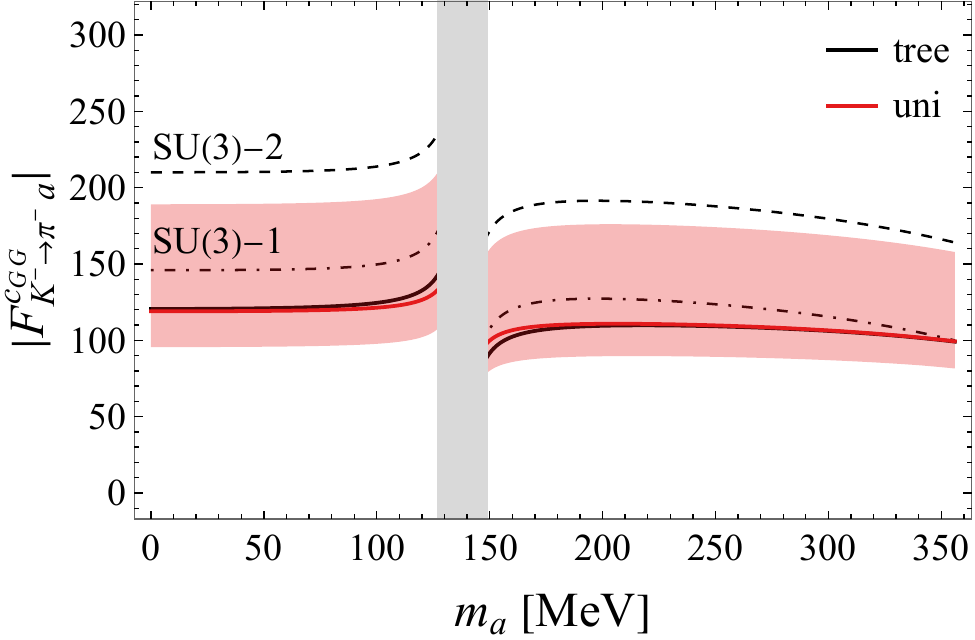}}
\quad
\subfigure[]
{\includegraphics[width=0.3\linewidth]{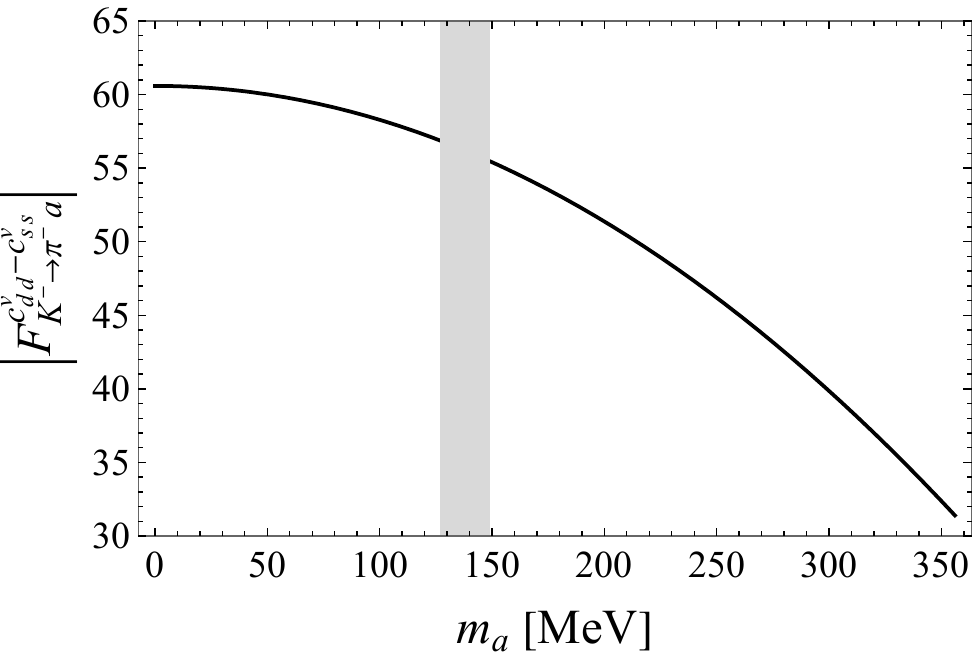}}
\\
\subfigure[]
{\includegraphics[width=0.3\linewidth]{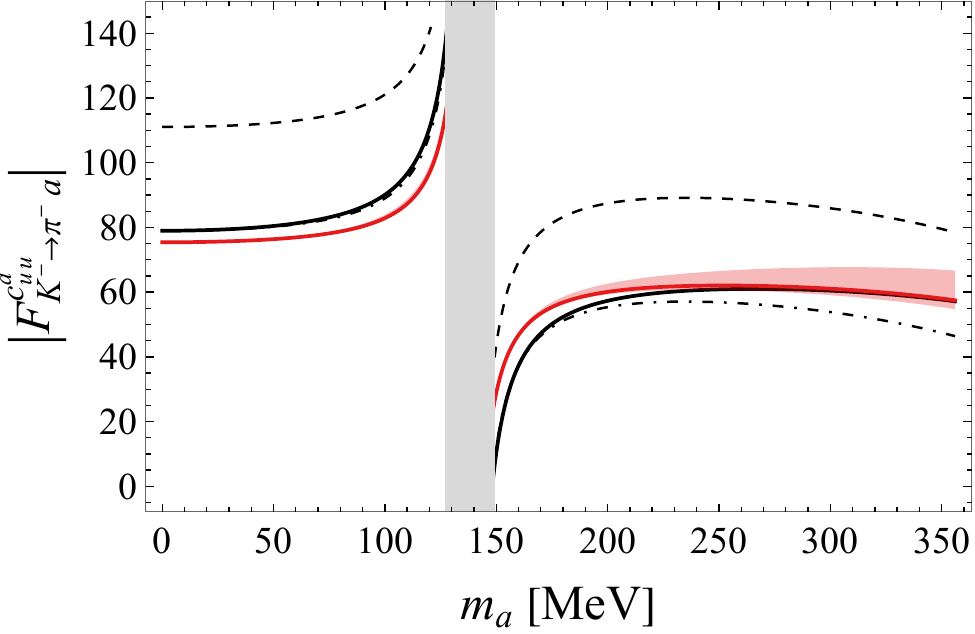}}
\quad
\subfigure[]
{\includegraphics[width=0.3\linewidth]{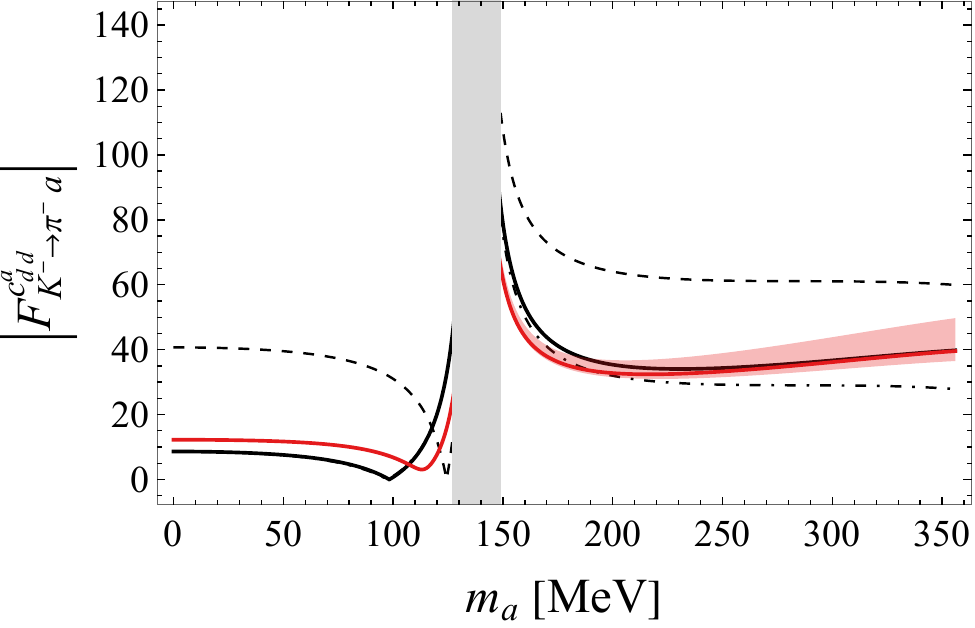}}
\quad
\subfigure[]
{\includegraphics[width=0.3\linewidth]{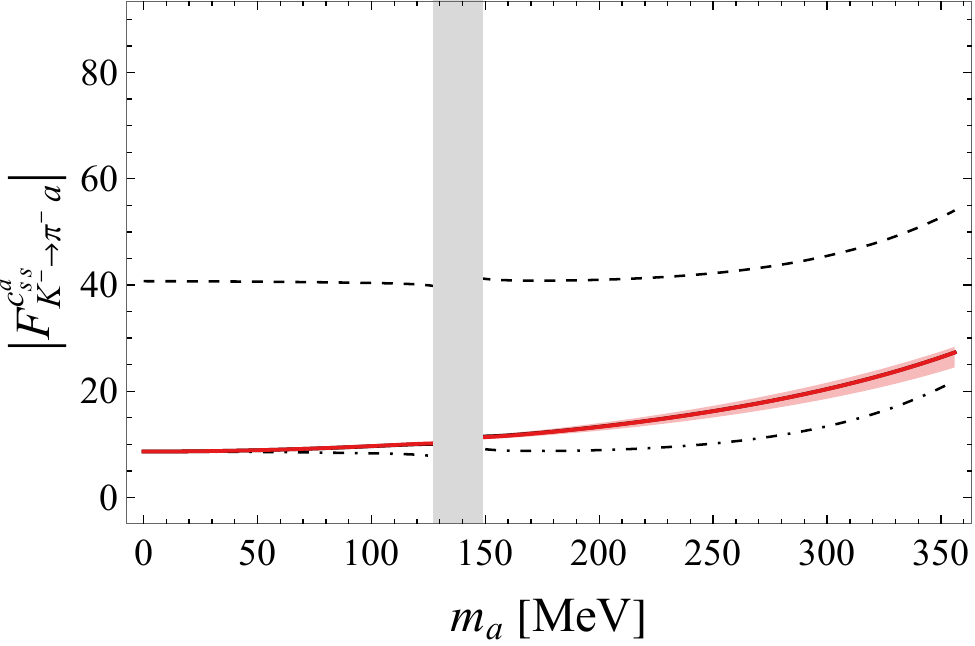}}
\caption{Variation of $|F_{K^-\to\pi^-a}^{\cALP}|$ with $m_a$. The black solid lines denote the tree-level U(3) results. The results labeled ``SU(3)-1'' (dash-dotted lines) are obtained from the tree-level U(3) results by taking $M_0\to\infty$. The results labeled ``SU(3)-2'' (dashed lines) are obtained from the SU(3)-1 results by further setting $g_8^{\psi}=0$. The red solid lines represent the U(3) results with $\pi\pi$ loop corrections included. The red shaded bands around the red solid lines show the distributions obtained by Gaussian sampling of the mixing-angle parameters. 
}\label{fig_Fac_Km-to-pim-a}
\end{figure}

In Fig.~\ref{fig_Fac_Km-to-pim-a}, we show various results of $|F_{K^-\to\pi^-a}^{\cALP}|$ defined in Eq.~\eqref{eq_A_Km_to_pim_a_WFC} for the $m_a$ in the range from $0$ to $m_{K^{-}}-m_{\pi}$. The tree-level U(3) results are displayed as black solid lines, which explicitly include the contributions from $\eta_0$ and the new weak chiral operator associated with $g_8^{\psi}$. 
We can decouple the singlet $\eta_0$ by taking the limit $M_0\to\infty$. 
The corresponding results from this procedure are shown as dash-dotted lines and labeled as ``SU(3)-1''. Note that the new weak operator still contributes to the ``SU(3)-1'' results. By further setting $g_8^{\psi}=0$ in the ``SU(3)-1'' case, we obtain the ``SU(3)-2'' results (dashed lines), which correspond to the conventional SU(3) results. By comparing the above three cases, we conclude that, in the $K^-\to\pi^-a$ decay, the $\eta_0$ degree of freedom itself does not induce significant corrections, whereas the new $g_8^{\psi}$ operator could give noticeable contributions. 

The results obtained from the unitarized $K^-\to\pi^-a$ decay amplitude in Eq.~(\ref{eq_Ktopia-WFC-uni}) are shown as red solid lines in Fig.~\ref{fig_Fac_Km-to-pim-a}. These results almost overlap with the tree-level U(3) results, which numerically confirms that the $\pi\pi$ loop corrections are not important in the $K^-\to\pi^-a$ decay. 

To assess the dependence of the $K^-\to\pi^-a$ decay amplitude on the $\eta$-$\eta'$ mixing parameters, we perform random Gaussian sampling of the two-mixing-angle parameters, by taking their central values and standard deviations from Ref.~\cite{Gu:2018swy}. The distributions obtained from the unitarized decay amplitude are shown as red bands in Fig.~\ref{fig_Fac_Km-to-pim-a}. It can be seen that the $c_{GG}$ contribution is sensitive to the $\eta$-$\eta'$ mixing, whereas the other contributions in the WFC part are not. We emphasize that these bands should not be understood as the full uncertainty bands of the unitarized results, as they represent only the partial uncertainty associated with the phenomenological $\eta$-$\eta'$ mixing parameters.

\begin{figure}[t]
\centering
\subfigure[]
{\includegraphics[width=0.3\linewidth]{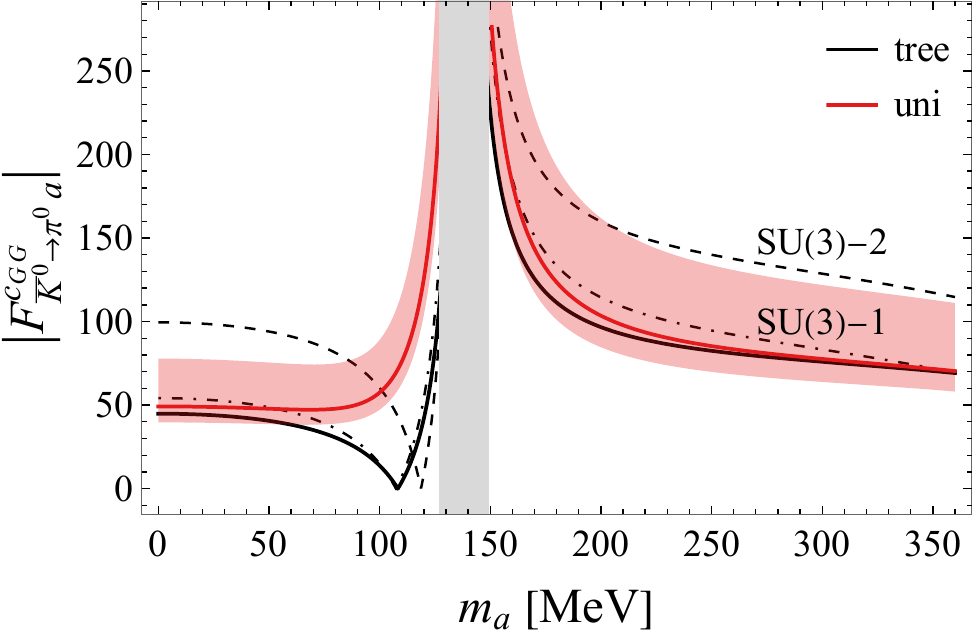}}
\quad
\subfigure[]
{\includegraphics[width=0.3\linewidth]{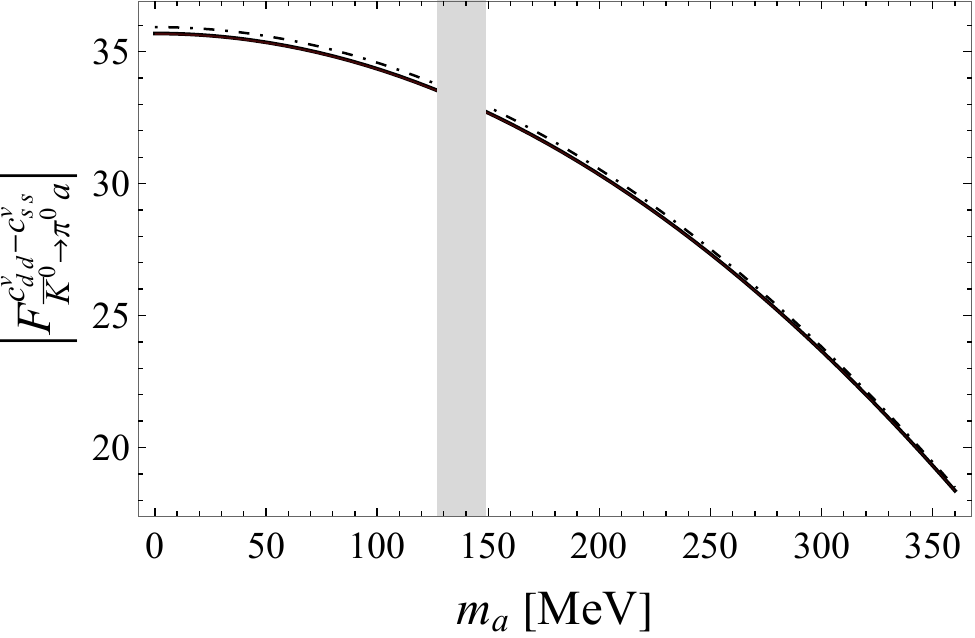}}
\\
\subfigure[]
{\includegraphics[width=0.3\linewidth]{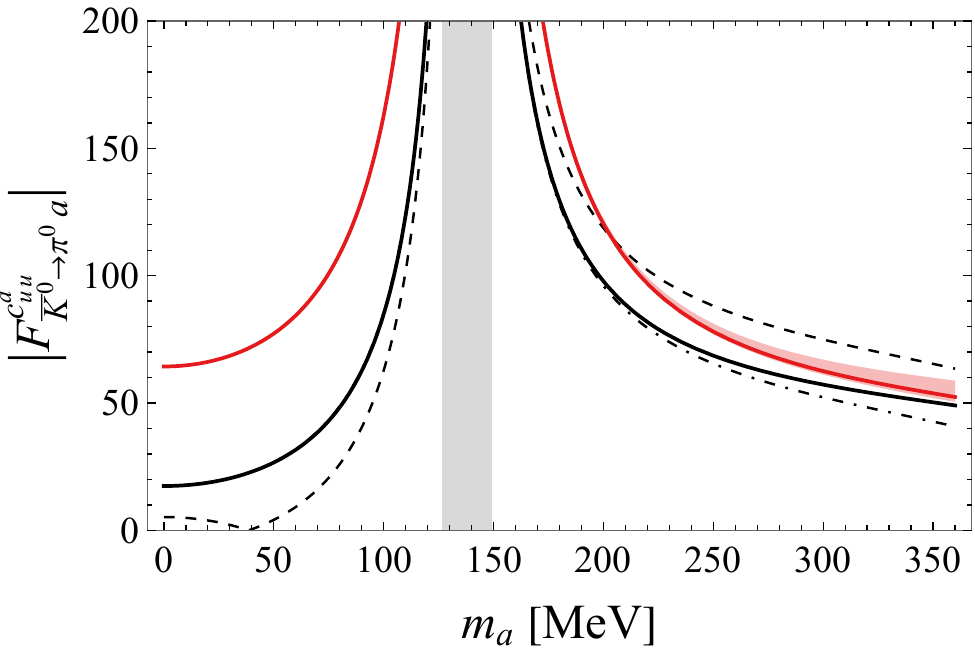}}
\quad
\subfigure[]
{\includegraphics[width=0.3\linewidth]{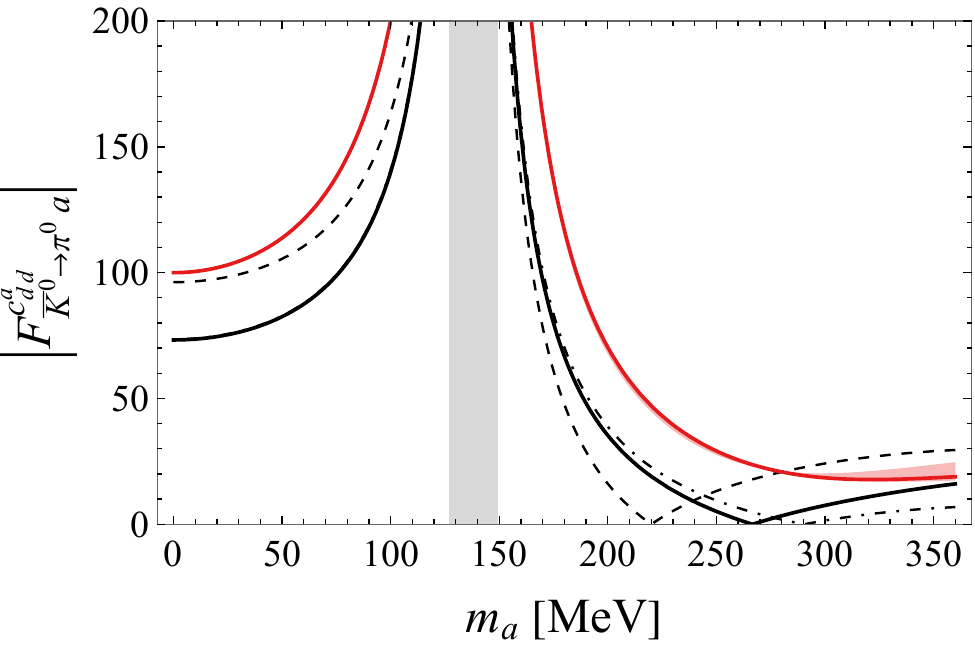}}
\quad
\subfigure[]
{\includegraphics[width=0.3\linewidth]{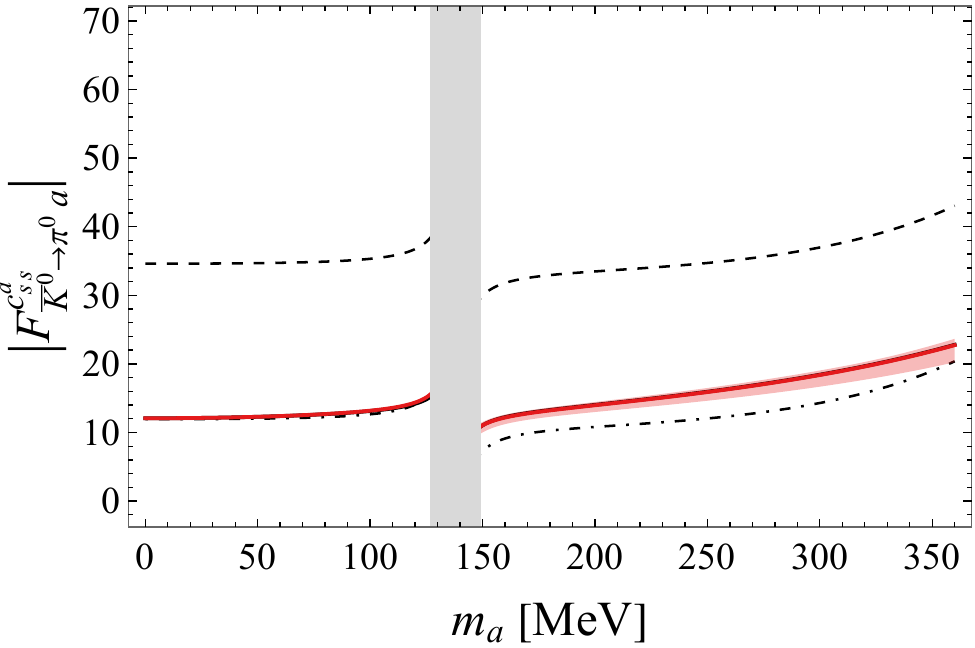}}
\caption{Variation of $|F_{\bar{K}^0\to\pi^0 a}^{\cALP}|$ with $m_a$. The line styles and shaded bands have the same meanings as in Fig.~\ref{fig_Fac_Km-to-pim-a}.}\label{fig_Fac_Kbar0-to-pi0-a}
\end{figure}

The various results of $|F_{\bar{K}^0\to\pi^0 a}^{\cALP}|$ are shown in Fig.~\ref{fig_Fac_Kbar0-to-pi0-a}. A similar conclusion can be drawn for the $\bar{K}^0\to\pi^0a$ decay: the $\eta_0$ degree of freedom does not induce significant corrections, whereas the U(3) $g_8^{\psi}$ operator gives a pronounced contribution. Comparing with the situation in the $K^-\to\pi^-a$ process, a different feature arises in the neutral channel. 
The noticeable different curves from the tree-level and unitarized amplitudes,  as shown in Fig.~\ref{fig_Fac_Kbar0-to-pi0-a}, indicate that the $\pi\pi$ loop corrections can be significant in the $\bar{K}^0\to\pi^0a$ decay. This is mainly due to the involvement of the $I=0$ $\pi\pi$ intermediate state in the $\bar{K}^0\to\pi^0a$ decay.

\subsection{Constraints on ALPs from the $K\to\pi a$ decays}

Recently, the NA62 collaboration updated the upper limit on the branching fraction of $K^+\to\pi^+X_{\rm inv}$ for $m_{X}$ in the ranges $0$--$100$~MeV and $150$--$260$~MeV, based on their 2016--2022 measurement of $K^+\to\pi^+\nu\bar{\nu}$~\cite{NA62:2025upx}. Here $X_{\rm inv}$ denotes a long-lived invisible particle or a particle that decays invisibly. This limit sets stringent constraints on ALPs with masses $m_a\lesssim300$~MeV.

\begin{figure}[tbph]
	\centering
	\subfigure[\label{fig_constraints1a}]
	{\includegraphics[width=0.3\linewidth]{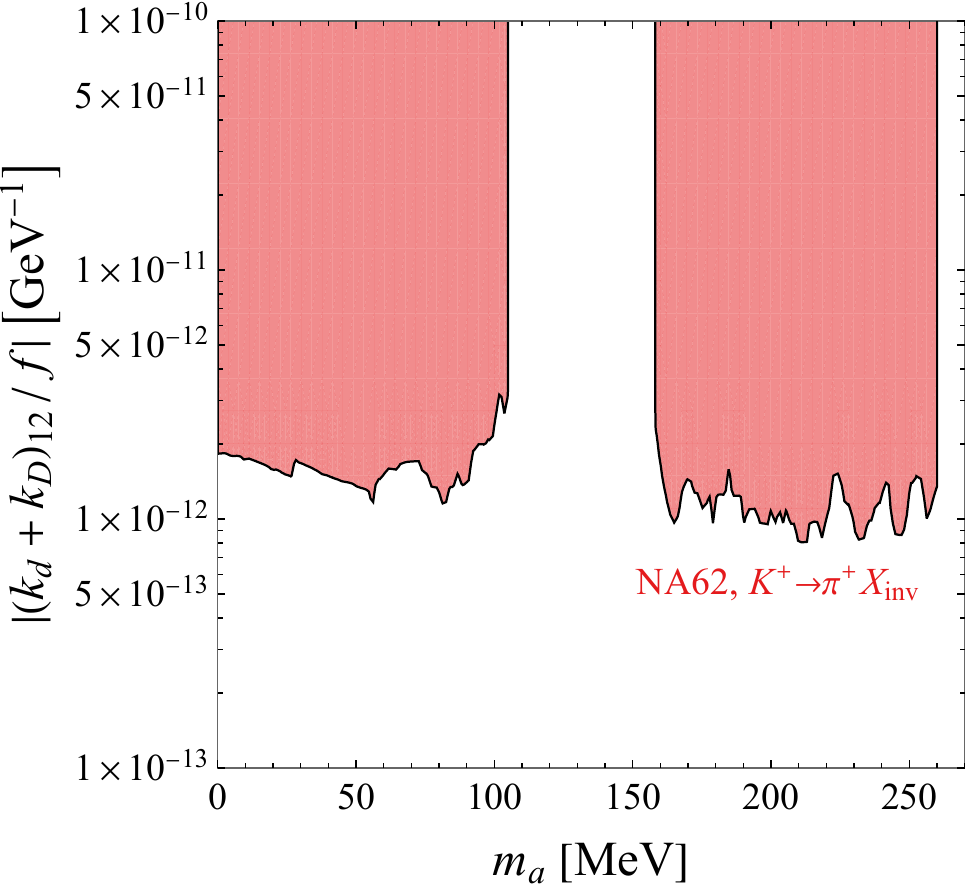}}
	\hfill
	\subfigure[\label{fig_constraints1b}]
	{\includegraphics[width=0.3\linewidth]{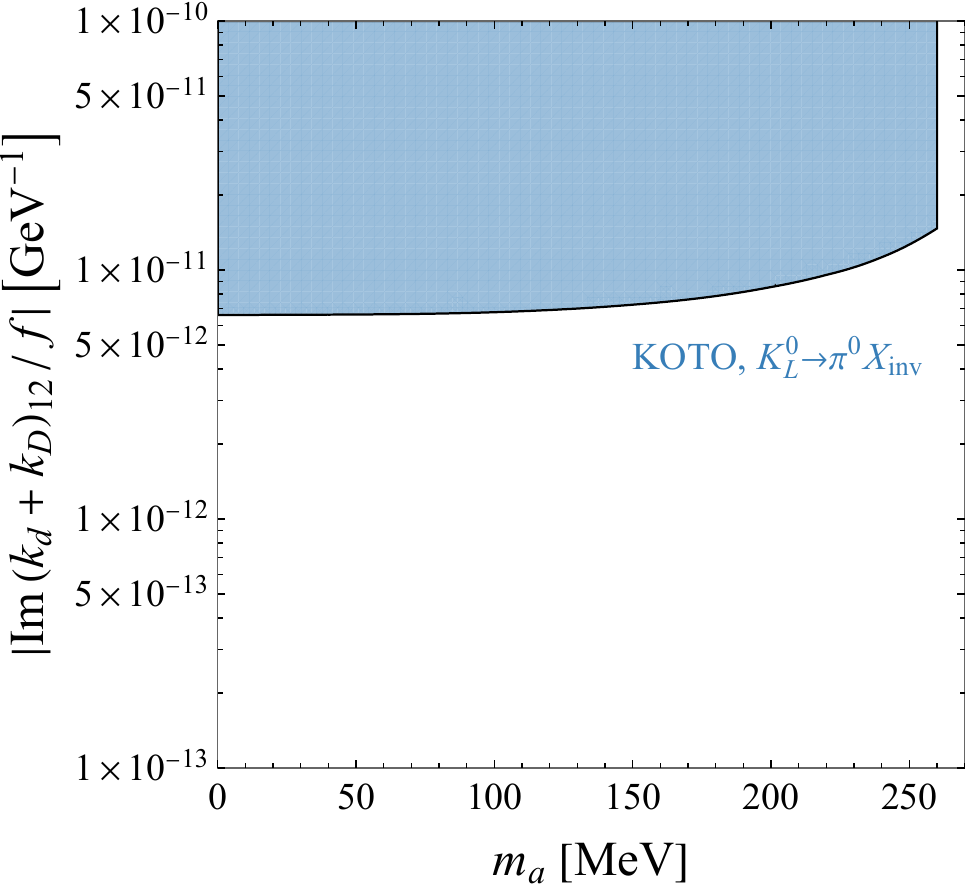}}
	\hfill
	\subfigure[\label{fig_constraints1c}]
	{\includegraphics[width=0.3\linewidth]{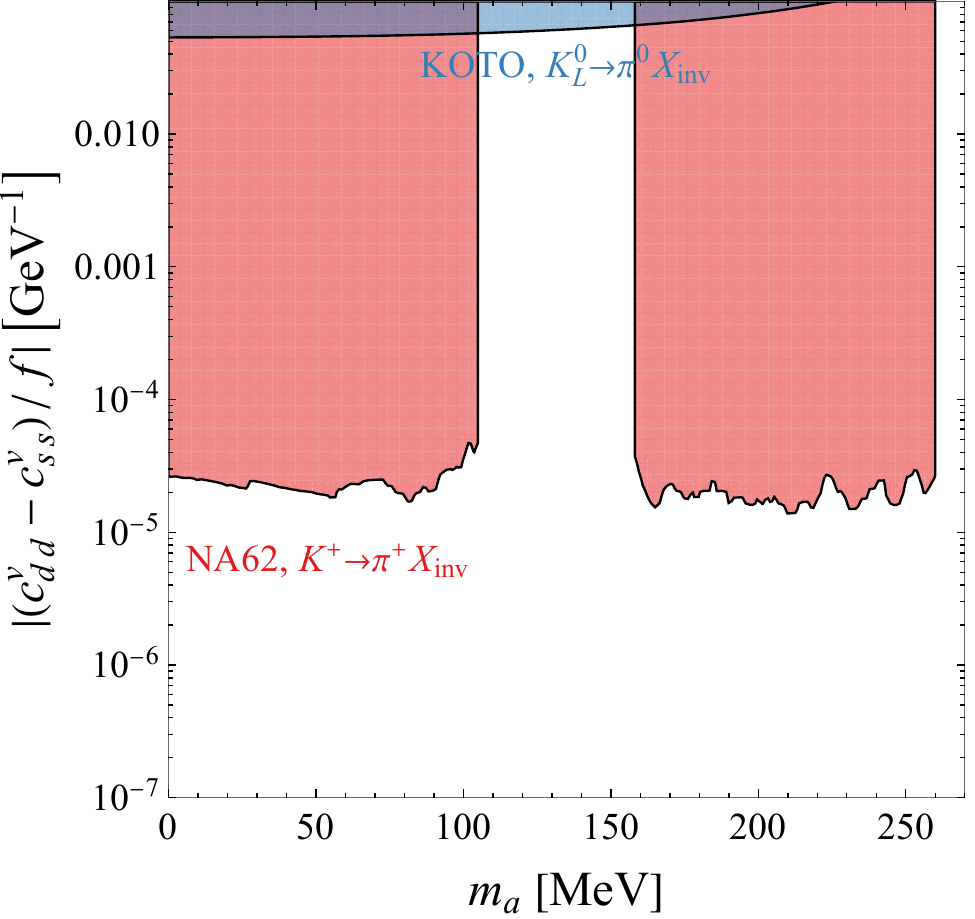}}
	\caption{Excluded regions from the upper limits on the branching ratios of $K^+\to\pi^+X_{\rm inv}$ (NA62~\cite{NA62:2025upx}, red) and $K^0_L\to\pi^0X_{\rm inv}$ (KOTO~\cite{KOTO:2024zbl}, blue), assuming that only $(k_d+k_D)_{12}$~(\ref{fig_constraints1a},\ref{fig_constraints1b}) or $c^v_{dd}-c^v_{ss}$~(\ref{fig_constraints1c}) is nonzero. The ALP is assumed to be a long-lived particle. }\label{fig_constraints1}
\end{figure}

To contribute to $\mathcal{B}(K^+\to\pi^+X_{\rm inv})$, the ALP produced in $K^+\to\pi^+a$ decays must either escape the detector before decaying into visible final states, or decay into invisible final states. Due to the phase space constraints, the ALP can only decay into $\gamma\gamma$, $e^+e^-$, $\mu^+\mu^-$, and $\nu\bar{\nu}$. In this work, we concentrate on the scenario that the ALP does not couple directly to leptons, and hence decays predominantly into $\gamma\gamma$, i.e., its proper lifetime is given by $\tau_0=\Gamma_{a\to\gamma\gamma}^{-1}$. The ALP lifetime is a crucial input for the following discussions. We provide in Appendix~\ref{app_atogaga} the calculation of the $a\to\gamma\gamma$ decay width within U(3) $\ChPT$.

For experiments in which the initial kaons are highly boosted in the longitudinal direction, such as NA62, NA48 and KOTO, the fraction of ALPs produced in the $K\to\pi a$ decays that can escape the detector is given by~\cite{Bauer:2021mvw}
\begin{equation}\label{eq.fescdef}
F_{\rm esc}(m_a,\tau_0,\beta_K,L_{max})
=\int_0^{\pi/2}\mathrm{d}\theta\,\sin\theta\exp\left(-\frac{L_{max}}{\Delta x_L}\right)\,,
\end{equation}
which depends on the effective detector length $L_{max}$ and the ALP longitudinal flight distance $\Delta x_L$, defined by
\begin{equation}
\Delta x_L = \frac{p_{a,L}\tau_0}{m_a}\,,
\quad
p_{a,L}=\gamma_K\left(\beta_K E_a^* + p_a^* \cos\theta\right)\,,
\end{equation}
where $\beta_K = p_K/E_K$ is the kaon velocity, $\gamma_K = \left(1-\beta_K^2\right)^{-1/2}$, and $E_a^*$ and $p_a^*$ denote the ALP energy and momentum in the kaon rest frame,
\begin{equation}
E_a^*=\frac{m_K^2+m_a^2-m_{\pi}^2}{2m_K}\,,
\quad
p_a^*=\frac{\sqrt{\lambda(m_K^2,m_{\pi}^2,m_a^2)}}{2m_K}\,,
\end{equation}
while $\theta$ is the angle between $\vec{p}_a^{\,*}$ and the longitudinal direction.

Accordingly, the upper limit on the branching ratio of $K^+\to\pi^+X_{\rm inv}$ can be mapped onto ALP parameter space through
\begin{equation}
F_{\rm esc}^{\rm NA62}\mathcal{B}(K^+\to\pi^+a)\leq \left[\mathcal{B}(K^+\to\pi^+X)\right]_{\rm UL}\,,
\label{eq_constraints_KptopipX}
\end{equation}
where
\begin{equation}
\mathcal{B}(K^+\to\pi^+a) = \frac{\tau_{K^+}\lambda^{1/2}\left(m_{K^{\pm}}^2,m_{\pi}^2,m_a^2\right)}{16\pi m_{K^{\pm}}^3}\left| A_{K^+\to\pi^+a} \right|^2\,,\label{eq_BR_Kptopipa}
\end{equation}
and we introduce the shorthand $F_{\rm esc}^{\rm exp}\equiv F_{\rm esc}(m_a,\tau_0,\beta_K^{\rm exp},L_{max}^{\rm exp})$. The inequality above depends on the ALP mass and couplings. The NA62 experimental measurement covers the mass ranges $m_a\in[0,110]\cup [150,260]$~MeV. 
By turning on a single coupling at a time while setting the others to zero, we can derive the excluded regions in the $m_a$--coupling plane. In the following, we use the unitarized $K\to\pi a$ decay amplitude as input and adopt the central values for the $\eta$-$\eta'$ mixing parameters. Varying these theoretical inputs does not alter the order of magnitude of the amplitude, and the resulting ALP bounds remain the same order of magnitude. 

\begin{figure}[tbph]
	\centering
	\subfigure[\label{fig_constraints2a}]
	{\includegraphics[width=0.45\linewidth]{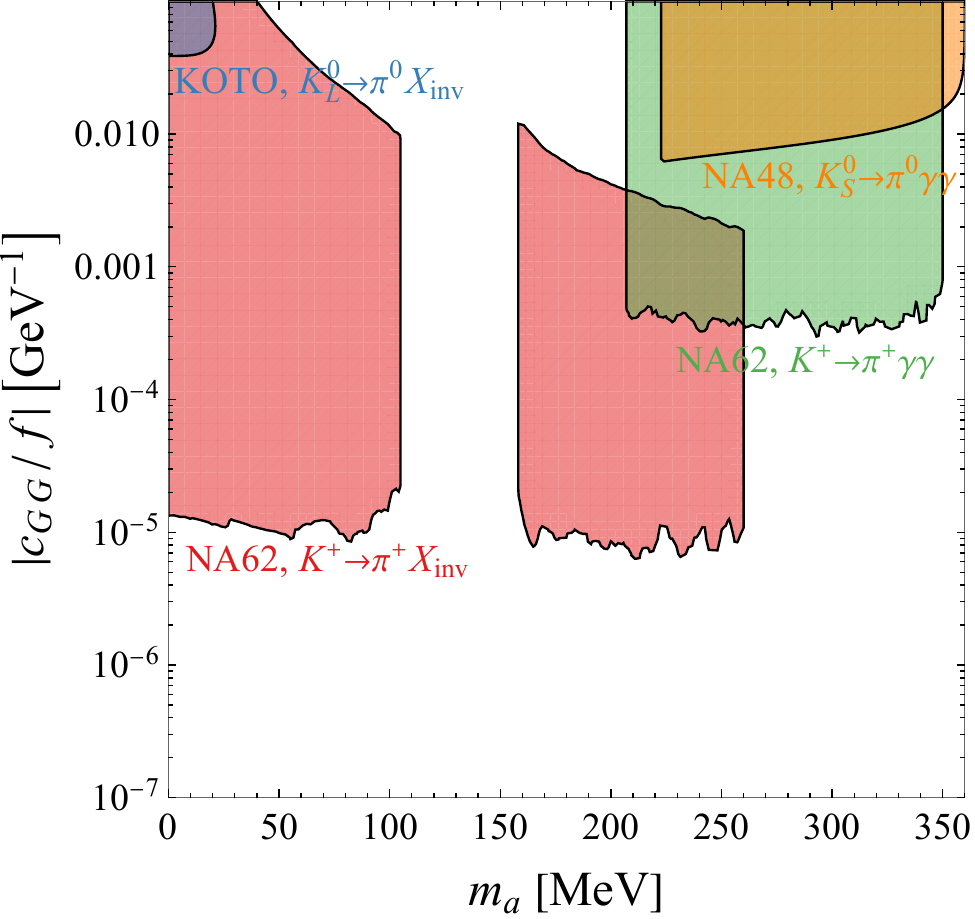}}
	\hfill
	\subfigure[\label{fig_constraints2b}]
	{\includegraphics[width=0.45\linewidth]{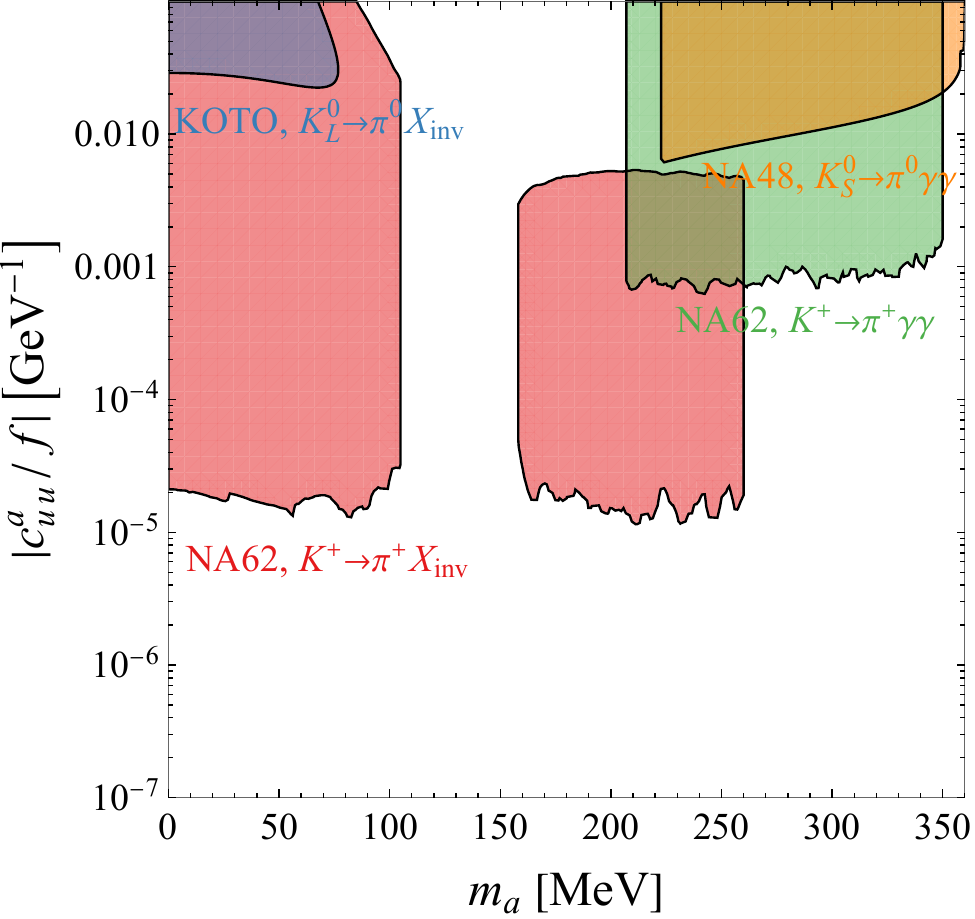}}
	\\
	\subfigure[\label{fig_constraints2c}]
	{\includegraphics[width=0.45\linewidth]{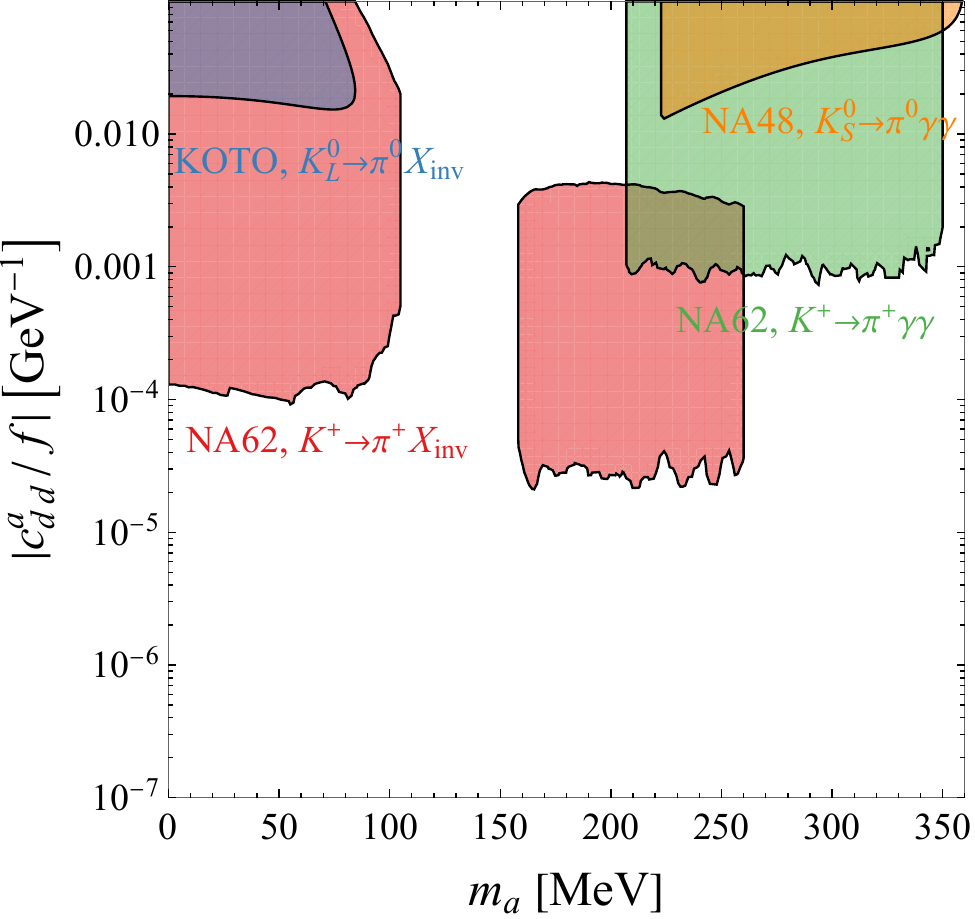}}
	\hfill
	\subfigure[\label{fig_constraints2d}]
	{\includegraphics[width=0.45\linewidth]{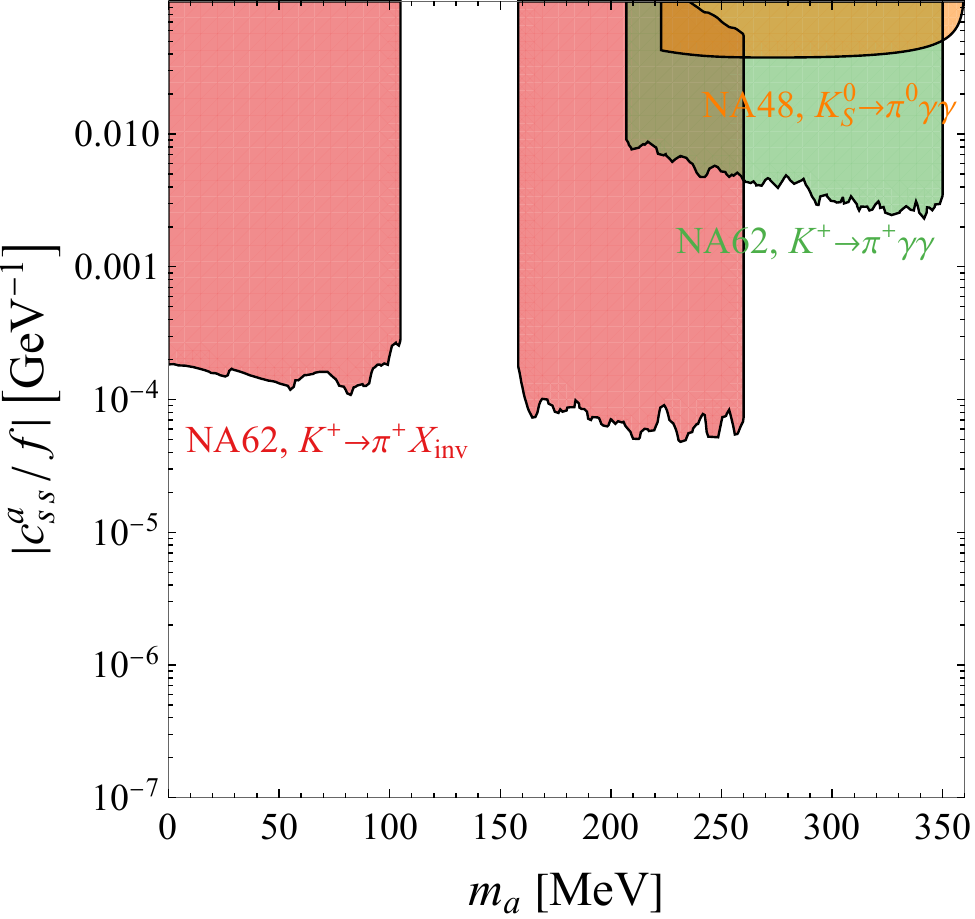}}
	\caption{Excluded regions from the upper limits on the branching ratios of $K^+\to\pi^+X_{\rm inv}$ (NA62~\cite{NA62:2025upx}, red), $K^0_L\to\pi^0X_{\rm inv}$ (KOTO~\cite{KOTO:2024zbl}, blue), $K^+\to\pi^+a\to\pi^+\gamma\gamma$ (NA62~\cite{NA62:2023olg}, green), and $K^0_S\to\pi^0a\to\pi^0\gamma\gamma$ (NA48~\cite{NA48:2003ydp}, orange), assuming that the ALP decays exclusively into $\gamma\gamma$ and that only one coupling among $c_{GG}$ (\ref{fig_constraints2a}), $c^a_{uu}$ (\ref{fig_constraints2b}), $c^a_{dd}$ (\ref{fig_constraints2c}), and $c^a_{ss}$ (\ref{fig_constraints2d}) is nonzero at a time.}\label{fig_constraints2}
\end{figure}

The ALP bounds inferred from the NA62 upper limits on the branching ratios of $K^+\to\pi^+X_{\rm inv}$ are denoted by the red regions in Figs.~\ref{fig_constraints1} and \ref{fig_constraints2}. When only the coupling $(k_d+k_D)_{12}$ or $c^v_{dd}-c^v_{ss}$ is switched on, the ALP cannot decay into $\gamma\gamma$. Hence, in these two cases $F_{\rm esc}^{\rm NA62}=1$ and Eq.~(\ref{eq_constraints_KptopipX}) yields upper limits on the effective couplings $(k_d+k_D)_{12}/f$ and $(c^v_{dd}-c^v_{ss})/f$ as functions of the ALP mass, as shown in Figs.~\ref{fig_constraints1a} and \ref{fig_constraints1c}, respectively. In Fig.~\ref{fig_constraints1a}, for $m_a\simeq 0$, the upper limit for the flavor-changing ALP coupling is $|(k_d+k_D)_{12}/f|<1.83\times 10^{-12}~\mathrm{GeV}^{-1}$. When the couplings of $c_{GG}$, $c^a_{uu}$, $c^a_{dd}$ and $c^a_{ss}$ are turned on, the ALP can decay into $\gamma\gamma$. In this case, $F_{\rm esc}^{\rm NA62}$ acts as an exponential suppression factor when the ALP coupling becomes sufficiently large, according to Eq.~\eqref{eq.fescdef}, since the ALP flight distance is inversely proportional to its decay width. The exclusion regions in the $m_a$--$|\cALP/f|$ planes obtained from Eq.~(\ref{eq_constraints_KptopipX}) are shown in Fig.~\ref{fig_constraints2}, where each panel gives the results by varying only a single coupling among $c_{GG}$, $c^a_{uu}$, $c^a_{dd}$ and $c^a_{ss}$ and fixing others at zero. 

\begin{table}[tbp]
	\centering
	\caption{Summary of the experimental measurements used to derive constraints on ALPs. 
		The effective detector lengths of the NA62, NA48 and KOTO experiments are taken from Ref.~\cite{Bauer:2021mvw}. The kaon momenta are taken to be 75~GeV for NA62 and NA48~\cite{NA62:2017rwk}, and 1.5~GeV for KOTO, close to the peak momentum of 1.42~GeV measured at the beam exit~\cite{Masuda:2015eta}. The same phenomenological setup is adopted in Ref.~\cite{Bauer:2021mvw}.}\label{tab_Exp}
	\begin{tabular}{llll}
		\hline\hline
		Decay channel&Experiment&ALP signature&Experimental parameters
		\\
		\hline
		$K^+\to\pi^++\mathrm{inv}$&NA62~\cite{NA62:2025upx}&long-lived&$p_K^{\rm NA62}=75~\GeV$, $L_{max}^{\rm NA62}=140~\mathrm{m}$
		\\
		$K_L^0\to\pi^0+\mathrm{inv}$&KOTO~\cite{KOTO:2024zbl}&long-lived&$p_K^{\rm KOTO}=1.5~\GeV$, $L_{max}^{\rm KOTO}=4.148~\mathrm{m}$
		\\
		$K^+\to\pi^+\gamma\gamma$&NA62~\cite{NA62:2023olg}&$a\to\gamma\gamma$&$p_K^{\rm NA62}=75~\GeV$, $L_{max}^{\rm NA62}=140~\mathrm{m}$
		\\
		$K^0_S\to\pi^0\gamma\gamma$&NA48~\cite{NA48:2003ydp}&$a\to\gamma\gamma$&$p_K^{\rm NA48}=75~\GeV$, $L_{max}^{\rm NA48}=140~\mathrm{m}$
		\\
		\hline\hline
	\end{tabular}
\end{table}

Apart from the NA62 constraints, we also consider the ALP bounds from three other experimental measurements, which are summarized in Table~\ref{tab_Exp} together with the relevant experimental parameters listed in the last column. 
The KOTO collaboration has set an upper limit on the branching fraction of $K_L^0\to\pi^0X_{\rm inv}$ for $m_X$ in the range $0$--$260$~MeV~\cite{KOTO:2024zbl}. The $K_L^0\to\pi^0a$ decay can be constrained by this measurement if the ALP escapes the detector, leading to
\begin{equation}
F_{\rm esc}^{\rm KOTO}\mathcal{B}(K^0_L\to\pi^0 a)\leq \left[\mathcal{B}(K^0_L\to\pi^0X)\right]_{\rm UL}\,,
\quad
\text{for $m_a\in [0,260]$~MeV,}\label{eq_constraint_KLtopi0X}
\end{equation}
with 
\begin{equation}
\mathcal{B}(K^0_L\to\pi^0 a) = \frac{\tau_{K^0_L}\lambda^{1/2}(m_{K^0}^2,m_{\pi}^2,m_a^2)}{16\pi m_{K^0}^3}\left|A_{K^0_L\to\pi^0 a}\right|^2\,.
\end{equation}

The $K_L^0\to\pi^0a$ decay is driven by the CP violation strength. In our convention, the CP transformations of $K^0$ and $\bar{K}^0$ are
\begin{equation}
CP|K^0\rangle = -|\bar{K}^0\rangle\,,
\quad
CP|\bar{K}^0\rangle = - |K^0\rangle\,.
\end{equation}
The mass eigenstates $K^0_L$ and $K^0_S$ resulting from $K^0$-$\bar{K}^0$ mixing are given by~\cite{Buras:2001pn}
\begin{equation}
|K^0_{L,S}\rangle = \frac{\left(1+\bar{\varepsilon}\right)|K^0\rangle \pm \left(1-\bar{\varepsilon}\right)|\bar{K}^0\rangle}{\sqrt{2\left(1+\left|\bar{\varepsilon}\right|^2\right)}}\,,
\end{equation}
where $\bar{\varepsilon}$ parameterizes the CP violation, with $\bar{\varepsilon}= |\bar{\varepsilon}|e^{i\phi_{\bar{\epsilon}}}$, $|\bar{\varepsilon}|=2.228\times 10^{-3}$ and $\phi_{\bar{\varepsilon}}=43.5\degree$~\cite{Cirigliano:2011ny}. To first order in CP violation, we have
\begin{equation}
A_{K^0_L\to\pi^0 a}=\frac{1}{\sqrt{2}}\left(A_{K^0\to\pi^0 a}+ A_{\bar{K}^0\to\pi^0 a}\right)+\frac{\bar{\varepsilon}}{\sqrt{2}}\left(A_{K^0\to\pi^0 a} - A_{\bar{K}^0\to\pi^0 a}\right)\,,
\end{equation}
where the first term originates from direct CP violation in the $\bar{K}^0/K^0\to\pi^0a$ decay amplitude, and the second one is from the CP violation in the $K^0$-$\bar{K}^0$ mixing. We parameterize the $\bar{K}^0/K^0\to\pi^0a$ decay amplitudes as
\begin{subequations}
	\begin{align}
	&A_{\bar{K}^0\to\pi^0a}=\frac{(k_d+k_D)_{12}}{f}K_{\rm AFC}+G_W\sum_{\cALP}\frac{\cALP}{f}\sum_g g K_{\cALP,g}\,,
	\\
	&A_{K^0\to\pi^0 a}=-\frac{(k_d+k_D)_{21}}{f}K_{\rm AFC}-G_W\sum_{\cALP}\frac{\cALP}{f}\sum_g g^* K_{\cALP,g}\,,
	\end{align}
\end{subequations}
with $g\in \{g_8,g_8^{\psi},g_{27}^{1/2},g_{27}^{3/2}\}$. 
Here $K_{\rm AFC}$ denotes the residue of the AFC amplitude $A_{\bar{K}^0\to\pi^0a}^{\rm AFC}$ after factoring out the ALP coupling $(k_d+k_D)_{12}/f$, while $K_{\cALP,g}$ denotes the coefficient of $g$ in the factor $F_{\bar{K}^0\to\pi^0 a}^{\cALP}$ defined in Eq.~\eqref{eq_A_Kbar0-to-pi0-a_WFC}.
By neglecting the second-order CP violation in the SM, we find
\begin{equation}
\begin{aligned}
A_{K^0_L\to\pi^0 a} =& \frac{\sqrt{2}i\mathrm{Im}(k_d+k_D)_{12}}{f}K_{\rm AFC} - \bar{\varepsilon}\frac{\sqrt{2}\mathrm{Re}(k_d+k_D)_{12}}{f}K_{\rm AFC}
\\
&+G_W\sum_{\cALP}\frac{\cALP}{f}\left(\sum_g\sqrt{2}i\mathrm{Im}g\, K_{\cALP,g} - \sqrt{2}\bar{\varepsilon}F_{\bar{K}^0\to\pi^0a}^{\cALP}\right) + \mO(\bar{\varepsilon}^2)\,.\label{eq_AKLtopi0a}
\end{aligned}
\end{equation}

By plugging the expression of Eq.~\eqref{eq_AKLtopi0a} into Eq.~\eqref{eq_constraint_KLtopi0X}, we can derive the ALP bounds from the KOTO upper limit on the branching fraction of $K^0_L\to\pi^0X_{\rm inv}$, and the results are shown as blue regions in Figs.~\ref{fig_constraints1} and \ref{fig_constraints2}. When only the coupling $(k_d+k_D)_{12}$ is switched on, we neglect the $\mO(\bar{\varepsilon})$ term and Eq.~(\ref{eq_constraint_KLtopi0X}) then yields an upper limit on $|\mathrm{Im}(k_d+k_D)_{12}/f|$ as a function of $m_a$, as shown in Fig.~\ref{fig_constraints1b}. For $m_a \simeq 0$, this limit becomes $|\mathrm{Im}(k_d+k_D)_{12}/f|<6.59\times 10^{-12}~\mathrm{GeV}^{-1}$. When only one of the flavor-conserving ALP couplings is turned on, the bounds of the $K^0_L\to\pi^0 a$ from KOTO turn out to be significantly weaker than those of the $K^+\to \pi^+ a$ from NA62, as the WFC contributions in the former case are suppressed by the SM CP violation, compared to the latter case. For an order-of-magnitude estimate, we consider, besides the indirect CP violation from $K^0$-$\bar{K}^0$ mixing, the direct CP violation solely originating from $g_8$, which can be approximated as~\cite{Gisbert:2017vvj} 
\begin{equation}
\mathrm{Im}g_8=-0.87\mathrm{Im}\tau\,,
\quad
\mathrm{Im}\tau \simeq -\lambda^4 A^2 \eta \simeq -6.16\times 10^{-4}\,.
\end{equation}
The resulting constraints are displayed in Fig.~\ref{fig_constraints1c} and Fig.~\ref{fig_constraints2}. The bound on $c^a_{ss}$ obtained from KOTO is too weak to be visible in Fig.~\ref{fig_constraints2d}, and is therefore not shown.

Measurements of visible final states can provide ALP bounds complementary to those from the invisible cases. The NA62 collaboration has set an upper limit on the product of branching fractions $\mathcal{B}(K^+\to\pi^+a)\mathcal{B}(a\to\gamma\gamma)$ for $m_a$ in the range $207$--$350$~MeV. For the ALP to contribute to this search, it must decay into $\gamma\gamma$ within the detector length. The constraint reads
\begin{equation}
\begin{aligned}
&\left(1-F_{\rm esc}^{\rm NA62}\right)\mathcal{B}(K^+\to\pi^+a)\mathcal{B}(a\to\gamma\gamma)\leq
\left[\mathcal{B}(K^+\to\pi^+a)\mathcal{B}(a\to\gamma\gamma)\right]_{\rm UL}\,,
\end{aligned}
\end{equation}
where we assume $\mathcal{B}(a\to\gamma\gamma) = 1$. When only the coupling $(k_d+k_D)_{12}$ or $c^v_{dd}-c^v_{ss}$ is switched on, the ALP cannot decay into $\gamma\gamma$ and this constraint is irrelevant. When one of the couplings $c_{GG}$, $c^a_{uu}$, $c^a_{dd}$, or $c^a_{ss}$ is turned on, the resulting bounds are denoted by the green regions in  Fig.~\ref{fig_constraints2}.

Similar bounds can also be obtained from the measurement of $K^0_S\to\pi^0\gamma\gamma$ in a specific kinematic region by NA48 in 2003~\cite{NA48:2003ydp} 
\begin{equation}
\mathcal{B}(K^0_S\to\pi^0\gamma\gamma)\big|^{\rm exp}_{m_{\gamma\gamma}^2>0.2m_{K}^2} = \left(4.9\pm1.8\right)\times 10^{-8}\,.
\end{equation}
By combining the $\ChPT$ prediction~\cite{Ecker:1987fm} 
\begin{equation}
\mathcal{B}(K^0_S\to\pi^0\gamma\gamma)\big|^{\chi}_{m_{\gamma\gamma}^2>0.2m_{K}^2} =3.8\times 10^{-8}\,,
\end{equation}
the constraint can be written as
\begin{equation}
\left(1-F_{\rm esc}^{\rm NA48}\right)\mathcal{B}(K^0_S\to\pi^0a)\mathcal{B}(a\to\gamma\gamma)\leq
\left(4.9+n\times 1.8 - 3.8\right)\times 10^{-8}\,,
\label{eq_constraints_KStopi0a}
\end{equation}
where the bound is applicable for the range $m_a^2>0.2 m_{K^0}^2$, and we take $n=1.64$, corresponding to the one-sided 95\% confidence level, to obtain a conservative estimate. The branching fraction of $K^0_S\to\pi^0a$ reads
\begin{equation}
\mathcal{B}(K^0_S\to\pi^0a)=\frac{\tau_{K^0_S}\lambda^{1/2}(m_{K^0}^2,m_{\pi}^2,m_a^2)}{16\pi m_{K^0}^3}\left|A_{K^0_S\to\pi^0a}\right|^2\,.
\end{equation}
By neglecting the SM CP violation, the $K^0_S\to\pi^0a$ decay amplitude is given by
\begin{equation}
\begin{aligned}
A_{K^0_S\to\pi^0a}=&\frac{1}{\sqrt{2}}\left(A_{K^0\to\pi^0a}-A_{\bar{K}^0\to\pi^0a}\right)
\\
=&-\frac{\sqrt{2}\mathrm{Re}(k_d+k_D)_{12}}{f}K_{\rm AFC} - \sqrt{2}G_W\sum_{\cALP}\frac{\cALP}{f}F_{\bar{K}^0\to\pi^0a}^{\cALP}\,.
\end{aligned}
\end{equation}
When one of the ALP couplings $c_{GG}$, $c^a_{uu}$, $c^a_{dd}$, or $c^a_{ss}$ is turned on at a time, the corresponding bounds are shown as orange regions in Fig.~\ref{fig_constraints2}. The constraints from the $K^0_S\to\pi^0\gamma\gamma$ measurement by NA48 are clearly weaker than those from the $K^+\to\pi^+\gamma\gamma$ result from the NA62.

\section{Summary and conclusions}\label{sec.sum}

The $K\to\pi a$ decay plays a vital role in constraining the ALP bounds for $m_a\lesssim 350$~MeV.
In this work, we calculate the $K\to\pi a$ amplitudes at LO in U(3) $\ChPT$. By systematically incorporating the ALP couplings from the redefinition of quark fields and the external sources, we obtain the physical decay amplitudes that do not depend on the redefinition parameter $\kappa_q$, which provides a nontrivial self-consistent check of our calculations. In the U(3) framework, the explicit inclusion of the singlet $\eta_0$ naturally entails the appearance of new weak operators. After taking into account these contributions, our amplitudes can fully reproduce the large-$N_C$ behavior of QCD. Moreover, taking the limit $M_0\to \infty$, our amplitudes recover the conventional SU(3) $\ChPT$ results in literature. As a first step toward incorporating higher-order effects, we construct a unitarized $K\to\pi a$ decay amplitude by including $\pi\pi$ final-state interactions, in which the subtraction constants in the $\pi\pi$-loop functions are determined by fitting the $\pi\pi$ phase shifts.

To ensure a consistent phenomenological analysis, we re-determine the relevant weak chiral couplings by fitting the experimental results of the $K\to\pi\pi$ decays within the same unitarization framework.
By comparing with the SU(3) results, we find that in the $K\to\pi a$ decay, although the $\eta_0$ degree of freedom itself does not induce significant corrections, the new U(3) weak chiral operator could give a pronounced contribution. We investigate the effects of the intermediate $\pi\pi$ interactions in the $K\to\pi a$ decays through the chiral unitarization approach. It is found that the $\pi\pi$ loops play marginal roles in the $K^-\to\pi^-a$ decay, since it is a $|\Delta I|=1/2$ process, while only the $\pi\pi$ intermediate states with $I=2$ (belonging to $|\Delta I|=3/2$) can appear in this process. By contrast, the $\pi\pi$ loop effects are significant in the $\bar{K}^0\to\pi^0a$ decay, where the $\pi\pi$ intermediate states with $I=0$ can enter.

Under the assumption that the ALPs exclusively decay into the $\gamma\gamma$ channel, we give various updated bounds on the ALP couplings from the $K\to\pi a$ decays at different ALP masses $m_a$. The most stringent bounds on the ALPs are derived from the experimental measurements of the 
$K^+\to\pi^++\,invisible$ processes by NA62 in the ALP mass range $0< m_a<110$~MeV and $150~{\rm MeV}<m_a<260~{\rm MeV}$. The $K_L\to\pi^0 a$ decay is constrained by the KOTO measurements of $K_L\to\pi^0+\, invisible$, providing a stringent bound on the imaginary part of the vector flavor-changing ALP coupling $(k_d+k_D)_{12}$. For the flavor-conserving ALP couplings, however, the corresponding bounds in $K_L\to\pi^0+\,invisible$ are much weaker than those from the $K^+\to\pi^++\, invisible$ channel, since the ALP signal is suppressed by the CP violation in the $K_L\to\pi^0 a$ decay.
The ALP bounds are also obtained in the visible decays of $K^+\to\pi^+\gamma\gamma$ and $K_S\to\pi^0\gamma\gamma$, for the ALP mass ranges around $210~{\rm MeV}\lesssim m_a \lesssim 350~{\rm MeV}$. Although such exclusion bounds from the visible decay processes are less stringent than those from the $K^+\to\pi^++\, invisible$ case, they can provide extended bounds for larger ALP masses.

\section*{Acknowledgements}

This work is partially supported by Hebei Natural Science Foundation under Grants No.~A2026205016, by National Natural Science Foundation of China (NSFC) under Grants No.~12475078, No.~12150013, No.~11975090, and also by the Science Foundation of Hebei Normal University with Contract No.~L2023B09.

\appendix

\section{Large-$N_C$ behavior of the $K\to\pi a$ decay amplitude}\label{sec.nckpia}

\begin{figure}[t]
	\centering
	\subfigure[\label{fig_Ktopia-LNC-a}]
	{\includegraphics[width=0.3\linewidth]{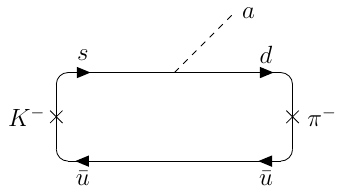}}
	\hfill
	\subfigure[\label{fig_Ktopia-LNC-b}]
	{\includegraphics[width=0.3\linewidth]{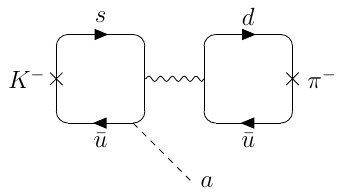}}
	\hfill
	\subfigure[\label{fig_Ktopia-LNC-c}]
	{\includegraphics[width=0.3\linewidth]{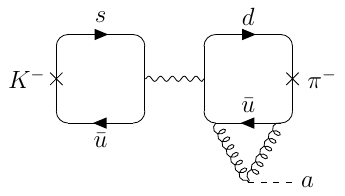}}
	\\
	\subfigure[\label{fig_Ktopia-LNC-d}]
	{\includegraphics[width=0.3\linewidth]{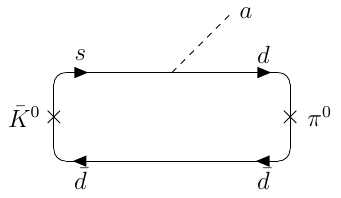}}
	\hfill
	\subfigure[\label{fig_Ktopia-LNC-e}]
	{\includegraphics[width=0.3\linewidth]{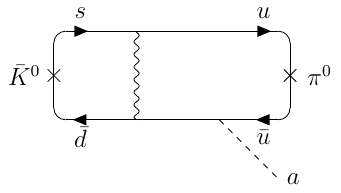}}
	\hfill
	\subfigure[\label{fig_Ktopia-LNC-f}]
	{\includegraphics[width=0.3\linewidth]{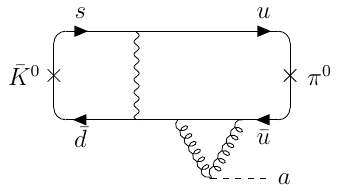}}
	\caption{Schematic diagrams illustrating the orders in $N_C$ of the $\bar{K}\to\pi a$ decay amplitudes.}\label{fig_Ktopia-LNC}
\end{figure}

The U(3) framework can fully reproduce the large-$N_C$ properties of QCD. The large-$N_C$ behavior of the $K\to\pi a$ decay amplitude can therefore serve as a consistency check of our calculation. This behavior can be read off directly from the quark-level Feynman diagrams, as illustrated in Fig.~\ref{fig_Ktopia-LNC}. For the AFC diagrams in Figs.~\ref{fig_Ktopia-LNC-a} and \ref{fig_Ktopia-LNC-d}, one obtains $A_{\bar{K}\to\pi a}^{\rm AFC} = \mO(N_C^0)$. It is obvious that the corresponding AFC amplitudes in Eqs.~(\ref{eq_A_Km_to_pim_a_AFC}) and (\ref{eq_A_Kbar0-to-pi0-a_AFC}) respect this large-$N_C$ scaling. 
The WFC part is more involved and in the following we elaborate in detail the $N_C$ scalings of the WFC amplitudes.

The $N_C$ orders of the ALP-quark and ALP-gluon couplings are given by
\begin{equation}
\vcenter{\hbox{\includegraphics[scale=1]{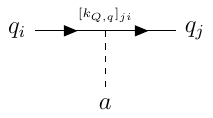}}}\,\sim \mO(N_C^0)\,,
\qquad
\vcenter{\hbox{\includegraphics[scale=1]{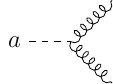}}}\,\sim \mO(N_C^{-1})\,,
\end{equation} 
with the latter being proportional to $\alpha_s$ and therefore suppressed by $N_C^{-1}$. Accordingly, the $c_{GG}$ contribution in $A_{\bar{K}\to\pi a}^{\rm WFC}$ is suppressed by $N_C^{-1}$, compared to the $k_{Q,q}$ contributions. According to Figs.~\ref{fig_Ktopia-LNC-b} and \ref{fig_Ktopia-LNC-c}, the $N_C$ orders of the $k_{Q,q}$ and $c_{GG}$ contributions to $A_{K^-\to\pi^-a}^{\rm WFC}$ are $\mO(N_C)$ and $\mO(N_C^0)$, respectively. Those in the $\bar{K}^0\to\pi^0 a$ decay are further suppressed by $N_C^{-1}$ according to Figs.~\ref{fig_Ktopia-LNC-e} and \ref{fig_Ktopia-LNC-f}. Because $G_W \sim \mO(N_C^2)$, the $N_C$ orders of $F_{\bar{K}\to\pi a}^{\cALP}$ are given by
\begin{equation}
\begin{array}{lll}
F_{K^-\to\pi^-a}^{\cALP}\big|_{\cALP\in\{c^a_{uu},c^a_{dd},c^a_{ss},c^v_{dd}-c^v_{ss}\}}\sim \mO(N_C^{-1})\,,&\quad&
F_{K^-\to\pi^-a}^{c_{GG}}\sim \mO(N_C^{-2})\,,
\\
F_{\bar{K}^0\to\pi^0 a}^{\cALP}\big|_{\cALP\in\{c^a_{uu},c^a_{dd},c^a_{ss},c^v_{dd}-c^v_{ss}\}}\sim \mO(N_C^{-2})\,,&\quad&
F_{\bar{K}^0\to\pi^0a}^{c_{GG}}\sim \mO(N_C^{-3})\,.
\end{array}\label{eq_LNC-Fac}
\end{equation}
Recall that our calculation for the WFC part includes incomplete NLO contributions associated with $g_8'$, $g_8^P$ and $g_8^{\theta}$. The relations between the leading-$N_C$ parts of these weak chiral couplings are unknown in this work. Therefore, we ignore $g_8'$, $g_8^P$ and $g_8^{\theta}$ in this check. Using the leading-$N_C$ coefficients of $g_8$, $g_8^{\psi}$, $g_{27}^{1/2}$ and $g_{27}^{3/2}$ listed in Eq.~(\ref{eq_gi_LN}), we verify that our amplitudes satisfy the large-$N_C$ scaling in Eq.~(\ref{eq_LNC-Fac}).

Since in several places we take the $M_0\to \infty$ limit in the U(3) amplitudes to recover the SU(3) ones,  we also make a tentative investigation of the large-$N_C$ behavior of the amplitudes after taking the limit of $M_0\to\infty$. In this limit, we find that the $k_{Q,q}$ contributions still display the correct large-$N_C$ behavior, while the $c_{GG}$ contribution exhibits an incorrect large-$N_C$ scaling. Specifically, in the limit $M_0\to\infty$, $F_{\bar{K}\to\pi a}^{c_{GG}}$ acquires the same $N_C$ order as the $k_{Q,q}$ contributions 
\begin{equation}
F_{K^-\to\pi^-a}^{c_{GG}}\big|_{M_0\to\infty}\sim\mO(N_C^{-1})\,,
\quad
F_{\bar{K}^0\to\pi^0a}^{c_{GG}}\big|_{M_0\to\infty}\sim\mO(N_C^{-2})\,.
\end{equation}
It is not surprising that one obtains such an incorrect large-$N_C$ behavior at $M_0\to\infty$, since in the strict large-$N_C$ limit, $M_0\to 0$. Alternatively, this result can also be understood via the PCAC relation.
Let us consider an ALP effective Lagrangian containing only the ALP-gluon coupling 
\begin{equation}
\mL=\frac{1}{2}\partial_{\mu}a\partial^{\mu}a-\frac{1}{2}m_{a,0}^2a^2-c_{GG}\frac{\alpha_s}{4\pi}\frac{a}{f}G_{\mu\nu}^a\widetilde{G}^{\mu\nu,a}+\mL_{\rm QCD}\,.
\end{equation}
The ALP couples to quarks through the ALP-gluon coupling, which behaves as $1/N_C$ and can be easily understood via the diagram 
\begin{equation}
\vcenter{\hbox{\includegraphics{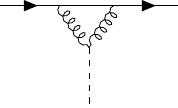}}}\,\sim \mO(N_C^{-1})\,.\label{eq_a-gg-qq}
\end{equation}
By redefining the quark fields through $q\to e^{-i\gamma_5c_{GG}\frac{a}{3f}}q$, one can rotate away the ALP-gluon coupling and obtain another basis 
\begin{equation}
\mL=\frac{1}{2}\partial_{\mu}a\partial^{\mu}a-\frac{1}{2}m_{a,0}^2a^2+\mL_{\rm QCD}
-\frac{a}{3f}c_{GG}\left[\partial_{\mu}\left(\bar{q}\gamma^{\mu}\gamma_5q\right)-2i\bar{q}M\gamma_5q\right]
+\mO\left(\frac{a^2}{f^2}\right)\,,
\end{equation}
in which the ALP couples directly to quark currents. The PCAC relation for the singlet axial-vector current, 
\begin{equation}
\partial_{\mu}\left(\bar{q}\gamma^{\mu}\gamma_5q\right)-2i\bar{q}M\gamma_5q=3\frac{\alpha_s}{4\pi}G\widetilde{G}=\mO(N_C^{-1})\,,\label{eq_PCAC}
\end{equation}
guarantees that the two $\mO(N_C^0)$ couplings cancel each other, so that the resulting large-$N_C$ scaling is consistent with Eq.~(\ref{eq_a-gg-qq}).
In the U(3) framework this cancellation is guaranteed, yielding the correct large-$N_C$ scaling of $F_{\bar{K}\to\pi a}^{c_{GG}}$. However, in the limit $M_0\to\infty$, the $\eta_0$ decouples, then Eq.~(\ref{eq_PCAC}) can no longer be fully reproduced at the meson level, which leads to the incorrect large-$N_C$ scaling of $F_{\bar{K}\to\pi a}^{c_{GG}}$.

\section{The values of phenomenological parameters in this work}
\label{app_para-values}

We summarize the phenomenological parameters used in this work. The meson decay constant is taken as $F = 92.1$~MeV, and the meson masses are
\begin{equation}
\begin{aligned}
&m_{\pi^{\pm}}=m_{\pi^0}=m_{\pi}=138~\MeV\,,
\\
&m_{\eta}=548~\MeV\,,\quad m_{\eta'}=958~\MeV\,,
\\
&m_{K^{\pm}}=494~\MeV\,,\quad m_{K^0}=498~\MeV\,.
\end{aligned}
\end{equation}
The isospin breaking parameter $\delta_I$ is estimated from the experimental squared masses as
\begin{equation}
\begin{aligned}
\delta_I = B_0(m_d - m_u)=&\left(m_{K^0}^2\right)_{\exp}-\left(m_{K^\pm}^2\right)_{\exp}-\left[\left(m_{\pi^0}^2\right)_{\exp}-\left(m_{\pi^\pm}^2\right)_{\exp}\right]
\simeq&5343~\MeV^2\,.
\end{aligned}
\end{equation}

The values of $M_0$ and the $\eta$-$\eta'$ mixing parameters are taken from Ref.~\cite{Gu:2018swy}. Specifically, $M_0$ is fixed at 820~MeV, and using the Fit~B ($F$) values from Tables~2 and 4 of that reference we obtain
\begin{equation}
\beta_{11} = 0.79\pm 0.03\,,
\quad
\beta_{22} = 0.78\pm 0.05\,,
\quad
\beta_{12} = -0.05\pm0.06\,,
\quad
\beta_{21} = 0.36\pm 0.06\,.
\end{equation}

The CKM matrix elements and the Fermi constant are taken from the PDG~\cite{ParticleDataGroup:2024cfk}
\begin{equation}
V_{ud}=0.97367\,,
\quad
V_{us}=0.22431\,,
\quad
G_F=1.1663788\times 10^{-11}~\MeV^{-2}\,,
\end{equation}
from which we obtain
\begin{equation}
G_W = -\frac{F^4G_F}{\sqrt{2}}V_{ud}^*V_{us}=-1.29606\times 10^{-4}~\MeV^2\,.
\end{equation}

\section{The $a\to\gamma\gamma$ decay within U(3) $\ChPT$}
\label{app_atogaga}

The $a\to\gamma\gamma$ decay amplitude can be calculated using U(3) $\ChPT$. The $a\to\gamma\gamma$ decay is an anomalous process. Besides the $a\gamma\gamma$ vertex in  Eq.~(\ref{eq_L_QCD_ALP_kappaq}), given by
\begin{equation}
\mL_{\rm ALP}\big|_{a\gamma\gamma}=-\hat{c}_{\gamma\gamma}\frac{\alpha}{4\pi}\frac{a}{f}F_{\mu\nu}\widetilde{F}^{\mu\nu}\,,
\quad
\hat{c}_{\gamma\gamma}=c_{\gamma\gamma} - 2N_C\mathrm{Tr}\left(\kappa_qQ_q^2\right)c_{GG}\,,\label{eq_vagaga}
\end{equation}
and the LO chiral Lagrangian $\mL_{\rm eff}^{(0)}$ in Eq.~(\ref{eq_L_eff_0}), we also need the unnatural-parity part of the chiral Lagrangian denoted by $\widetilde{\mL}_{\rm eff}$ that contains the Levi-Civita tensor $\varepsilon_{\mu\nu\rho\sigma}$.
In the $\delta$ expansion, the LO piece of $\widetilde{\mL}_{\rm eff}$ is the well known Wess-Zumino-Witten (WZW) term, and its explicit expression in the U(3) case can be written as~\cite{Gao:2022xqz,Gao:2024vkw,Bickert:2020kbn} 
\begin{equation}\label{eq.lagwzw}
\mL_{\rm WZW}^{\phi\gamma\gamma} = -\frac{N_C\alpha}{6\pi F}\varepsilon^{\mu\nu\rho\sigma}\partial_{\mu}A_{\nu}\partial_{\rho}A_{\sigma}
\left(\pi^3+\frac{\beta_{11}+2\sqrt{2}\beta_{21}}{\sqrt{3}}\eta + \frac{\beta_{12}+2\sqrt{2}\beta_{22}}{\sqrt{3}}\eta'\right)\,,
\end{equation}
which is counted as $\mO(N_C p^4)\sim \mO(\delta)$.

\begin{figure}[t]
\centering
\subfigure[\label{fig_atogaga_a}]
{\includegraphics[width=0.3\linewidth]{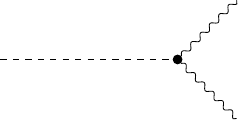}}
\hfill
\subfigure[\label{fig_atogaga_b}]
{\includegraphics[width=0.3\linewidth]{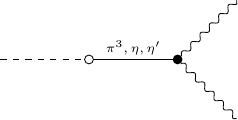}}
\hfill
\subfigure[\label{fig_atogaga_c}]
{\includegraphics[width=0.3\linewidth]{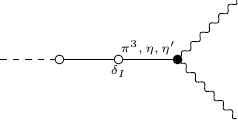}}
\caption{Feynman diagrams for $a\to\gamma\gamma$ decay. The black dot denotes unnatural parity vertex including $\varepsilon_{\mu\nu\rho\sigma}$, while empty dot denotes natural parity vertex from $\mL_{\rm eff}^{(0)}$.}\label{fig_atogaga}
\end{figure}

The Feynman diagrams for $a\to\gamma\gamma$ involve an odd number of unnatural-parity vertices and an arbitrary number of natural-parity vertices (provided by $\mL_{\rm eff}^{(n)}$). We consider the LO contribution, i.e., tree-level diagrams containing either a single $a\gamma\gamma$ vertex from Eq.~(\ref{eq_vagaga}) or a single $\mL_{\rm WZW}$ vertex, together with an arbitrary number of  vertices from $\mL_{\rm eff}^{(0)}$. The relevant Feynman diagrams are shown in Fig.~\ref{fig_atogaga}. The direct $a\gamma\gamma$ vertex in Fig.~\ref{fig_atogaga_a} arises only from Eq.~(\ref{eq_vagaga}), while the $\phi\gamma\gamma$ vertices, with $\phi=\pi^3,\eta,\eta'$, in Figs.~\ref{fig_atogaga_b} and \ref{fig_atogaga_c} are provided in Eq.~\eqref{eq.lagwzw}.

We define the $a\to\gamma\gamma$ decay amplitude as
\begin{subequations}
	\begin{align}
	&\langle\gamma(\lambda_1,k_1)\gamma(\lambda_2,k_2)|i\hat{T}|a(p)\rangle = \left(2\pi\right)^4\delta^4\left(p-k_1-k_2\right)i\mathcal{M}_{a\to\gamma\gamma}\,, 
	\\
	&i\mathcal{M}_{a\to\gamma\gamma}=i\mathcal{M}_{a\to\gamma\gamma}^{\mu\nu}\varepsilon_{\mu}^*(\lambda_1,k_1)\varepsilon_{\nu}^*(\lambda_2,k_2)\,,
	\end{align}
\end{subequations}
where $\lambda_{1,2}$ and $k_{1,2}$ denote the photon polarizations and momenta, and $\varepsilon_{\mu}^*(\lambda_1,k_1)$, $\varepsilon_{\nu}^*(\lambda_2,k_2)$ are the corresponding polarization vectors. Using the LSZ reduction formula, the LO $a\to\gamma\gamma$ decay amplitude is obtained as
\begin{align}
&i\M_{a\to\gamma\gamma}^{\mu\nu}=A_a^{\mu\nu} + x_{a\pi^3}A_{\pi^3}^{\mu\nu} + x_{a\eta}A_{\eta}^{\mu\nu} + x_{a\eta'}A_{\eta'}^{\mu\nu}\,,\label{eq_Matogaga}
\\
&x_{a\pi^3}=\frac{\sigma_{a\pi^3}(m_a^2)}{m_a^2-m_{\pi}^2}+\frac{\sigma_{a\eta}(m_a^2)\sigma_{\pi^3\eta}}{(m_a^2-m_{\pi}^2)(m_a^2-m_{\eta}^2)}+\frac{\sigma_{a\eta'}(m_a^2)\sigma_{\pi^3\eta'}}{(m_a^2-m_{\pi}^2)(m_a^2-m_{\eta'}^2)}\,,
\\
&x_{a\eta}=\frac{\sigma_{a\eta}(m_a^2)}{m_a^2-m_{\eta}^2}+\frac{\sigma_{a\pi^3}(m_a^2)\sigma_{\pi^3\eta}}{(m_a^2-m_{\pi}^2)(m_a^2-m_{\eta}^2)}\,,
\\
&x_{a\eta'}=\frac{\sigma_{a\eta'}(m_a^2)}{m_a^2-m_{\eta'}^2}+\frac{\sigma_{a\pi^3}(m_a^2)\sigma_{\pi^3\eta'}}{(m_a^2-m_{\pi}^2)(m_a^2-m_{\eta'}^2)}\,,
\end{align}
where $\sigma_{a\pi^3,a\eta,a\eta'}(m_a^2)$ and $\sigma_{\pi^3\eta,\pi^3\eta'}$ are the mixing matrix elements derived from the LO quadratic terms in Eq.~(\ref{eq_L_quadratic}) 
\begin{subequations}
	\begin{align}
	&\sigma_{a\pi^3}(m_a^2)=\frac{F}{f}\left(k_{a\pi^3}m_a^2-m_{a\pi^3}^2\right)\,,
	\\
	&\sigma_{a\eta}(m_a^2)=\frac{F}{f}\left[\left(\beta_{11}k_{a\eta_8}+\beta_{21}k_{a\eta_0}\right)m_a^2-\left(\beta_{11}m_{a\eta_8}^2+\beta_{21}m_{a\eta_0}^2\right)\right]\,,
	\\
	&\sigma_{a\eta'}(m_a^2)=\frac{F}{f}\left[\left(\beta_{22}k_{a\eta_0}+\beta_{12}k_{a\eta_8}\right)m_a^2-\left(\beta_{22}m_{a\eta_0}^2+\beta_{12}m_{a\eta_8}^2\right)\right]\,,
	\\
	&\sigma_{\pi^3\eta}=-\delta_I\left(\frac{1}{\sqrt{3}}\beta_{11}+\sqrt{\frac{2}{3}}\beta_{21}\right)\,,
	\\
	&\sigma_{\pi^3\eta'}=-\delta_I\left(\sqrt{\frac{2}{3}}\beta_{22}+\frac{1}{\sqrt{3}}\beta_{12}\right)\,,
	\end{align}
\end{subequations}
and $A_{a,\pi^3,\eta,\eta'}^{\mu\nu}$ are the amputated amplitudes 
\begin{subequations}
	\begin{align}
	&A_a^{\mu\nu}=i\hat{c}_{\gamma\gamma}\frac{\alpha}{\pi f}\varepsilon^{\rho\mu\sigma\nu}k_{1\rho}k_{2\sigma}\,,
	\\
	&A_{\pi^3}^{\mu\nu}=i\frac{N_C\alpha}{3\pi F}\varepsilon^{\rho\mu\sigma\nu}k_{1\rho}k_{2\sigma}\,,
	\\
	&A_{\eta}^{\mu\nu}=i\frac{N_C\alpha}{3\pi F}\frac{\beta_{11}+2\sqrt{2}\beta_{21}}{\sqrt{3}}\varepsilon^{\rho\mu\sigma\nu}k_{1\rho}k_{2\sigma}\,,
	\\
	&A_{\eta'}^{\mu\nu}=i\frac{N_C\alpha}{3\pi F}\frac{\beta_{12}+2\sqrt{2}\beta_{22}}{\sqrt{3}}\varepsilon^{\rho\mu\sigma\nu}k_{1\rho}k_{2\sigma}\,.
	\end{align}
\end{subequations}
Each of the four terms in Eq.~(\ref{eq_Matogaga}) depends on $\kappa_q$ individually. However, their $\kappa_q$ dependence cancels completely at LO, yielding a physical amplitude that is independent of $\kappa_q$.

The amplitude $\M_{a\to\gamma\gamma}^{\mu\nu}$ can be further written as
\begin{equation}
\M_{a\to\gamma\gamma}^{\mu\nu}=\frac{\alpha}{\pi f}\varepsilon^{\mu\nu\rho\sigma}k_{1\rho}k_{2\sigma}F_{a\to\gamma\gamma}\,,
\end{equation}
where the factor $F_{a\to\gamma\gamma}$ is obtained at LO in U(3) $\ChPT$ as
\begin{equation}\label{eq_Fatogaga}
\begin{aligned}
F_{a\to\gamma\gamma} = &
-c_{\gamma\gamma}-c_{GG} \left[\frac{2 N_C}{9}x_2(M_0^2) \left(2-\frac{\delta_I}{m_a^2-m_{\pi}^2}\right)+\frac{\sqrt{2} N_C}{9}x_3(M_0^2) \left(1-\frac{\delta_I}{m_a^2-m_{\pi}^2}\right)\right]
\\
&-c^a_{uu} \Bigg[
\frac{N_Cm_a^2 }{6 \left(m_a^2-m_{\pi}^2\right)}+\frac{N_C}{18}  x_1(m_a^2)+\frac{2 N_C}{9} x_2(m_a^2)+\frac{N_C }{3 \sqrt{2}}x_3(m_a^2)
\\
&-\frac{\delta_I }{m_a^2-m_{\pi}^2}\left(\frac{N_C}{9}  x_1(m_a^2)+\frac{N_C}{3}  x_2(m_a^2)+\frac{5 N_C}{9 \sqrt{2}}x_3(m_a^2)\right)
\Bigg]
\\
&+c^a_{dd} \Bigg[
\frac{N_Cm_a^2 }{6 \left(m_a^2-m_{\pi}^2\right)}-\frac{N_C}{18}  x_1(m_a^2)-\frac{2 N_C}{9} x_2(m_a^2)-\frac{N_C }{3 \sqrt{2}}x_3(m_a^2)
\\
&-\frac{\delta_I }{m_a^2-m_{\pi}^2}\left(\frac{N_C}{9}  x_2(m_a^2)+\frac{N_C }{9 \sqrt{2}}x_3(m_a^2)\right)
\Bigg]
\\
&+c^a_{ss} \Bigg[
\frac{N_C}{9}  x_1(m_a^2)-\frac{2 N_C}{9}x_2(m_a^2)+\frac{N_C }{3 \sqrt{2}}x_3(m_a^2)
\\
&-\frac{\delta_I }{m_a^2-m_{\pi}^2}\left(\frac{N_C}{9}  x_1(m_a^2)-\frac{N_C}{9}  x_2(m_a^2)+\frac{N_C }{9 \sqrt{2}}x_3(m_a^2)\right)
\Bigg]\,,
\end{aligned}
\end{equation}
with the functions $x_{1,2,3}(m^2)$ defined in Eq.~(\ref{eq_x}). The $a\to\gamma\gamma$ decay width then follows as
\begin{equation}
\Gamma_{a\to\gamma\gamma} = \frac{\alpha^2m_a^3}{64\pi^3f^2}\left|F_{a\to\gamma\gamma}\right|^2\,.
\end{equation}
In the limits $M_0\to\infty$ and $m_s\to\infty$,  Eq.~(\ref{eq_Fatogaga}) is reduced to be 
\begin{equation}
F_{a\to\gamma\gamma}\big|_{M_0\to\infty,m_s\to\infty}=-c_{\gamma\gamma}+c_{GG}\left(\frac{5}{3}+\frac{\delta_I}{m_{\pi}^2}\right)+\frac{m_a^2}{m_{\pi}^2-m_a^2}\left(c_{GG}\frac{\delta_I}{m_{\pi}^2}+\frac{c^a_{uu}-c^a_{dd}}{2}\right)\,,
\end{equation}
which is consistent with the SU(2) result of Ref.~\cite{Bauer:2021mvw}. 

\bibliography{reference}
\bibliographystyle{apsrev4-2}
	
\end{document}